\documentclass[seceq]{ptptex}
\usepackage[dvips]{graphicx}
\usepackage{wrapft}

\def\lesim{\m@thcombine<\sim}
\def\gesim{\m@thcombine>\sim}
\def\lessgtr{\m@thcombine<>}
\def\gtrless{\m@thcombine><}

\newcommand{\bra}[1]{\left\langle #1 \right|}

\newcommand{\ket}[1]{\left| #1 \right\rangle}

\newcommand{\adag}{a^{\dagger}}

\newcommand{\Ahat}{\hat{A}}

\newcommand{\Ahatdag}{\hat{A}^{\dagger}}
\newcommand{\Bdag}{B^{\dagger}}
\newcommand{\Bhat}{\hat{B}}
\newcommand{\Bhatdag}{\hat{B}^{\dagger}}

\newcommand{\cdag}{c^{\dagger}}

\newcommand{\Hhat}{\hat{H}}

\newcommand{\hhat}{\hat{h}}
\newcommand{\fb}{{\bf f}}
\newcommand{\fp}{f^{(+)}}

\newcommand{\fm}{f^{(-)}}
\newcommand{\Fhat}{\hat{F}}
\newcommand{\Fhatdag}{\hat{F}^\dagger}
\newcommand{\Fhatp}{\hat{F}^{(+)}}
\newcommand{\Fhatm}{\hat{F}^{(-)}}

\newcommand{\Hc}{{\cal H}}

\newcommand{\jp}{j^{\prime}}

\newcommand{\Nhat}{\hat{N}}
\newcommand{\nL}[1]{n_{L_{#1}}}
\newcommand{\nK}[1]{n_{K_{#1}}}
\newcommand{\nKb}{\mbox{\boldmath $n_K$}}
\newcommand{\nLb}{\mbox{\boldmath $n_L$}}

\newcommand{\Dhat}{\hat{D}}
\newcommand{\Ghat}{\hat{G}}

\newcommand{\Qhat}{\hat{Q}}
\newcommand{\Rhat}{\hat{R}}
\newcommand{\Phat}{\hat{P}}

\newcommand{\That}{\hat{\Theta}}

\newcommand{\Fp}{F^{(+)}}
\newcommand{\Fm}{F^{(-)}}
\newcommand{\Rp}{R^{(+)}}

\newcommand{\Ab}{\mbox{\boldmath $A$}}
\newcommand{\Abdag}{\mbox{\boldmath $A$}^{\dagger}}

\newcommand{\Nb}{\mbox{\boldmath $N$}}

\newcommand{\del}{\partial}

\newcommand{\beq}{\begin{equation}}
\newcommand{\beqa}{\begin{eqnarray}}
\newcommand{\eeq}{\end{equation}}
\newcommand{\eeqa}{\end{eqnarray}}

\newcommand{\Zhat}{\hat{Z}}

\newcommand{\Kp}{K^{+}}
\newcommand{\Km}{K^{-}}
\newcommand{\Kz}{K^0}

\newcommand{\Lp}{L^{+}}
\newcommand{\Lm}{L^{-}}
\newcommand{\Lz}{L^0}

\newcommand{\SB}{{\bf S}}

\title{
Effects of Time-Odd Components in Mean Field on
Large Amplitude Collective Dynamics
}

\author{Nobuo HINOHARA,$^1$~~ Takashi NAKATSUKASA,$^2$~~ 
Masayuki MATSUO$^3$ \\ 
and~ Kenichi MATSUYANAGI$^1$}
\inst{
{\small\it $^1$ Department of Physics, Graduate School of Science,}\\
{\small\it Kyoto University, Kitashirakawa, Kyoto 606-8502, Japan}\\
{\small\it $^2$ Institute of Physics and Center for Computational Sciences, 
University of Tsukuba,}\\
{\small\it Tsukuba 305-8571, Japan}\\
{\small\it $^3$ Department of Physics, Faculty of Science,}\\
{\small\it Niigata University, Niigata 950-2181, Japan}
}

\abst{
We apply the adiabatic self-consistent collective coordinate (ASCC) 
method to the multi-$O(4)$ model and study collective mass (inertia function)
of the many-body tunneling motion.
Comparing results with those of the exact diagonalization,
we show that the ASCC method succeeds in describing gradual
change of excitation spectra from an anharmonic vibration about the
spherical shape to a doublet pattern 
associated with a deformed double-well potential
possessing the oblate-prolate symmetry.
The collective mass is significantly increased
by the quadrupole-pairing contribution to time-odd components of
the moving mean field.
In contrast, the cranking (Inglis-Belyaev) mass based on
the constrained mean field, which ignores the time-odd components, 
is smaller than the ASCC mass and fails to reproduce the exact spectra.
}

\begin{document}

\maketitle

\section{Introduction}

Microscopic theories of large-amplitude collective dynamics
are a long-standing and fundamental subject of nuclear structure physics
\cite{rin80,bla86,abe83}.
Though many theories have been proposed and tested for a variety of 
phenomena of the large-amplitude collective motion
\cite{row76,bri76,vil77,mar77,bar78,goe78,mar80,gia80,dob81,goe81,muk81,
row82,fio83,rei84,kur83,yam84,mat85,mat86,shi87,yam87,
bul89,wal91,kle91a,kan94,nak98a,nak98b,nak00,lib99,yul99,pro04,alm04a,alm04b},
still many theoretical problems remains not being fully clarified 
\cite{kle91b,dan00,kur01}.
In particular, the microscopic determination of collective mass 
(inertia of the collective motion) is being a difficult and controversial
issue. 

The Inglis-Belyaev cranking mass \cite{rin80}, which is derived by using the 
adiabatic perturbation theory, has been widely used in the literature.
It has been known that values of the cranking mass are systematically 
too low to reproduce experimental data of the low-frequency 
$K^{\pi}=0^+$ vibrational modes in deformed nuclei\cite{dud80}.
The Inglis-Belyaev cranking mass does not take account of
contributions of the residual interactions.
Especially, effects of the time-odd components of the moving mean-field are
completely neglected.
Importance of such time-odd contributions to the collective 
mass has been discussed
in connection with the Thouless-Valatin moment of inertia 
for rotational motions\cite{tho62,dob95} and in the context of 
the adiabatic time-dependent Hartree-Fock (ATDHF) theory
\cite{bri76,bar78,gia80,dob81,rei84}.
It is also well known that, without taking into account 
the time-odd components, one cannot obtain the correct mass 
for the center of mass motion of a nucleus. 
In Ref.~\citen{gia80}, the contribution from the time-odd components 
were evaluated on the basis of the ATDHF theory 
and its importance was demonstrated for the isoscalar giant quadrupole modes.
In this work, however, the pairing correlations were not taken into account.
It should be noted that, 
in the time-dependent Hartree-Fock-Bogoliubov (TDHFB) theory 
for nuclei with superfluidity, the pairing correlations also cause the time-odd 
components in the moving mean field.
To the best of our knowledge, however, there are very few papers\cite{dob81} 
trying to evaluate the time-odd effects of the pairing correlations 
on large-amplitude collective dynamics.
This is rather surprising because we know that the pairing correlations
play crucial roles in low-energy nuclear collective dynamics
(see e.g., Refs.~\citen{bar90,ber94}).   
Thus, in this paper, we develop a microscopic theory of large-amplitude 
collective motions in nuclei with superfluidity 
and discuss the time-odd mean-field effects, associated with 
the pairing correlations, on the inertia mass of collective motion.   

We approach to this goal on the basis of
the self-consistent collective coordinate (SCC) method\cite{mar80}
and its extension to include the pairing correlations\cite{mat86}.
This method is based on the TDHFB theory and
enables us to extract, in a fully self-consistent manner, 
the optimum collective coordinate and the collective momentum 
from the huge-dimensional TDHFB phase space.
This feature of the SCC method is in marked contrast to the widely used 
generator coordinate method\cite{rin80} in which the collective coordinate
(generator coordinate) is chosen in a phenomenological manner.
The SCC method has been successfully applied 
to various kinds of low-frequency anharmonic vibrations and high-spin 
rotational motions\cite{mat84,mat85a,mat85b,tak89,yam89a,yam89b,aib90,
yam91,ter91,ter92,mat92,shi01}. 
On the other hand, for genuine large-amplitude cases,
such as the nuclear fission and the shape coexistence
phenomena\cite{woo92,fis00,bou03},
a practical scheme of solving its basic equations had not been available.
A possible solution was proposed in Ref. \citen{mat00};
the adiabatic approximation of the SCC method, 
called the adiabatic SCC (ASCC) method.
The ASCC method was first applied to the solvable multi-$O(4)$ model
and its feasibility was tested~\cite{kob03}.
The multi-$O(4)$ model may be regarded as a simplified version of the 
pairing plus quadrupole (P+Q) interaction model\cite{bar65,bar68,bes69}, 
and it has been used as 
a testing ground of microscopic theories of nuclear collective motions
\cite{mat82,miz81,suz88,fuk91}.
Recently, Kobayasi {\it et al.} applied the ASCC method to the
oblate-prolate shape coexistence phenomena in $^{68}$Se and $^{72}$Kr
\cite{fis00,bou03},
and successfully extracted a collective path 
connecting the oblate and prolate equilibrium points\cite{kob04}.
It turned out that the collective path runs through the triaxially deformed 
region and it almost coincides with the valley line
in the deformation-energy surface obtained by the constrained 
Hartree-Bogoliubov (HB) calculation.
Similar studies have been also carried out by Almehed and Walet
\cite{alm04a,alm04b}.
In these works, the P+Q interaction Hamiltonian was adopted.
In order to study the time-odd contributions to the collective mass, 
however, we need to use a more general Hamiltonian including, 
for instance, the quadrupole-pairing interaction, 
because the time-odd contributions from the P+Q interactions 
are known to vanish\cite{bar78,dob81,bel65}. 
The importance of the quadrupole-pairing interaction for low-frequency 
collective excitations has been well known (see references cited 
in the review~\citen{shi01}).

In this paper, we extend the multi-$O(4)$ model to include the
quadrupole-type pairing interaction.
Varying its strength within a range consistent with experimental data
for the ratio $\Delta_2/\Delta_0$ of the quadrupole and the monopole 
pairing gaps\cite{shi01}, we evaluate its effect on the collective mass 
through the time-odd part of the mean field.

This paper is organized as follows: 
In \S2, the basic equations of the ASCC method are briefly reviewed.
In \S3, we apply the ASCC method to a new version of 
the multi-$O(4)$ model with the quadrupole-type pairing interaction.
A numerical algorithm to solve the ASCC equations for the multi-$O(4)$ model
is presented.
In \S4, we present results of numerical analysis.
Conclusions are given in \S5. 
A preliminary version of this work was reported previously 
in Ref.~\citen{hin05}.

\section{Basic equations of the ASCC method}
\label{sec:basic-eq}

The SCC method enables us to extract the ``optimal'' collective 
subspace from the huge-dimensional TDHFB space\cite{mar80}.
The large amplitude collective motion takes place in this collective
subspace. In the case that the collective subspace is parametrized by 
a pair of collective variables,
a single collective coordinate $q$ and a momentum $p$,
it is called the collective path.
In superconducting nuclei, 
we need an additional set of collective variables to take into account
the pairing rotational degrees of freedom, 
which recovers the particle number symmetry broken 
by the mean field approximation\cite{mat86}.
The pairing rotation is described by the particle-number variable $N$
and its canonically conjugate gauge angle $\varphi$.
Thus, a TDHFB state vector on the collective path
is parametrized by four collective variables ($q, p$,
$\varphi$, $N$).
In the SCC method,
the large-amplitude collective motion is assumed to be approximately 
decoupled with other non-collective degrees of freedom.
This assumption is called 
``the maximal decoupling condition of the collective submanifold.'' 
The time dependence of the TDHFB state on the collective path is then 
determined by this set of collective variables. 

The time-dependent variational principle for the collective motion is 
described as 
\begin{align}
\delta\bra{\phi(q,p,\varphi, N)} i\frac{\del}{\del t} 
-{\hat H} \ket{\phi(q,p,\varphi, N)}=0,
\label{eq:TDVP}
\end{align}
where $\ket{\phi(q,p,\varphi,N)}$ represents 
a TDHFB state vector on the collective path 
and the variation is taken for all possible deviations around it.
The intrinsic state $\ket{\phi(q,p,N)}$ 
in the gauge space (associated with the pairing rotation) is defined by
\begin{align}
\label{eq:pair_rot}
\ket{\phi(q,p,\varphi,N)}=e^{-i \varphi \Nhat}\ket{\phi(q,p,N)}.
\end{align}
These four collective variables are required to satisfy the 
following canonical variable conditions,
\begin{align}
&\bra{\phi(q,p,N)}i\frac{\del}{ \del q}\ket{\phi(q,p,N)} = p, 
&\bra{\phi(q,p,N)} \frac{\del}{i\del p}\ket{\phi(q,p,N)} = 0, 
\label{eq:cvc1}
\\ 
&\bra{\phi(q,p,N)}        \Nhat        \ket{\phi(q,p,N)} = N 
\equiv N_0 + n,
&\bra{\phi(q,p,N)} \frac{\del}{i\del N}\ket{\phi(q,p,N)} = 0. 
\label{eq:cvc2}
\end{align}
Here, $N_0$ denotes the number of particles of the system
~(we assume a single kind of fermions, for simplicity),
and the difference $n=N-N_0$ represents the dynamical number fluctuation.

We now introduce the adiabatic approximation to the 
TDHFB state $\ket{\phi(q,p,N)}$:
We assume that the collective momentum and the number fluctuation are 
small and that the moving state
$\ket{\phi(q,p,N)}$ is close to
$\ket{\phi(q)}=\ket{\phi(q,p=0,N=N_0)}$.
Using the fact that an arbitrary TDHFB state vector can be constructed 
from a given TDHFB state vector by a unitary transformation 
(generalized Thouless theorem)\cite{rin80,bla86},
we set up the moving TDHFB state vector
$\ket{\phi(q,p,N)}$ from the static state vector $\ket{\phi(q)}$
in the following form: 
\begin{align}
\ket{\phi(q,p,N)}=e^{i\Ghat(q,p,N)}\ket{\phi(q)}=
e^{ip{\hat Q(q)}+in{\hat \Theta(q)}}|\phi(q)\rangle,
\label{eq:ascc_state}
\end{align}
where $\Qhat(q)$ and $\That(q)$ are one-body operators.
Substituting the moving state vector $\ket{\phi(q,p,N)}$ of this form into 
Eqs.~(\ref{eq:cvc1}) and (\ref{eq:cvc2}),
and comparing the coefficients of the first order in $p$ and $n$, 
we get 
\begin{align}
\bra{\phi(q)}[\Qhat(q),\Phat(q)]\ket{\phi(q)} &= i, \label{eq:cvcqp} \\
\bra{\phi(q)}[\That(q),\Nhat   ]\ket{\phi(q)} &= i, \label{eq:cvctn}
\end{align}
where $\Phat(q)$ is the local shift operator of the collective
coordinate $q$, defined by
\begin{align}
 \Phat(q)\ket{\phi(q)} = i \frac{\del}{\del q}\ket{\phi(q)}.
\end{align}
Other commutation relations such as 
$\bra{\phi(q)}[\Qhat(q), \Nhat]\ket{\phi(q)}$ and $\bra{\phi(q)}[\Nhat,\Phat(q)]\ket{\phi(q)}$ are zero.

The collective Hamiltonian is also expanded up to the second order in $p$,
\begin{align}
 \Hc(q,p,N) &\equiv \bra{\phi(q,p,N)}\Hhat\ket{\phi(q,p,N)} \nonumber \\
            &= V(q) + \frac{1}{2}B(q) p^2 + \lambda(q)n,
\label{eq:coll_h}
\end{align}
where
\begin{align}
 V(q) &= \Hc(q,p,N)|_{p=0,N=N_0} = \bra{\phi(q)}\Hhat\ket{\phi(q)},
 \label{eq:pot}\\
 B(q) &= \frac{\del^2\Hc}{\del p^2}\Big\arrowvert_{p=0,N=N_0}
 =
           -\bra{\phi(q)}[[\Hhat,\Qhat(q)],\Qhat(q)]\ket{\phi(q)},
 \label{eq:B(q)}\\
 \lambda(q) &=\frac{\del\Hc(q,p,N)}{\del N}\Big\arrowvert_{p=0,N=N_0}
                 =
 \bra{\phi(q)}[\Hhat,i\That(q)]\ket{\phi(q)}. \label{eq:lambda}
\end{align}
The collective potential $V(q)$ expresses static properties of the
collective path, while the mass function $B(q)$ represents the dynamical
property, i.e., inertia of the collective motion.
The quantity $\lambda(q)$ is regarded as the locally defined chemical potential.

The basic equations of the ASCC method are obtained by performing an
adiabatic expansion of the equation of collective path (\ref{eq:TDVP})
with respect to the collective momentum $p$, 
and requiring the variations to be zero for each order in $p$.
In the zero-th order, we obtain the moving-frame HFB equation:
\begin{align} 
\delta\bra{\phi(q)}\Hhat_M(q) \ket{\phi(q)} = 0, \label{eq:ascc1}
\end{align}
where 
\begin{align}
\Hhat_M(q)=\Hhat - \lambda(q)\Nhat-\frac{\del V}{\del q}\Qhat(q) \label{eq:Hhat_M}
\end{align}
is the moving-frame Hamiltonian.

The first and second orders of the adiabatic expansion of
Eq. (\ref{eq:TDVP}) yield the local harmonic equations 
(also called  the moving-frame quasiparticle RPA equation)\cite{mat00}. 
They are composed of the following two equations:
\begin{align}
\delta\bra{\phi(q)}[\Hhat_M(q), \Qhat(q) ] -\frac{1}{i}B(q)\Phat(q)
\ket{\phi(q)} = 0, \label{eq:ascc2}
\end{align}
\begin{align}
\delta\bra{\phi(q)} [\Hhat_M(q), \frac{1}{i}\Phat(q)] -C(q)\Qhat(q)
&-\frac{1}{2B(q)}[[\Hhat_M(q), \frac{\del V}{\del q}\Qhat(q)], \Qhat(q)]
\nonumber\\
&-\frac{\del\lambda}{\del q} \Nhat \ket{\phi(q)} = 0,  
\label{eq:ascc3}
\end{align}
where $C(q)$ represents local stiffness defined on the collective path,
\begin{align}
C(q) = \frac{\del^2 V}{\del q^2} 
+ \frac{1}{2B(q)}\frac{\del B}{\del q}\frac{\del V}{\del q}.
\end{align}
The collective variables $(q,p)$, and the collective Hamiltonian $\Hc(q,p,N)$
are determined by solving the ASCC equations
(\ref{eq:ascc1}), (\ref{eq:ascc2}), and (\ref{eq:ascc3}).

\section{Application of the ASCC to the multi-$O(4)$ model} 
\label{sec:multi-O4-ASCC}

In this section, we present an explicit form
of the ASCC equations for the multi-$O(4)$ model Hamiltonian.
The ASCC equations become a very simple form for separable interactions.
A numerical algorithm to find a collective path is also discussed.

\subsection{Multi-$O(4)$ model}

The multi-$O(4)$ model Hamiltonian has been used to test the 
validity of various kinds of theories of nuclear collective motion
\cite{mat82,miz81,suz88,fuk91}.
In this paper, we extend this model Hamiltonian such that it 
includes the quadrupole-type pairing interaction in addition to the
monopole-pairing interaction.
We write the model Hamiltonian in the following form:
\begin{align}
\Hhat &= \hhat_0 - \frac{1}{2} G_0 (\Ahatdag \Ahat + \Ahat\Ahatdag)
                      - \frac{1}{2} G_2 (\Bhatdag \Bhat + \Bhat\Bhatdag)
                      - \frac{1}{2} \chi {\Dhat}^{2}, \label{eq:o4Hamiltonian}
\\
\hhat_0 &= \sum_j e^0_j \Nhat_j. \nonumber
\end{align}
The first term in the right-hand side of Eq.~(\ref{eq:o4Hamiltonian}) denotes
the single-particle Hamiltonian giving the spherical single-particle energy
$e^0_j$ for each $j$-shell which possesses
$(2\Omega_j)$-fold degeneracy ($2\Omega_j=2j+1$).
Other terms represent the residual two-body interactions; in due order,
the monopole-pairing interaction, the quadrupole-type pairing interaction, 
and the quadrupole-type particle-hole interaction. 
Their interaction strengths are denoted by $G_0$, $G_2$, and $\chi$, 
respectively.
The operators appearing in this model Hamiltonian are 
defined in terms of the nucleon creation and annihilation operators 
($c_{jm}^{\dag}, c_{jm}$) by
\begin{align}
\Ahatdag = \sum_j \Ahatdag_j,~~
\Bhatdag = \sum_j d_j \Bhatdag_j,~~
\Nhat = \sum_j \Nhat_j,~~
\Dhat = \sum_j d_j \Dhat_j,
\end{align}
where
\begin{align}
\Ahat_j^\dag &= \sum_{m>0} c_{jm}^{\dag} c_{j-m}^{\dag}, &\quad
\Bhat_j^\dag &= \sum_{m>0} {\sigma_{jm}}c_{jm}^{\dag} c_{j-m}^{\dag},\\
\Nhat_j &= \sum_m c_{jm}^{\dag} c_{jm}, &\quad
\Dhat_j &= \sum_m {\sigma_{jm}} c_{jm}^{\dag} c_{jm},
\end{align}
with
\begin{align}
{\sigma_{jm}} =
\left \{
\begin{array}{cc}
 1 &  |m| < \Omega_{j}/2, \\
-1 &  |m|  >   \Omega_{j}/2. 
\end{array}
\right.
\end{align}
Here, the operators $\Ahat$ and $\Nhat$ denote the monopole-pair
and the number operators, while $\Bhat$ and $\Dhat$
represent the simplified quadrupole-pair and quadrupole 
particle-hole operators, respectively.
These operators contain the factors $d_j\sigma_{jm}$ 
which simulate the basic property of 
the quadrupole matrix elements $\bra{jm}r^2 Y_{20}\ket{jm}$ 
in a schematic way. 
Although they are not real quadrupole operators, 
we call them ``quadrupole" for brevity. 
Exact solutions (eigen-energies and eigen-functions) of 
the multi-$O(4)$ model are easily obtained by means of 
the matrix diagonalization method (see Appendix A).

\subsection{Quasiparticle representation}

To solve the ASCC equations, 
it is convenient to work on the quasiparticle basis
locally defined with respect to the state $\ket{\phi(q)}$ 
on the collective path. 
For the multi-$O(4)$ model,
the Bogoliubov transformation to quasiparticle creation and annihilation 
operators, $a_{i}^{\dag}(q)$ and $a_{i}(q)$, satisfying the vacuum condition, 
$a_{i}(q)\ket{\phi(q)}=0$, is written as
\begin{align}           
\left(
\begin{array}{c}
      {\adag_{i}}(q)\\
      {a_{-i}}(q)
\end{array}
\right)
 \equiv
\begin{pmatrix}
           u_{i}(q) & -v_{i}(q)  \\
           v_{i}(q) & u_{i}(q) 
\end{pmatrix}
\left(
\begin{array}{c}
      {\cdag_{i}}\\
      {c_{-i}}
\end{array}
\right).
\label{eq:bogoliubov}
\end{align}
Here, the indices $\pm i$ represent the set of angular momentum quantum
numbers $(j,\pm m)$.

Using the quasiparticle bilinear operators,
\begin{align}
\Abdag_i(q) &= \adag_i(q)\adag_{-i}(q), \\
\Nb_i(q) &= \adag_i(q)a_i(q) + \adag_{-i}(q)a_{-i}(q),
\end{align}
the nucleon bilinear operators $\Ahatdag_{i}(q)$ and $\Nhat_{i}(q)$ 
are rewritten as 
\begin{align}
\Ahatdag_{i}(q) &= u_{i}(q) v_{i}(q) + u_{i}^2(q) \Abdag_i(q) 
 - v_{i}^{2}(q) {\Ab}_{i}(q)
 - u_{i}(q) v_{i}(q) \Nb_{i}(q), \\
\Nhat_{i}(q) &= 2v_{i}^{2}(q)+2u_{i}(q)v_{i}(q)(\Abdag_{i}(q)+\Ab_{i}(q))
 +(u_{i}^{2}(q)-v_{i}^{2}(q))\Nb_{i}(q).
\end{align}
The quasiparticle bilinear operators, $\Abdag_i(q),\Ab_i(q)$,
and $\Nb_i(q)$, satisfy the following commutation relations 
\begin{align}
\left[ \Ab_i(q),\Abdag_{i'}(q) \right]
     &= \delta_{ii'}(1-\Nb_{i}(q)), \\
\left[ \Nb_{i}(q),\Abdag_{i'}(q)\right]
     &= 2\delta_{ii'} \Abdag_{i'}(q).
\end{align}
The particle number $N_0$, the quadrupole deformation $D(q)$, 
the monopole-pairing gap $\Delta_0(q)$, 
and the quadrupole-pairing gap $\Delta_2(q)$ are given by 
the expectation values with respect to the 
mean-field state vector $\ket{\phi(q)}$:
\begin{align}
N_0  & =  \bra{\phi(q)}\Nhat\ket{\phi(q)} = 2\sum_{i>0} v_{i}^{2}(q),
\label{eq:number}\\
D(q) & = \langle\phi(q)|\Dhat|\phi(q)\rangle
       =  2\sum_{i>0} d_{i} \sigma_{i}v_{i}^{2}(q), 
\label{eq:deformation} \\
\Delta_0(q) &= G_0 \bra{\phi(q)}\Ahatdag\ket{\phi(q)} = G_0 \sum_{i>0}
 u_{i}(q)v_{i}(q),
\label{eq:delta0} \\
\Delta_2(q) &= G_2 \bra{\phi(q)}\Bhatdag\ket{\phi(q)} = G_2 \sum_{i>0}
d_{i} \sigma_{i} u_{i}(q) v_{i}(q).
\label{eq:delta2}
\end{align}

Below, we often omit the $q$-dependence in expressions,
like $\Ab_i(q)\rightarrow \Ab_i$.
It should be kept in mind that all these quantities are
locally defined with respect to the quasiparticle vacuum $\ket{\phi(q)}$
and depend on $q$.

\subsection{ASCC equations for separable interactions}

The ASCC equations can be easily solved  
when the effective interactions in the microscopic Hamiltonian are separable.
We can always write such a separable Hamiltonian in the following form:
\begin{align}
\Hhat = \hhat_0
  - \frac{1}{2}\sum_{s}\kappa_{s}\Fhatp_{s}\Fhatp_{s}
  + \frac{1}{2}\sum_{s}\kappa_{s}\Fhatm_{s}\Fhatm_{s},
\label{eq:separable}
\end{align}
where 
\begin{align}
\Fhat_s^{(\pm)} \equiv (\Fhat_s \pm \Fhatdag_s)/2 = \pm \Fhat_s^{(\pm)\dagger}.
\end{align}
The superscripts $(\pm)$ indicate the Hermitian or anti-Hermitian characters of
the bilinear operator $\Fhat$.
The multi-$O(4)$ model Hamiltonian under consideration contains three kind of 
residual interactions.
The indices, $s=$1, 2, and 3, to the operators
$\Fhat$ and the interaction strengths $\kappa_{s}$
represent the monopole-pairing, the quadrupole-pairing
and the quadrupole particle-hole interactions, respectively: 
$\Fhat_{s=1}=A, \Fhat_{s=2}=B$, $\Fhat_{s=3}=\Dhat$,
$\kappa_1=2G_0, \kappa_2=2G_2$, and $\kappa_3=\chi$.

For the separable Hamiltonian, 
it is possible to directly derive the ASCC equations 
from the time-dependent variational principle,
\begin{align}
 \delta\bra{\phi(t)}i\frac{\del}{\del t} - \hhat(t)\ket{\phi(t)}=0, 
\label{eq:mftdvp}
\end{align}
for the self-consistent mean-field Hamiltonian defined by
\begin{align}  
 \hhat(t) = \hhat_0 
        - \sum_s\kappa_s \Fhatp_s \bra{\phi(t)}\Fhatp_s\ket{\phi(t)}
        + \sum_s\kappa_s \Fhatm_s \bra{\phi(t)}\Fhatm_s\ket{\phi(t)}.
\end{align}
Here, the exchange terms are neglected as usual for the separable
interactions \cite{bar65,bes69}.
The second and third terms in the right-hand side of $\hhat(t)$
represent the time-even and time-odd components, respectively.
Note that the expectation values $\bra{\phi(t)}\Fhatm_s\ket{\phi(t)}$ 
are pure imaginary, indicating that the third term is odd 
under time reversal.
Substituting $\ket{\phi(t)} = e^{ip\Qhat(q)}\ket{\phi(q)}$ and 
expanding Eq.~(\ref{eq:mftdvp}) up to the second order in $p$,
we obtain the ASCC equations for the separable Hamiltonian: 
\begin{align}
\delta\bra{\phi(q)}\hhat_M(q) \ket{\phi(q)} = 0,
\label{eq:ascc1_sep}
\end{align}
\begin{align}
\delta\bra{\phi(q)}[\hhat_M(q), \Qhat(q) ] - \sum_s f^{(-)}_{Q,s} \Fhatm_s 
- {1\over i} B(q) \Phat(q) 
     \ket{\phi(q)} = 0,
\label{eq:ascc2_sep}
\end{align}
\begin{align}
\delta\bra{\phi(q)} [\hhat_M(q), {1\over i}B(q)\Phat(q)] 
    - \sum_s f^{(+)}_{P,s}(q) \Fhatp_s 
    - B(q)C(q)\Qhat(q)
    - \sum_s f^{(+)}_{R,s}(q) \Fhatp_s \nonumber \\
    + \sum_s [\Fhatm_s, (\hhat(q)-\lambda(q)\Nhat)_{A}] f^{(-)}_{Q,s} 
    -f_N(q) \Nhat
    \ket{\phi(q)} =0, 
\label{eq:ascc3_sep}
\end{align}
where $\hhat_M(q)$ denotes the self-consistent mean-field Hamiltonian 
in the moving frame, defined by
\begin{equation}
\hhat_M(q) = \hhat(q)  - \lambda(q)\Nhat
                     - {\del V\over \del q}\Qhat(q)
\end{equation}
with
\begin{align}
\hhat(q) &= \hhat_0  
- \sum_s \kappa_s \Fhatp_s \bra{\phi(q)}\Fhatp_s\ket{\phi(q)} .
\end{align}
$(\hhat(q)-\lambda(q)\Nhat)_{A}$ represents the $\Abdag_i$ and $\Ab_i$ 
parts of the operator in the parentheses, and
\begin{align}
 &\fm_{Q,s}(q) = -\kappa_s 
              \bra{\phi(q)}[\Fhatm_s, \Qhat(q)] \ket{\phi(q)}, \label{eq:fQs}\\
 &\fp_{P,s}(q)= 
   \kappa_s \bra{\phi(q)}[\Fhatp_s,\frac{1}{i} B(q)\Phat(q)] \ket{\phi(q)}, 
   \label{eq:fPs}\\
 &\fp_{R,s}(q)= -\frac{1}{2}\kappa_s 
 \bra{\phi(q)}[[\Fhatp_s,(\hhat(q) -\lambda(q)\Nhat)_{A}],\Qhat(q)] 
 \ket{\phi(q)}, \label{eq:fRs}\\
 &f_N(q) = B(q)\frac{\del \lambda}{\del q}. \label{eq:fN}
\end{align}
Note that all matrix elements are real, 
so that $\bra{\phi(q)}\Fhatm_s\ket{\phi(q)}=0$.
In the above equations, the quantities $\fm_{Q,s}$ represent 
the effects of the time-odd components of the mean field, 
which will bring about important effects discussed in the next section.
 
The above ASCC equations can be obtained also by 
inserting the expression (\ref{eq:separable})
into Eqs.~(\ref{eq:ascc2}) and (\ref{eq:ascc3}).
The exchange terms of the residual interactions should be omitted
\cite{bar65,bes69}.
It means that we adopt the Hartree-Bogoliubov (HB) approximation
in place of the HFB approximation. 

\subsection{The moving-frame HB equation}

The moving-frame HB equation (\ref{eq:ascc1_sep}) at a given $q$
determines the static TDHB state $\ket{\phi(q)}$. 
If we know the operator $\Qhat(q)$,
we can solve it using the gradient method.
The quantities $\lambda(q)$ and $\displaystyle\frac{\del V}{\del q}$ 
are regarded as Lagrange multipliers which will be determined by the
following two constraints.
The first constraint is the particle number:
\begin{align}
 \bra{\phi(q)}\Nhat\ket{\phi(q)} = N_0 . \label{eq:Ncond}
\end{align}
This constraint specifies the location in the particle number space.
The second one is 
\begin{align}
 \bra{\phi(q)}\Qhat(q-\delta q)\ket{\phi(q)} = \delta q, \label{eq:Qcond}
\end{align}
which can be readily derived from the canonical variable conditions,
\begin{align}
 \bra{\phi(q)}\Qhat(q)\ket{\phi(q)} = 0
\end{align}
and 
\begin{align}
 \bra{\phi(q)}\frac{\del \Qhat}{\del q}(q)\ket{\phi(q)} = -1.
\end{align}
Equation (\ref{eq:Qcond}) is the condition for the collective coordinate $q$ 
that the ``distance'' from $\ket{\phi(q-\delta q)}$ to $\ket{\phi(q)}$
is equal to $\delta q$.
In numerical calculations,
this quantity, $\delta q$, corresponds to a mesh size
of the discretized collective path.
Provided that $\Qhat(q-\delta q)$ and $\Qhat(q)$ are known, 
we can solve the moving-frame HB equation.
Because the variation in Eq.~(\ref{eq:ascc1_sep}) is taken with respect to
arbitrary two-quasiparticle creations,
\begin{align}
 \delta \ket{\phi(q)} = \adag_{i}\adag_{j} \ket{\phi(q)},
\end{align}
the two-quasiparticle terms proportional to $\Abdag_i$ and $\Ab_i$ 
in the moving-frame Hamiltonian $\hhat_M(q)$ should vanish.

We solve the moving-frame HB equation with the above two constraints 
by means of an algorithm analogous to the gradient method\cite{rin80}.
Details of this algorithm are summarized in Appendix~\ref{sec:grad}.

\subsection{The local harmonic equations}
\label{sec:LHE}

In order to obtain the collective path, we need to solve 
the local harmonic equations, (\ref{eq:ascc2_sep}) and (\ref{eq:ascc3_sep}),
and find the operators, $\Qhat(q)$ and $\Phat(q)$,
which determine the direction of the collective path in the TDHB space.
In solving the local harmonic equations, we note that the moving-frame 
Hamiltonian $\hhat_{M}(q)$ is expressed in terms of the
quasiparticle bilinear operators, $\Abdag_i, \Ab_i$, and $\Nb_i$, in the
following manner:
\begin{align}
\hhat_{M}(q) &= V(q) + \sum_{i>0}E_{i}(q){\Nb}_{i}, \label{eq:hm} \\
\Fhatp_{s} &= \bra{\phi(q)}\Fhatp_{s}\ket{\phi(q)} + 
 \Fhatp_{A,s} + \Fhatp_{B,s} \nonumber \\
&= \bra{\phi(q)} \Fhatp_{s} \ket{\phi(q)}
+\sum_{i>0} \Fp_{A,s}(i)
(\Abdag_{i} + \Ab_{i})+\sum_{i>0} \Fp_{B,s}(i){\Nb}_{i},
 \label{eq:Fhatp} \\
\Fhatm_s  &= \sum_{i>0} \Fm_{A,s}(i)(\Abdag_{i}-\Ab_{i})\label{eq:Fhatm}.
\end{align}
Here, 
\begin{align}  
 \Fp_{A,1}(i)&=\frac{1}{2}(u_{i}^{2}-v_{i}^{2}), &
 \Fp_{A,2}(i)&=\frac{1}{2}d_{i} \sigma_{i} (u_{i}^{2} - v_{i}^{2}), &
 \Fp_{A,3}(i)&=2d_{i}\sigma_{i}u_{i}v_{i}, \\  
 \Fm_{A,1}(i)&=-\frac{1}{2}, &
 \Fm_{A,2}(i)&=-\frac{1}{2}d_{i}\sigma_{i}, &
 \Fm_{A,3}(i)&=0, \\ 
 \Fp_{B,1}(i)&=-u_{i}v_{i},&
 \Fp_{B,2}(i)&=-d_{i}\sigma_{i}u_{i}v_{i} &
 \Fp_{B,3}(i)&= d_{i}\sigma_{i}(u_{i}^{2}-v_{i}^{2}). 
\end{align}
\begin{align}
 E_i(q) = (u^2_i - v^2_i)(e_i -\frac{\chi}{2}d_i\sigma_i
 D(q)-\lambda(q))
-2(\Delta_0(q)+d_i\sigma_i\Delta_2(q))u_i v_i.
\end{align}
These quantities are determined 
by solving the moving-frame HB equation (\ref{eq:ascc1_sep}).
For later convenience, we define the following quasiparticle bilinear operator:
\begin{align}
 \Rhat_s^{(+)}  \equiv 
 [ \Fhat_{B,s}^{(+)}, (\hhat(q)-\lambda(q)\Nhat)_{A}] 
                = 2 \sum_{i>0}R_s^{(+)}(i) ({\Abdag_i} - {\Ab_i}) 
\label{eq:Rs}
\end{align}
with
\begin{align}
R_{s}^{(+)}(i)= &\left\{2u_{i}(q)v_{i}(q)(e_i - \chi d_i \sigma_i D(q)
 -\lambda(q)) \right.\nonumber\\
 &\left.\ - 
(\Delta_0(q)+d_i\sigma_i\Delta_2(q))(u_i^2(q)-v_i^2(q))\right\} F_{B,s}^{(+)}(i).
\end{align}
The infinitesimal generators can be written as
\begin{align}
{\hat Q(q)} &= \sum_{i>0} Q_i({\Abdag_{i}}+{\Ab_{i}}), \label{eq:Qinqp}\\
{\hat P(q)} &= i \sum_{i>0} P_i({\Ab_{i}^{\dag}}-{\Ab_{i}}). \label{eq:Pinqp}
\end{align}
We can write down the matrix elements $Q_i$ and $P_i$ in terms 
of $\fm_{Q,s}$, $\fp_{P,s}$, $\fp_{R,s}$ and $f_N$ 
by substituting Eqs.~(\ref{eq:Qinqp}) and (\ref{eq:Pinqp}) 
into Eqs.~(\ref{eq:ascc2_sep}) and (\ref{eq:ascc3_sep}):
\begin{align}
Q_i &= \frac{2E_{i}}{(2E_{i})^{2}-\omega^2}
\sum_{s} F_{A,s}^{(-)}(i)f_{Q,s}^{(-)}+\frac{1}{(2E_{i})^{2}-\omega^2}
\sum_{s}(F_{A,s}^{(+)}(i)f_{PR,s}^{(+)} + N_{i}f_{N}), 
\label{eq:Qi} \\
P_i &= \frac{2E_{i}}{(2E_{i})^{2}-\omega^2}
\sum_{s}(F_{A,s}^{(+)}(i)f_{PR,s}^{(+)} + N_{i}f_{N})+\frac{\omega^2}{(2E_{i})^{2}-\omega^2}\sum_{s} F_{A,s}^{(-)}(i)f_{Q,s}^{(-)},
\label{eq:Pi}
\end{align}
where
\begin{align}
&N_i = 2 u_i(q) v_i(q), \\
&f_{PR,s}^{(+)}=f_{P,s}^{(+)}(q)+f_{R,s}^{(+)}(q), \\
&\omega =\sqrt{B(q)C(q)}.
\end{align}
Substituting Eqs.~(\ref{eq:Rs}), (\ref{eq:Qinqp}) and
(\ref{eq:Pinqp}) into Eqs.~(\ref{eq:fQs}), (\ref{eq:fPs}) and (\ref{eq:fRs}),
we also have the following relations,
\begin{align}
\fm_{Q,s} &= 2\kappa_{s} \sum_{i>0} \Fm_{A,s}(i)Q_i,  
\label{eq:disp1} \\
\fp_{PR,s} &= 2\kappa_{s} \sum_{i>0} \left\{ \Fp_{A,s}(i)P_i+
\Rp_{s}(i)Q_i\right\} .
\label{eq:disp2}
\end{align}
Note that $\fm_{Q,3}=0$.
From the canonical variable condition, 
the orthogonality between the collective and number fluctuation modes
is required;
\begin{align}
\langle\phi(q)|[\Nhat,\Phat(q)]|\phi(q)\rangle 
          = 2i\sum_{i>0} N_i P_i = 0. \label{eq:disp3}
\end{align}
Eliminating $Q_i$ and $P_i$ from
Eqs.~(\ref{eq:disp1}), (\ref{eq:disp2}), and (\ref{eq:disp3}) 
with the use of Eqs.~(\ref{eq:Qi}) and (\ref{eq:Pi}),
we finally reach a dispersion equation,
\begin{align}
 \SB (\omega^2)\cdot \fb = 0, \label{eq:disp}
\end{align}
for the quantity
$\fb=\fb(q)=\{\fm_{Q,1}$, $\fm_{Q,2}$, $ \fp_{PR,1}$, $\fp_{PR,2}$, 
$\fp_{PR,3}$, $f_N\}$.
Here $\SB=\{S_{ij}\}$ is a $6\times 6$ matrix whose elements are given by
\begin{subequations}
\begin{align}
 S_{11} &= 4 G_0 S^{(1)}(\Fm_{A,1}, \Fm_{A,1}) - 1, &\quad
 S_{12} &= 4 G_0 S^{(1)}(\Fm_{A,1}, \Fm_{A,2}), \\
 S_{13} &= 4 G_0 S^{(2)}(\Fm_{A,1}, \Fp_{A,1}), &\quad
 S_{14} &= 4 G_0 S^{(2)}(\Fm_{A,1}, \Fp_{A,2}), \\
 S_{15} &= 4 G_0 S^{(2)}(\Fm_{A,1}, \Fp_{A,3}), &\quad
 S_{16} &= 4 G_0 S^{(2)}(\Fm_{A,1}, N), 
\end{align}
\end{subequations}
\begin{subequations}
\begin{align}
 S_{21} &= 4 G_2 S^{(1)}(\Fm_{A,2}, \Fm_{A,1}), &\quad
 S_{22} &= 4 G_2 S^{(1)}(\Fm_{A,2}, \Fm_{A,2}) - 1,\\
 S_{23} &= 4 G_2 S^{(2)}(\Fm_{A,2}, \Fp_{A,1}), &\quad
 S_{24} &= 4 G_2 S^{(2)}(\Fm_{A,2}, \Fp_{A,2}), \\
 S_{25} &= 4 G_2 S^{(2)}(\Fm_{A,2}, \Fp_{A,3}), &\quad
 S_{26} &= 4 G_2 S^{(2)}(\Fm_{A,2}, N), 
\end{align}
\end{subequations}
\begin{subequations}
\begin{align}
 S_{31} &= 4 G_0 \{ S^{(1)}(\Rp_1, \Fm_{A,1})     +\omega^2 S^{(2)}(\Fp_{A,1},
 \Fm_{A,1})\}, \\
 S_{32} &= 4 G_0 \{ S^{(1)}(\Rp_1, \Fm_{A,2})     +\omega^2 S^{(2)}(\Fp_{A,1},
 \Fm_{A,2})\}, \\
 S_{33} &= 4 G_0 \{ S^{(1)}(\Fp_{A,1}, \Fp_{A,1}) + S^{(2)}(\Rp_1,
 \Fp_{A,1})\} - 1, \\
 S_{34} &= 4 G_0 \{ S^{(1)}(\Fp_{A,1}, \Fp_{A,2}) + S^{(2)}(\Rp_1,
 \Fp_{A,2})\}, \\
 S_{35} &= 4 G_0 \{ S^{(1)}(\Fp_{A,1}, \Fp_{A,3}) + S^{(2)}(\Rp_1,
 \Fp_{A,3})\}, \\
 S_{36} &= 4 G_0 \{ S^{(1)}(\Fp_{A,1}, N        ) + S^{(2)}(\Rp_1,
 N)\}, 
\end{align}
\end{subequations}
\begin{subequations}
\begin{align}
 S_{41} &= 4 G_2 \{ S^{(1)}(\Rp_2, \Fm_{A,1}) +\omega^2 S^{(2)}(\Fp_{A,2},
 \Fm_{A,1})\}, \\
 S_{42} &= 4 G_2 \{ S^{(1)}(\Rp_2, \Fm_{A,2}) +\omega^2 S^{(2)}(\Fp_{A,2},
 \Fm_{A,2})\}, \\
 S_{43} &= 4 G_2 \{ S^{(1)}(\Fp_{A,2}, \Fp_{A,1}) + S^{(2)}(\Rp_2,
 \Fp_{A,1})\}, \\
 S_{44} &= 4 G_2 \{ S^{(1)}(\Fp_{A,2}, \Fp_{A,2}) + S^{(2)}(\Rp_2,
 \Fp_{A,2})\} - 1, \\
 S_{45} &= 4 G_2 \{ S^{(1)}(\Fp_{A,2}, \Fp_{A,3}) + S^{(2)}(\Rp_2,
 \Fp_{A,3})\}, \\
 S_{46} &= 4 G_2 \{ S^{(1)}(\Fp_{A,2}, N) + S^{(2)}(\Rp_2,
 N)\}, 
\end{align}
\end{subequations}
\begin{subequations}
\begin{align}
 S_{51} &= 2 \chi \{ S^{(1)}(\Rp_3, \Fm_{A,1}) +\omega^2 S^{(2)}(\Fp_{A,3},
 \Fm_{A,1})\}, \\
 S_{52} &= 2 \chi \{ S^{(1)}(\Rp_3, \Fm_{A,2}) +\omega^2 S^{(2)}(\Fp_{A,3},
 \Fm_{A,2})\}, \\
 S_{53} &= 2 \chi \{ S^{(1)}(\Fp_{A,3}, \Fp_{A,1}) + S^{(2)}(\Rp_3,
 \Fp_{A,1})\}, \\
 S_{54} &= 2 \chi \{ S^{(1)}(\Fp_{A,3}, \Fp_{A,2}) + S^{(2)}(\Rp_3,
 \Fp_{A,2})\}, \\
 S_{55} &= 2 \chi \{ S^{(1)}(\Fp_{A,3}, \Fp_{A,3}) + S^{(2)}(\Rp_3,
 \Fp_{A,3})\} - 1,\\
 S_{56} &= 2 \chi \{ S^{(1)}(\Fp_{A,3}, N) + S^{(2)}(\Rp_3,
 N)\}, 
\end{align}
\end{subequations}
\begin{subequations}
\begin{align}
 S_{61} &= \omega^2 S^{(2)}(N, \Fm_{A,1}), &\quad
 S_{62} &= \omega^2 S^{(2)}(N, \Fm_{A,2}), \\
 S_{63} &= S^{(1)}(N, \Fp_{A,1}), &\quad
 S_{64} &= S^{(1)}(N, \Fp_{A,2}), \\
 S_{65} &= S^{(1)}(N, \Fp_{A,3}), &\quad
 S_{66} &= S^{(1)}(N, N). 
\end{align}
\end{subequations}
The quantities $S^{(1)}$ and $S^{(2)}$ are defined by
\begin{align}
 S^{(1)}(X,Y) &= \sum_{i>0} \frac{2E_i(q)}{(2E_i(q))^2 - \omega^2(q)} X_i
 Y_i, \\
 S^{(2)}(X,Y) &= \sum_{i>0} \frac{1}{(2E_i(q))^2 - \omega^2(q)} X_i Y_i.
\end{align}
The unknown quantities in the dispersion equation (\ref{eq:disp})
are $\fb(q)$ and $\omega^2(q)$.
The squared frequency $\omega^2(q)$ can be determined by the 
condition that the matrix $\SB(\omega^2(q))$ cannot have its inverse,
\begin{align}
\det \SB(\omega^2(q)) = 0. \label{eq:det}
\end{align}
When there are many solutions $\omega^2(q)$ satisfying this equation,
we choose the smallest $\omega^2(q)$ solution (including negative value) 
as the collective mode.
Once the $\omega^2(q)$ value and, consequently, 
the matrix  $\SB(q)$ is specified,
the direction of the vector $\fb(q)$ is found. 
On the other hand, its absolute value is fixed 
by the normalization condition for the collective mode, i.e.,
\begin{align}
\langle\phi(q)|[\hat{Q}(q),\Phat(q)]|\phi(q)\rangle 
          = 2i\sum_{i>0} Q_i(q)P_i(q) = i.
\end{align}
Let us note that we can take an arbitrary scale 
for the collective coordinate $q$. 
This means that we can set $B(q)=1$ on the collective path 
without loss of generality.
We adopt this choice in this paper.
Then, $\omega^2(q)$ represents nothing but the curvature
of the collective potential:
\begin{align}
\omega^2(q)=\frac{\partial^2 V(q)}{\partial q^2}.
\label{eq:freq}
\end{align}
There is still an arbitrary choice for the sign of
$\Qhat(q)$ and $\Phat(q)$.
This sign specifies the ``rear'' and ``front'' of one-dimensional
collective path.

\subsection{Numerical algorithm for solving the ASCC equations} 
\label{sec:algorithm}

The infinitesimal generator $\Qhat(q)$ and $\Phat(q)$,
which depends on the quasiparticle vacuum $\ket{\phi(q)}$,
is a solution of the local harmonic equations, while
the quasiparticle vacuum $\ket{\phi(q)}$,
which depends on $\Qhat(q)$,
is a solution of the moving-frame HB equation.
Thus, the set of ASCC equations require self-consistency and an
iterative solution.
In fact, we need a double iteration for each value of $q$, 
because the gradient method itself for solving the moving-frame HB equation   
is an iterative procedure.
The numerical algorithm utilized in the present work is
summarized as follows:

\begin{description}
\item{\it Step 0: HB state (starting point)}\newline
Solve the static HB equation and choose one of the solution.
Then, solve the  quasiparticle RPA equations
and select a collective excitation mode which has the lowest frequency
$\omega(q_0)$.
This provides the solution of the ASCC equations at $q=q_0$.

\item{\it Step 1: Initial setting}\newline
Assume that we have solved the ASCC equations at a position $q$
having the self-consistent generator $\Qhat(q)$ and the state $\ket{\phi(q)}$.

Set $\Qhat^{(0)}(q+\delta q) = \Qhat(q)$ 
as the initial guess for $\Qhat(q+\delta q)$,
then start the following iteration
to find the self-consistent solution at $q+\delta q$.
Here, the superscript $(n)$ indicates the number of iterations.

\item{\it Step 2: Moving-frame HB equation}\newline
Using the operator $\Qhat^{(n-1)}(q+\delta q)$ ($n\geq 1$),
solve the moving-frame HB equation at $q+\delta q$,
\begin{align}
\label{eq:MfHB}
\delta \bra{\phi^{(n)}(q+\delta q)}\Hhat - \lambda^{(n)}\Nhat
- \mu^{(n)} \Qhat^{(n-1)}(q+\delta q) \ket{\phi^{(n)}(q+\delta q)}
= 0 .
\end{align}
Here, the constraints, Eqs. (\ref{eq:Ncond}) and (\ref{eq:Qcond}),
determine the Lagrange multipliers, $\lambda^{(n)}$ and $\mu^{(n)}$.
We use the gradient method described in Appendix \ref{sec:grad} to solve
Eq. (\ref{eq:MfHB}).
This determines the moving-frame HB state $\ket{\phi^{(n)}(q+\delta q)}$.

\item{\it Step 3: Local harmonic equation}\newline
Using $\ket{\phi^{(n)}(q+\delta q)}$
with the Lagrange multipliers $\lambda(q+\delta q)=\lambda^{(n)}$, 
solve the local harmonic equations,
(\ref{eq:ascc2_sep}) and (\ref{eq:ascc3_sep}).
This determines the infinitesimal generator $\Qhat^{(n)}(q+\delta q)$.

\item{\it Step 4: Self-consistency}\newline
Updating the infinitesimal generator $\Qhat^{(n)}(q+\delta q)$,
go back to {\it Step 2}, and repeat {\it Steps 2} and {\it 3}
until all quantities at $q+\delta q$ converge.

\item{\it Step 5: Repetition}\newline
Change $q$ into $q+\delta q$ and go back to {\it Step 1}.
\end{description}

Carrying out these iterations, {\it Steps 1-5}, 
we obtain the collective path starting from the HB equilibrium point 
toward one direction ($q>q_0$).
The collective path toward the opposite direction is obtained by changing 
the sign of $\delta q$ and repeating the same procedure, {\it Steps 1-5}.
In this way, we determine 
the self-consistent collective path passing through the HB equilibrium point.

We should give an additional remark on {\it Step 3}.
When we solve the local harmonic equations,
we set $\fm_{Q,1}(q)=0$ in the iteration procedure to avoid 
a numerical instability problem\cite{hin05}.
We have confirmed that the solutions obtained under this assumption 
satisfy the required self-consistency.
Quite recently, we have found that this prescription can be justified
without loss of generality. 
This recent progress in the ASCC method, including this proof,
will be reported in another paper\cite{hin06}. 

The solution of the ASCC equations yields the classical collective Hamiltonian
(for a constant particle number $N=N_0$):
\begin{align}
 \Hc(q,p) = \frac{1}{2}p^2 + V(q).
\end{align}
We then obtain a quantum collective Hamiltonian by canonical quantization: 
$\Hc(q,p)\to\Hc(q,\displaystyle\frac{1}{i}\frac{\del}{\del q})$.
Note that, in this quantization step, 
there is no ambiguity associated with the ordering of $q$ and $p$, 
because the coordinate scale is chosen 
such that the inverse mass function is unity, i.e., $B(q)=1$.

\section{Numerical calculations and discussions}
\label{sec:numerical-cal}

\subsection{Details of numerical calculation}
\label{sec:parameter}

We solve the ASCC equations following the algorithm in the 
previous section and determine the collective path embedded in 
the TDHB phase space.
In order to investigate the effects of the quadrupole-pairing interaction
on the large-amplitude collective dynamics,
we use the same parameters as those used in Ref.~{\citen{kob03}}
except for the quadrupole-pairing strength $G_2$.
We consider a system composed of three shells with the spherical 
single-particle energies,
$e^0_{j_1} = 0,\ e^0_{j_2} = 1.0,\ e^0_{j_3} = 3.5$, 
the pair degeneracies,
$\Omega_{j_1}=14,\ \Omega_{j_2}=10,\ \Omega_{j_3}=4$,
and the single-particle quadrupole moment,
$d_{j_1} = 2,\ d_{j_2} = d_{j_3} = 1 $ for each shell.
One kind of Fermion is considered in this model and 
the number of particles $N_0$ is set to be 28. This value is 
the half of the total number of shell model states. 
The quadrupole particle-hole interaction strength $\chi$ is fixed 
to 0.04, while three values (0.20, 0.16, 0.14) are employed
for the monopole pairing interaction strength $G_0$.
For each case, effects of the the quadrupole-pairing interaction 
are studied varying its strength $G_2$ (0.00, 0.02, 0.04).
These parameter range are adopted so that the calculated ratios of 
the monopole and quadrupole pairing gaps simulate the result of 
realistic analysis of them\cite{shi01}.
The HB calculation yields a single spherical minimum
for the $G_0=0.20$ case, and two local minima 
corresponding to the oblate and prolate equilibrium shapes 
for the $G_0=0.16$ and $0.14$ cases.
For the latter cases, we present below the results obtained by starting 
from the prolate equilibrium point ($D>0$). 
Of course, we obtain the identical collective path even if we start
the calculation from the oblate equilibrium point ($D<0$).
Note that the multi-$O(4)$ Hamiltonian 
possesses the ``parity'' symmetry (invariance under the transformation 
$\sigma_{jm}\to-\sigma_{jm}$, that is, $\Dhat\to-\Dhat$ and
$\Bhat\to-\Bhat$). 
Therefore, all the quantities  are even or odd functions of the
quadrupole deformation parameter $D$.

\subsection{Collective path and collective potential}
\label{sec:path}

\begin{figure}[htbp]
\begin{center}
\includegraphics[width=100mm]{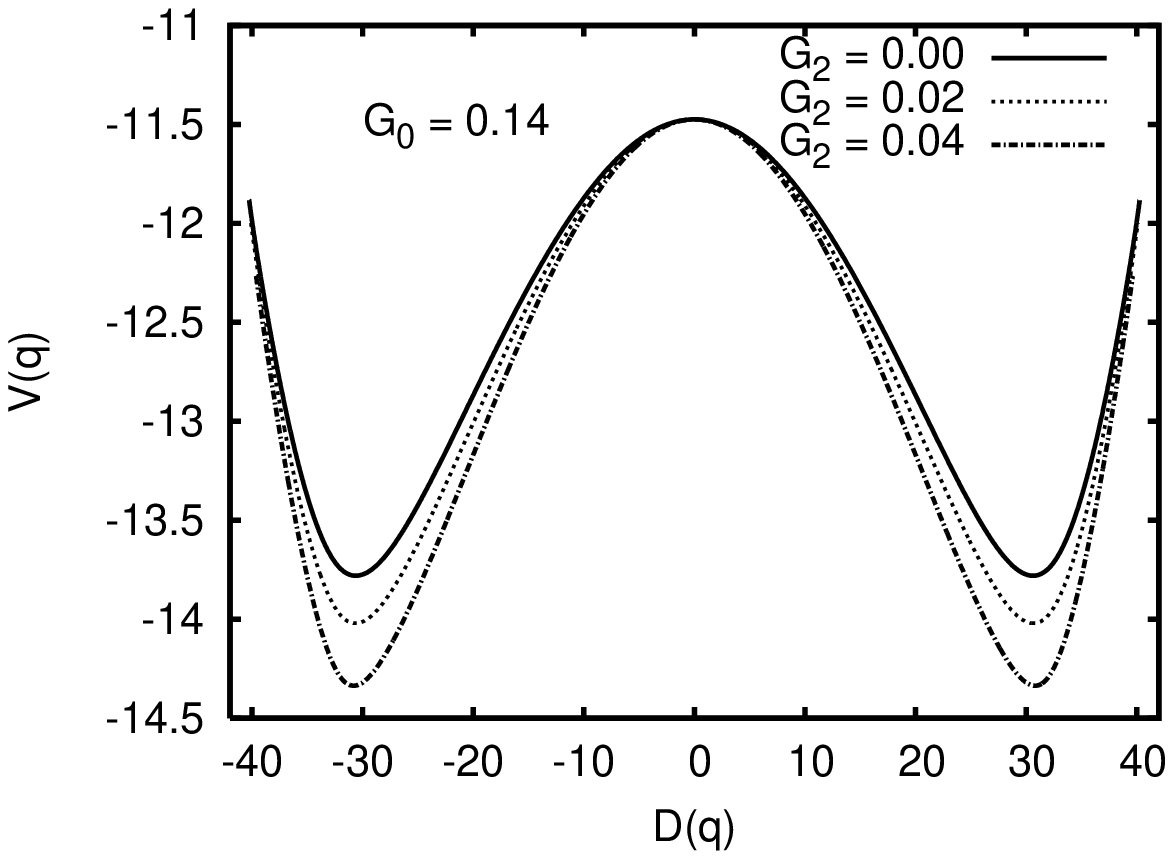} \\
\includegraphics[width=100mm]{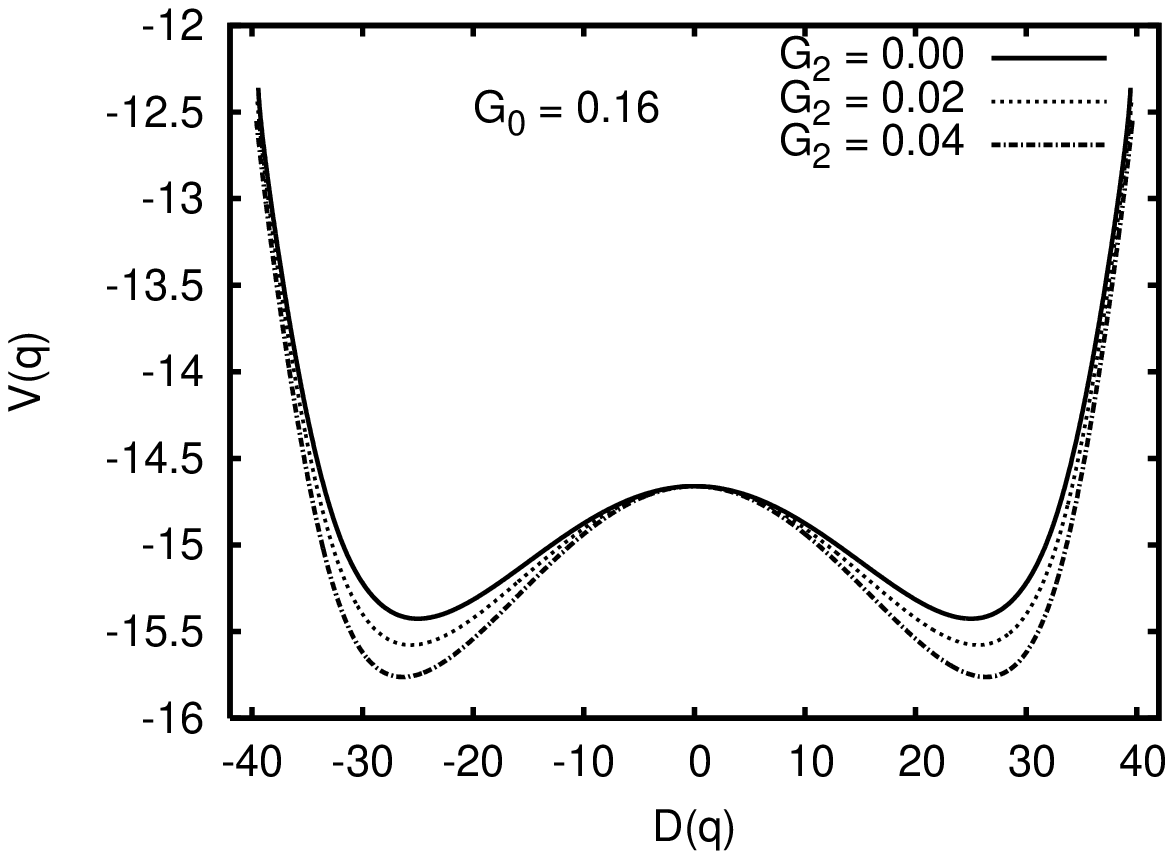} \\
\includegraphics[width=100mm]{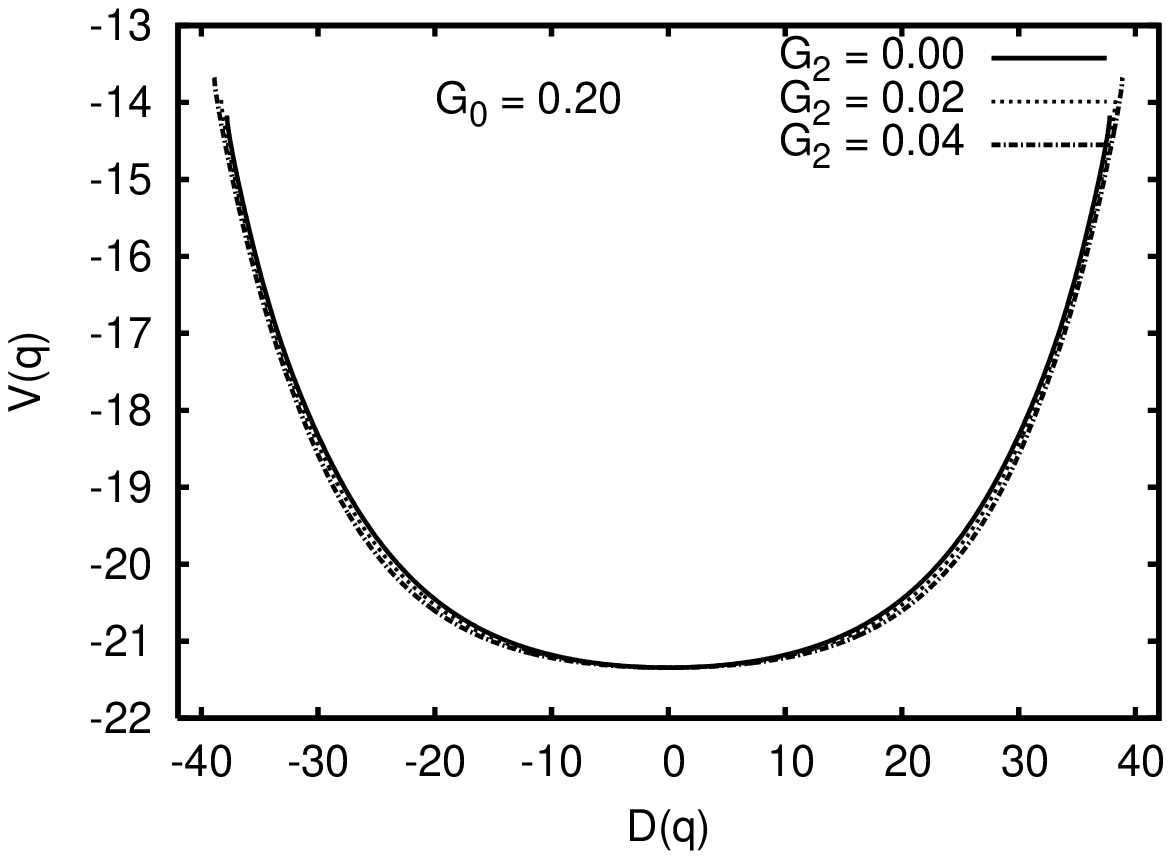} \\
\end{center}
\caption{Collective potentials 
plotted as functions of the quadrupole deformation $D$.
The upper, middle and lower panels display the results for
$G_0=0.14, 0.16$, and $0.20$, respectively.
In each panel, the results for $G_2= 0.00, 0.02, 0.04$ are compared.
}
\label{fig:VD0D2a}
\end{figure}

\begin{figure}[htbp]
\begin{center}
\begin{tabular}{cc}
\includegraphics[width=70mm]{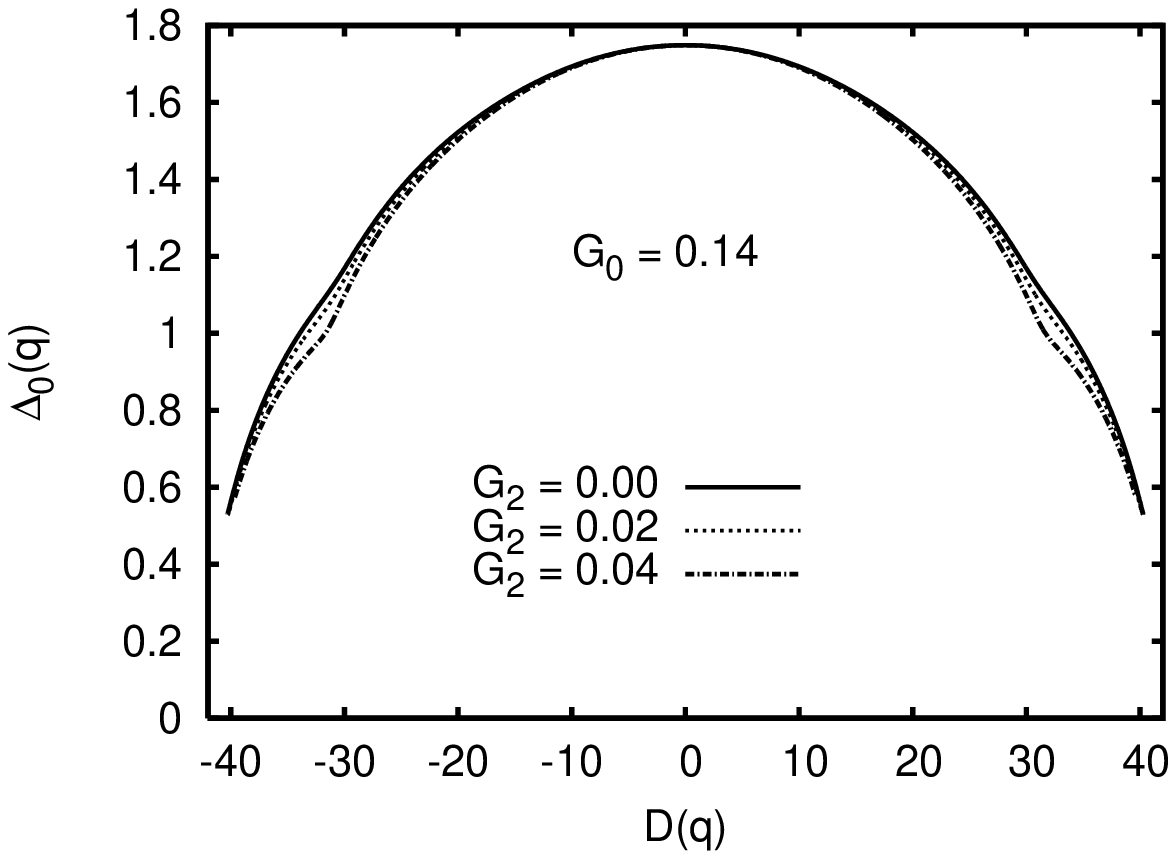} &
\includegraphics[width=70mm]{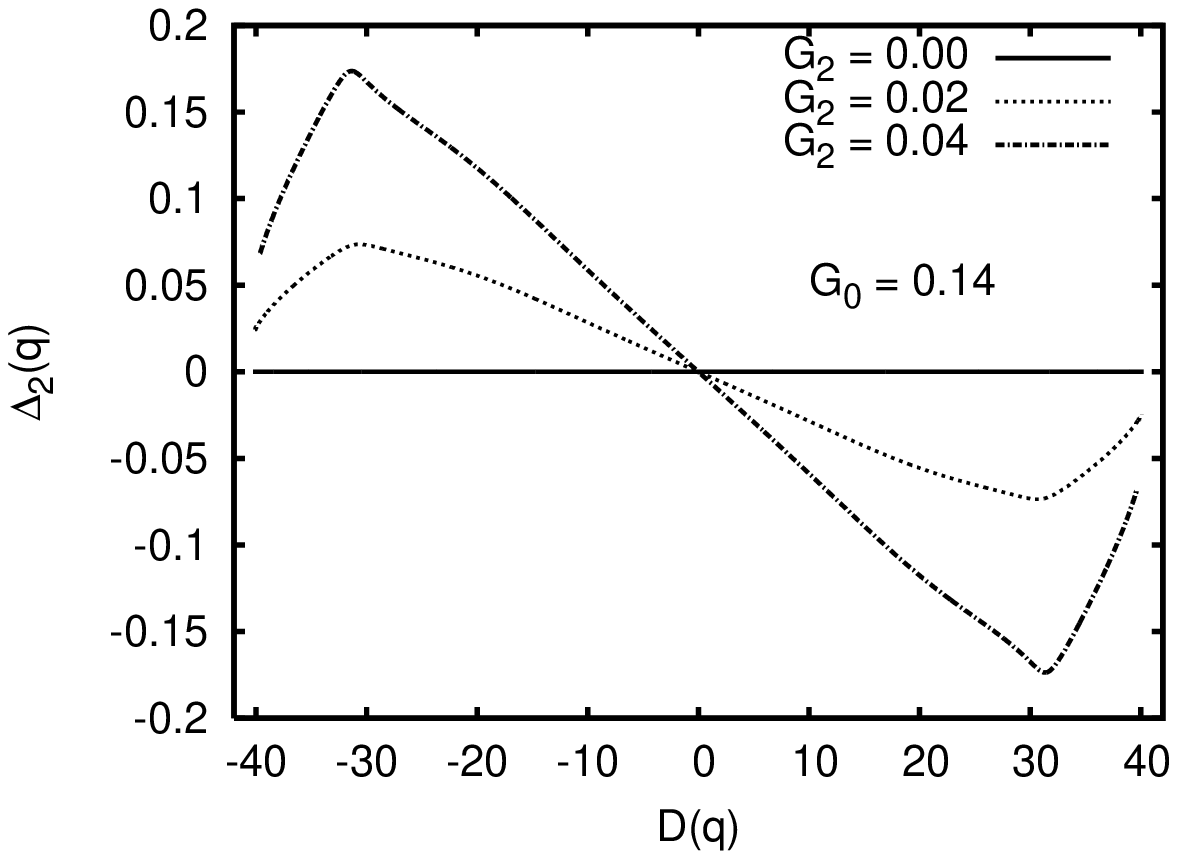} \\
\includegraphics[width=70mm]{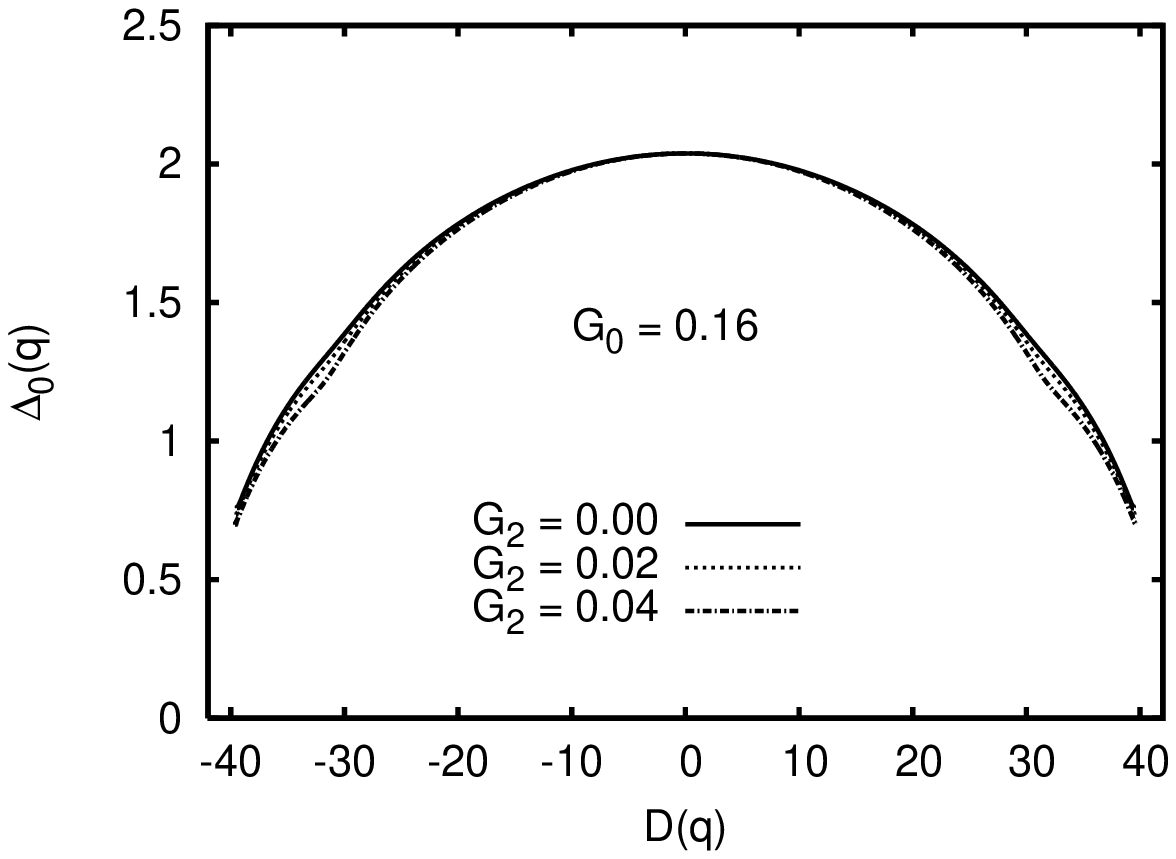} &
\includegraphics[width=70mm]{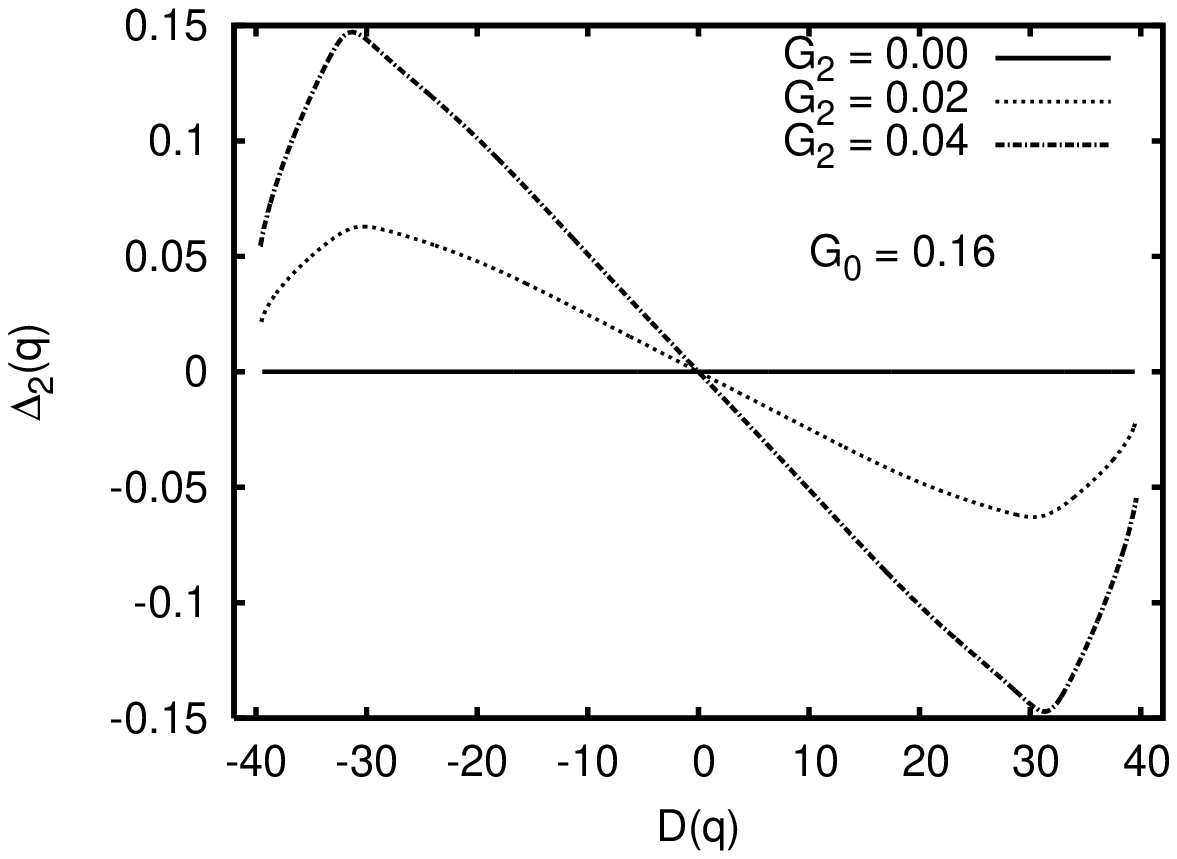} \\
\includegraphics[width=70mm]{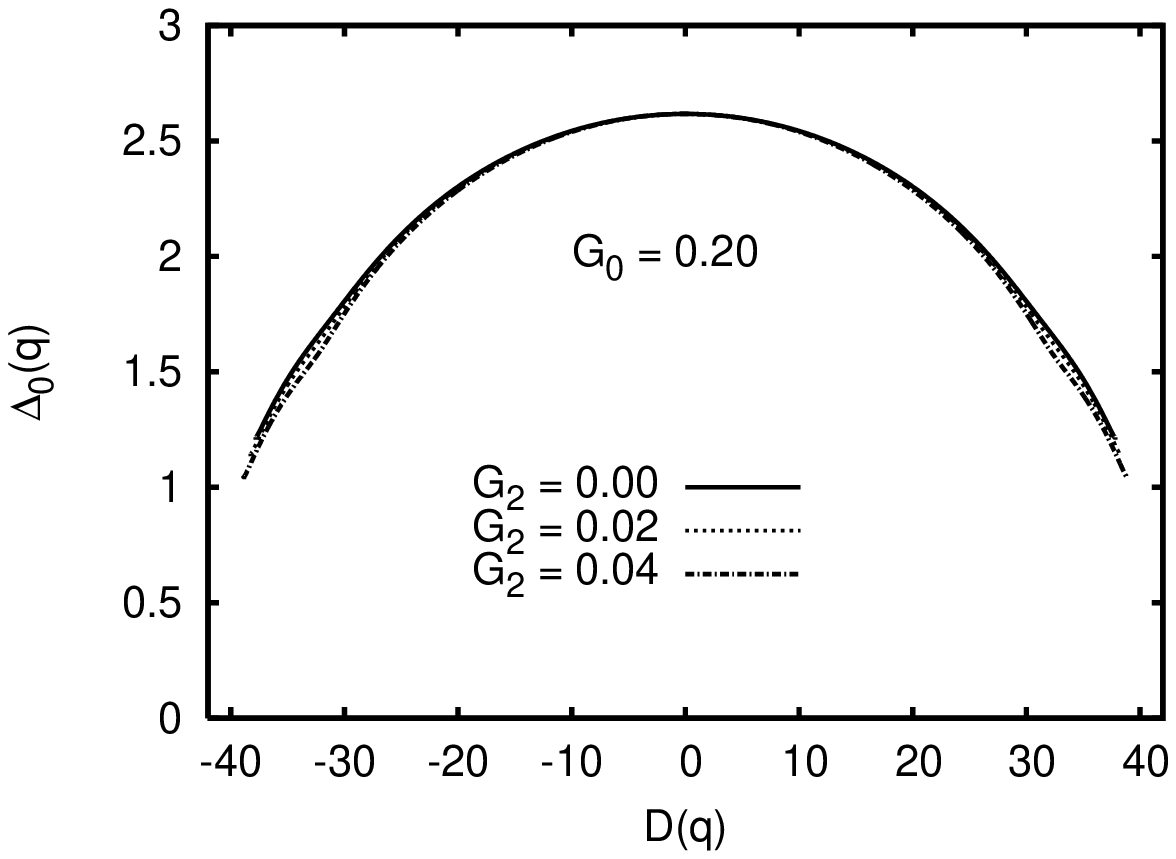} &
\includegraphics[width=70mm]{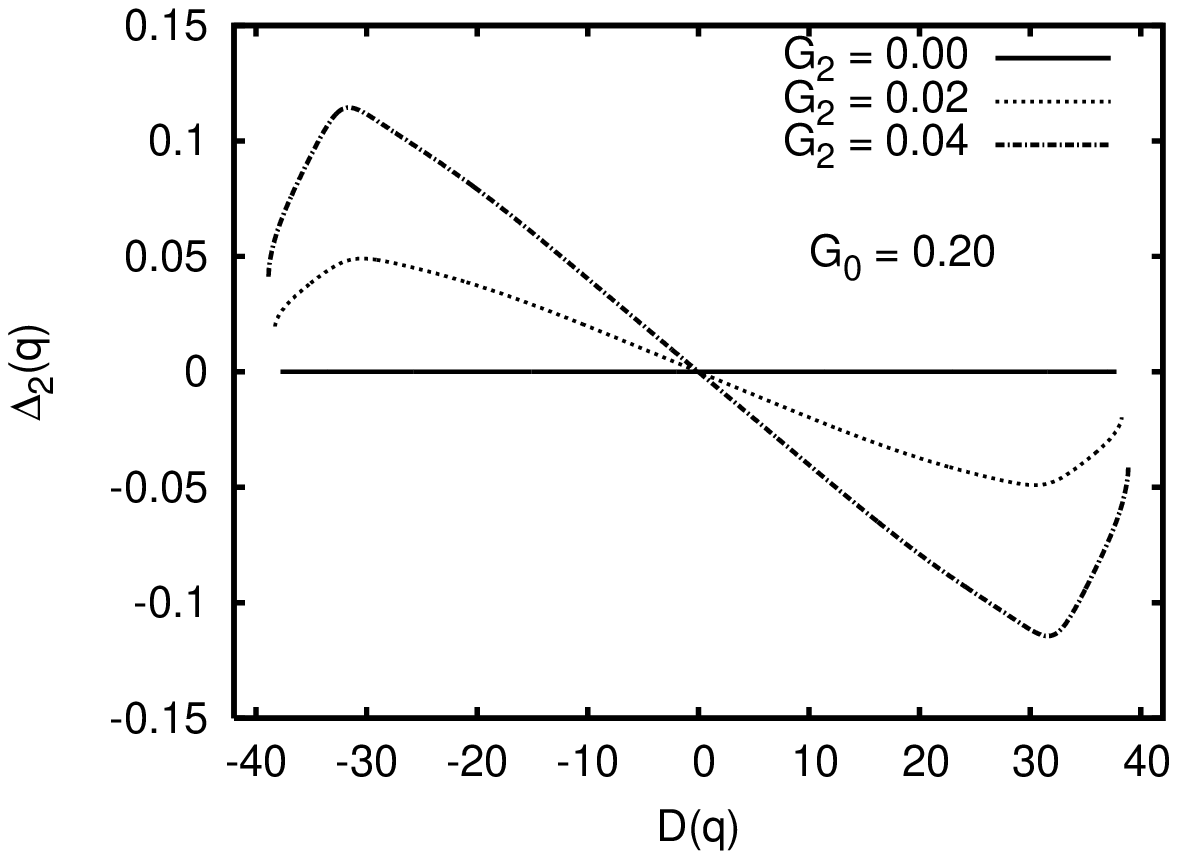}
\end{tabular}
\end{center}
\caption{
Monopole-pairing gaps $\Delta_0$ ({\it left column}) and
quadrupole-pairing gaps $\Delta_2$ ({\it right column}), 
plotted as functions of $D$.
The upper, middle and lower rows display the results for
$G_0=0.14, 0.16$, and $0.20$, respectively.
In each panel, the results for $G_2= 0.00, 0.02, 0.04$ are compared.
}
\label{fig:VD0D2b}
\end{figure}

\begin{figure}[htbp]
\begin{center}
\begin{tabular}{cc}
\includegraphics[width=70mm]{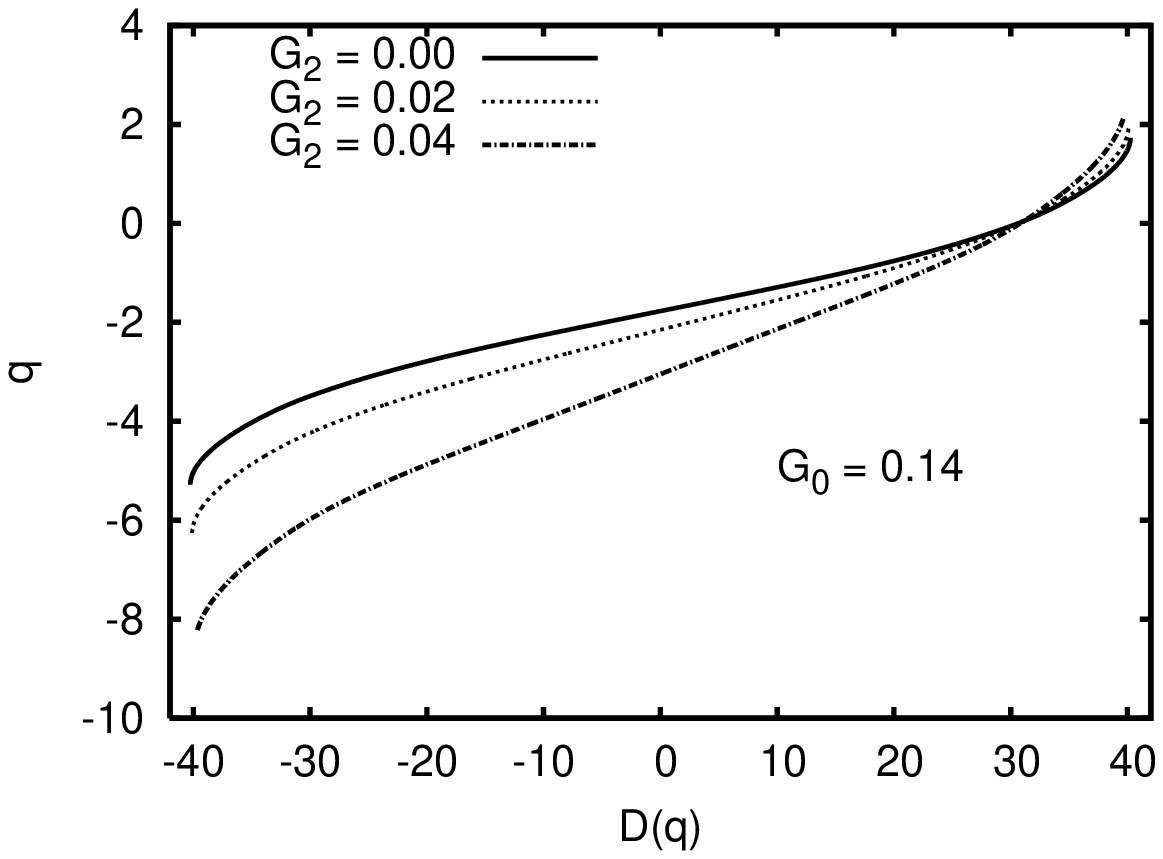} &
\includegraphics[width=70mm]{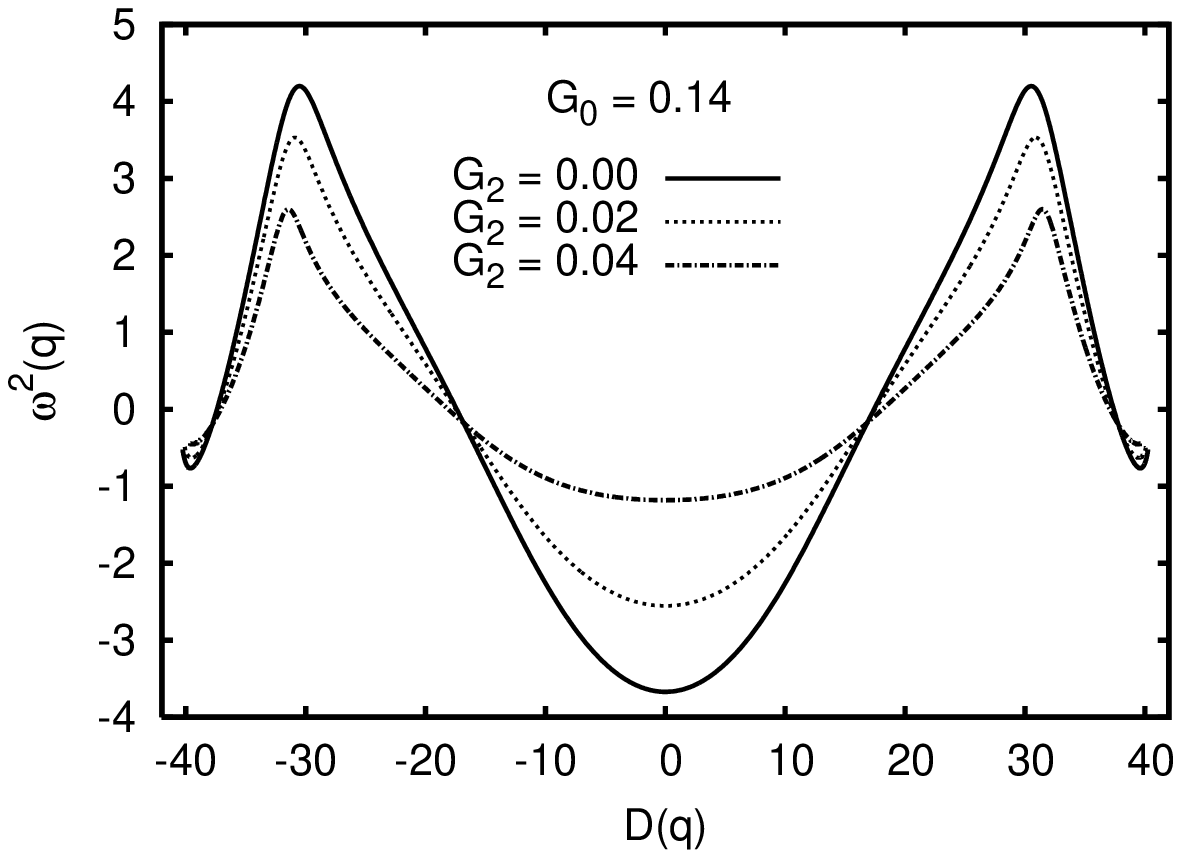} \\
\includegraphics[width=70mm]{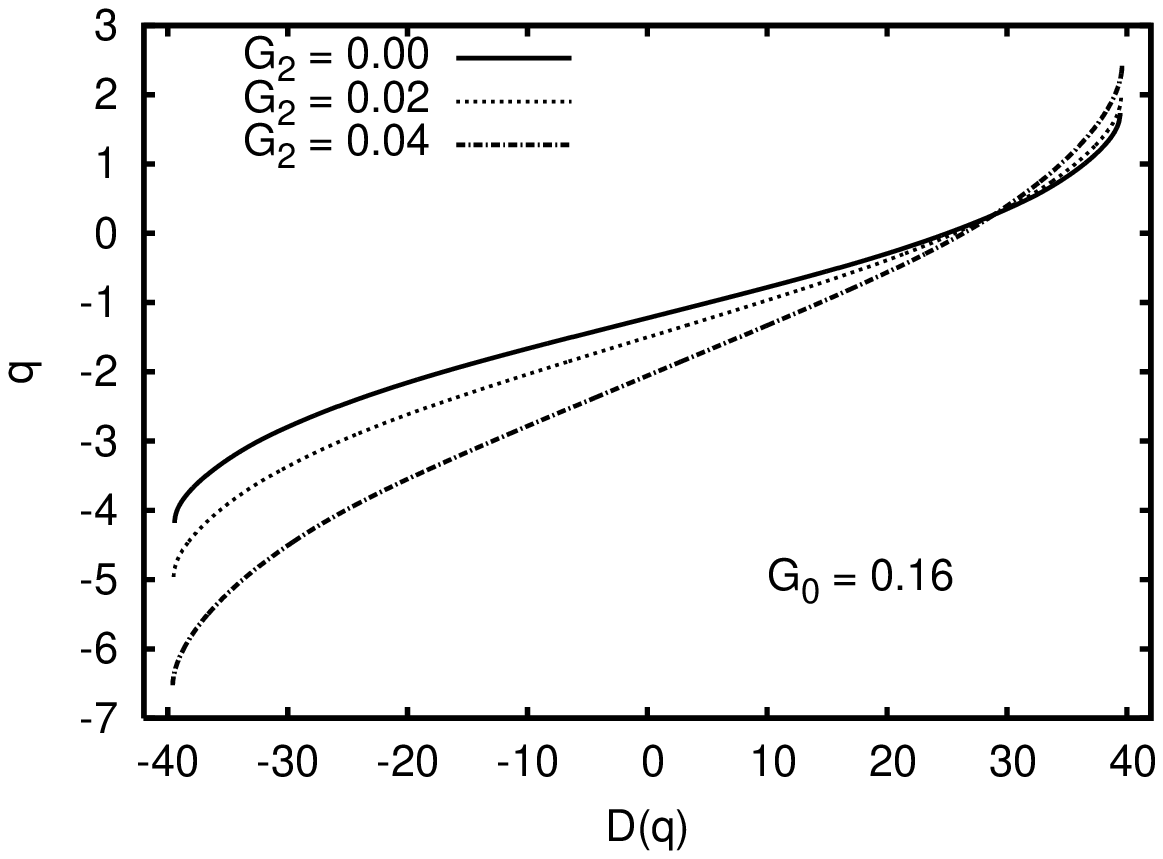} &
\includegraphics[width=70mm]{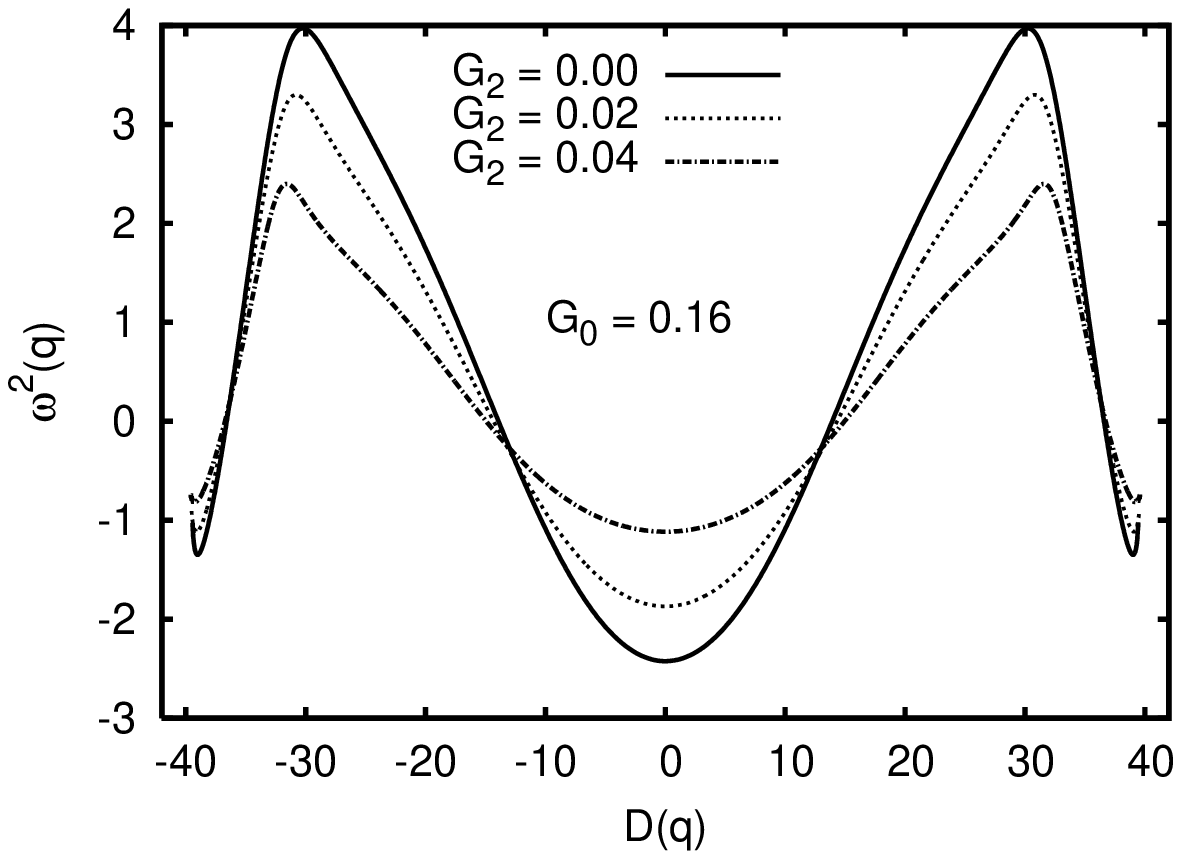} \\
\includegraphics[width=70mm]{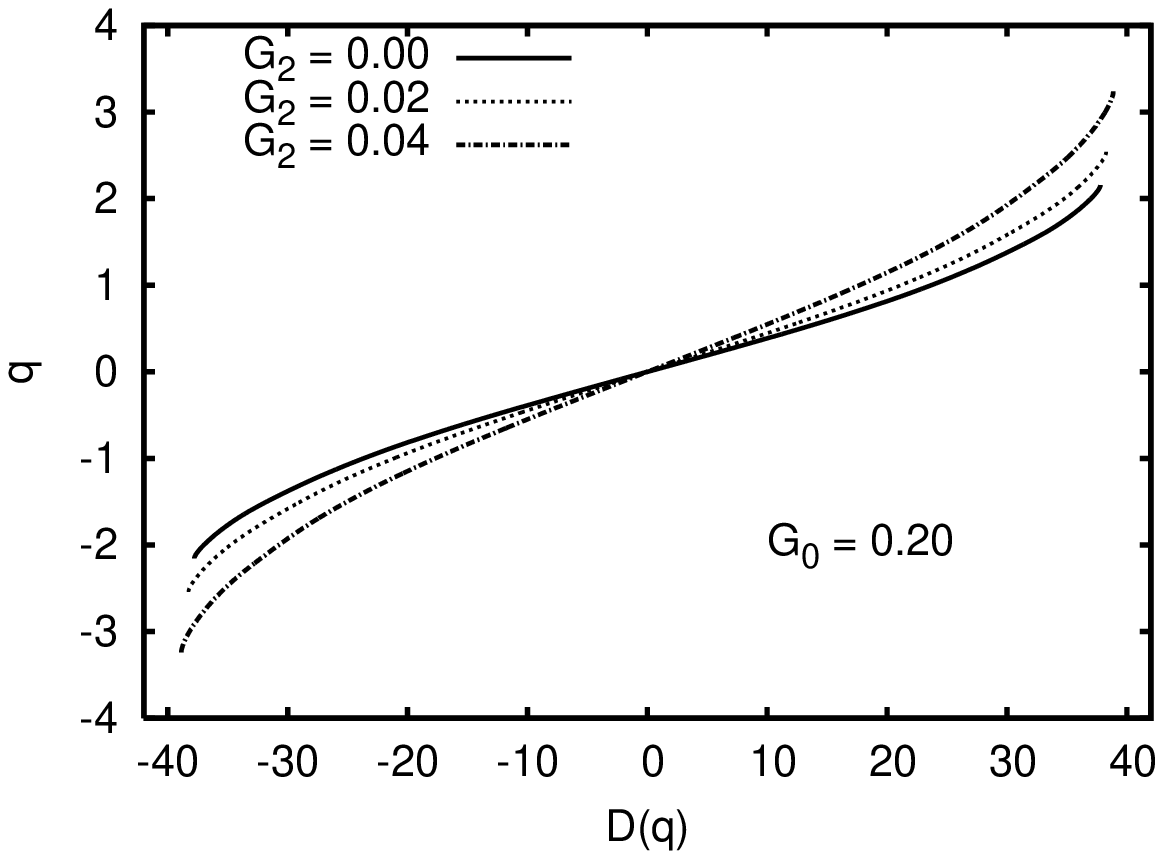} &
\includegraphics[width=70mm]{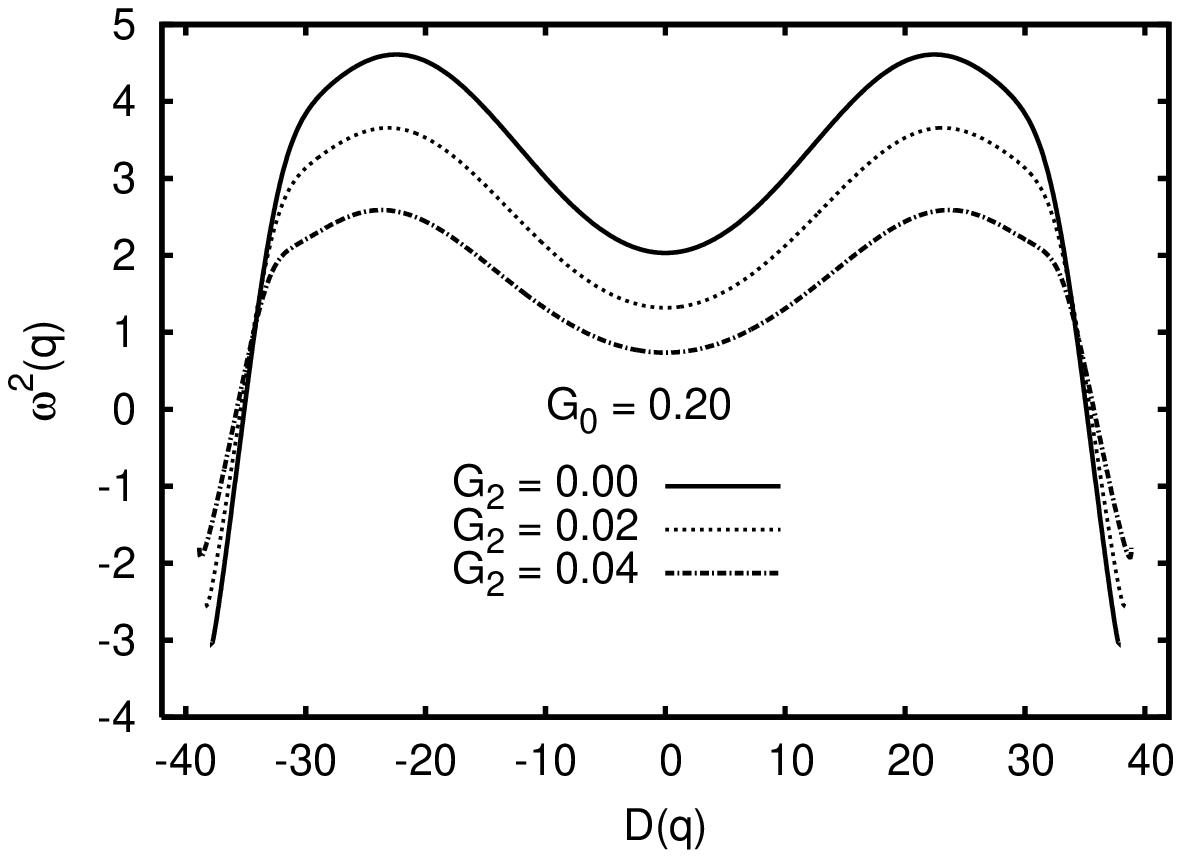} 
\end{tabular}
\end{center}
\caption{
{\it Left column}:
Relation between the collective coordinate $q$ 
and the quadrupole deformation $D(q) = \bra{\phi(q)}\Dhat\ket{\phi(q)}$.
The point $q=0$ corresponds to the HB equilibrium, 
which is the starting point of the numerical calculation. 
{\it Right column}:
Squared frequencies $\omega^2(q)$ of the local harmonic equation,
plotted as a function of $D$.
Note that they are negative, i.e., $\omega(q)$ is pure imaginary, 
in the region where the curvature of the collective potential is negative.
The upper, middle and lower rows display the results for
$G_0=0.14, 0.16$, and $0.20$, respectively.
In each panel, the results for $G_2= 0.00, 0.02, 0.04$ are compared.
}
\label{fig:q-D}
\end{figure}

A nice property of the multi-$O(4)$ model is that
we can simulate, by changing the ratio $G_0/\chi$, 
the phase transition in a finite quantum system 
from the single-well to the double-well potential.
This is analogous to nuclear shape phase transition and shape coexistence
phenomena.
The collective potential $V$ and the pairing gaps ($\Delta_0$, $\Delta_2$)
are displayed in Figs.~\ref{fig:VD0D2a} and \ref{fig:VD0D2b}.
In the ASCC method, these quantities are calculated as functions of
the collective coordinate $q$, but they are easily converted to those as
functions of $D$.
See the relation between $q$ and $D$ shown in the left panels of
Fig.~\ref{fig:q-D}.
We see that the collective potential change from the single well to the
double well as the ratio $G_0/\chi$ decreases, i.e., as the effect of
the monopole-pairing interaction is weakened.
When the oblate-prolate shape coexistence occurs ($G_0=0.14, 0.16$),
the collective path calculated from one local minimum 
passes through the other local minimum.
It is seen that $\Delta_0(D)$ decreases while $|\Delta_2(D)|$ increases 
as $D$ increases.
Both gaps tend to vanish near the limits, 
$D_{\rm min}=-42$ and $D_{\rm max}=42$.
This is due to the small model space
and is absent in realistic situations.
The quadrupole-pairing gap $|\Delta_2(D)|$ takes the maximum value 
at the prolate and oblate HB equilibrium points. 
Because of energy gain associated with the quadrupole pairing,
the oblate and prolate local minima in the collective potential $V(D)$ 
become deeper as $G_2$ increases. 

\begin{figure}[htbp]
\begin{center}
\begin{tabular}{cc}
\includegraphics[width=70mm]{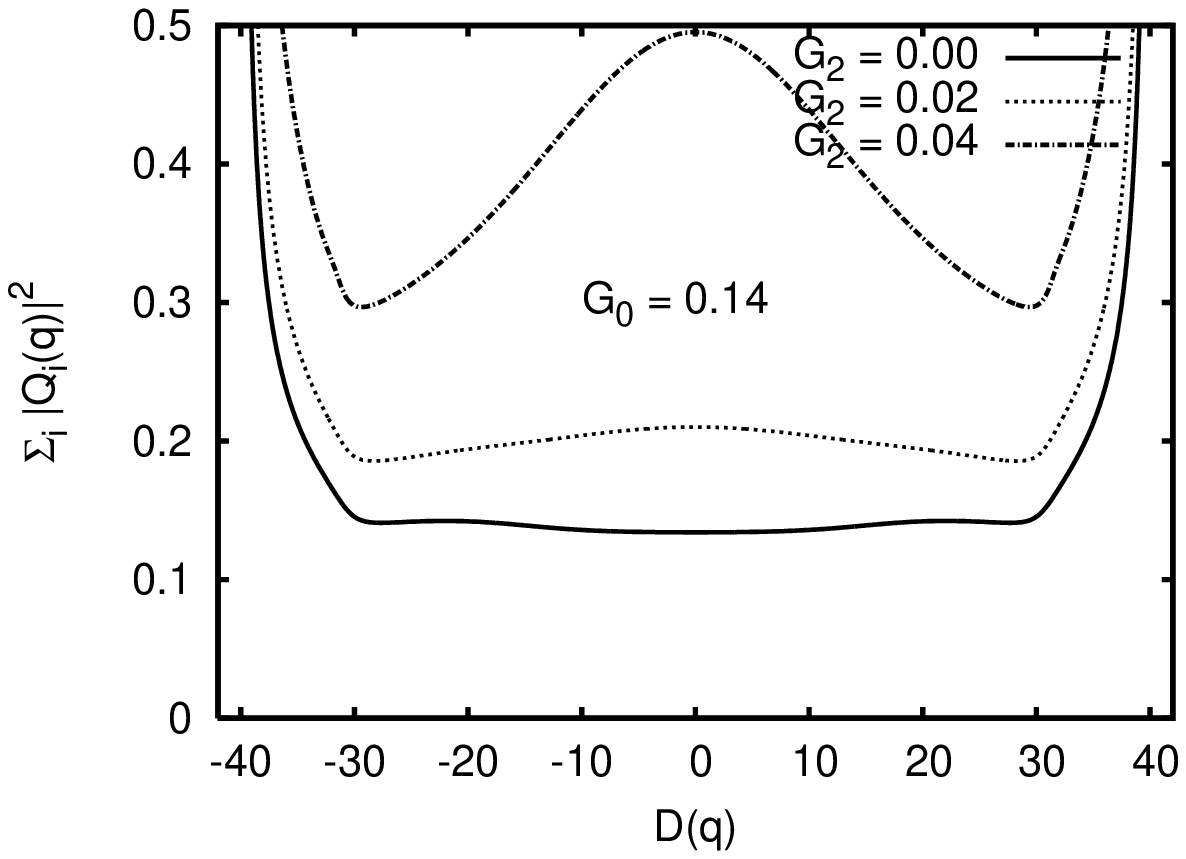} &
\includegraphics[width=70mm]{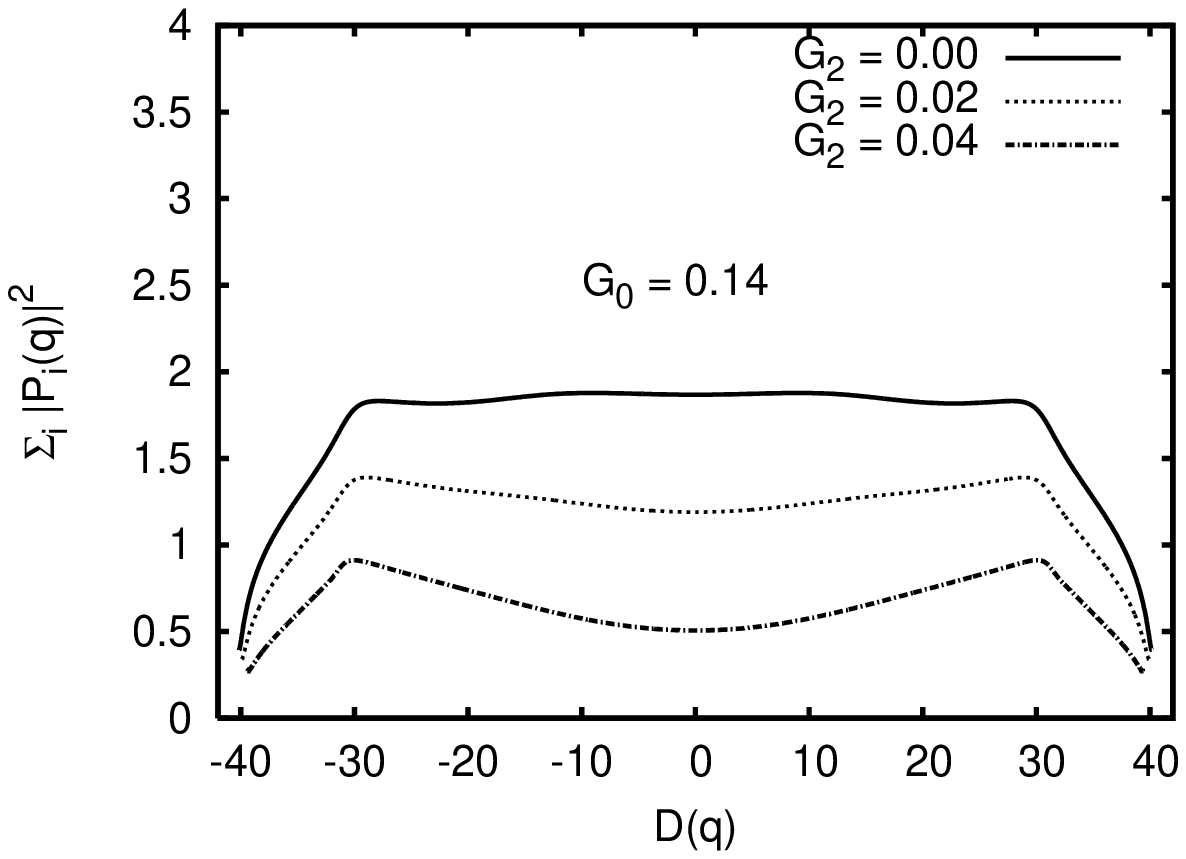} \\
\includegraphics[width=70mm]{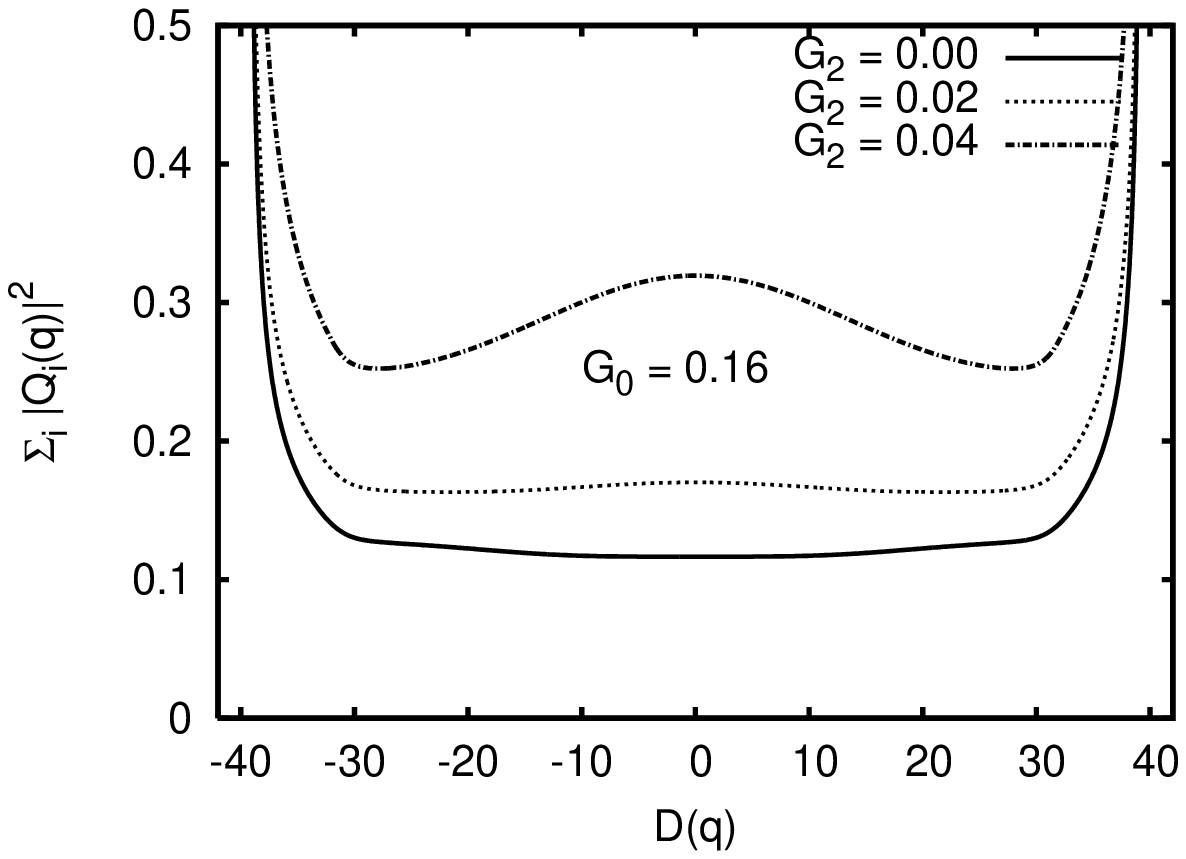} &
\includegraphics[width=70mm]{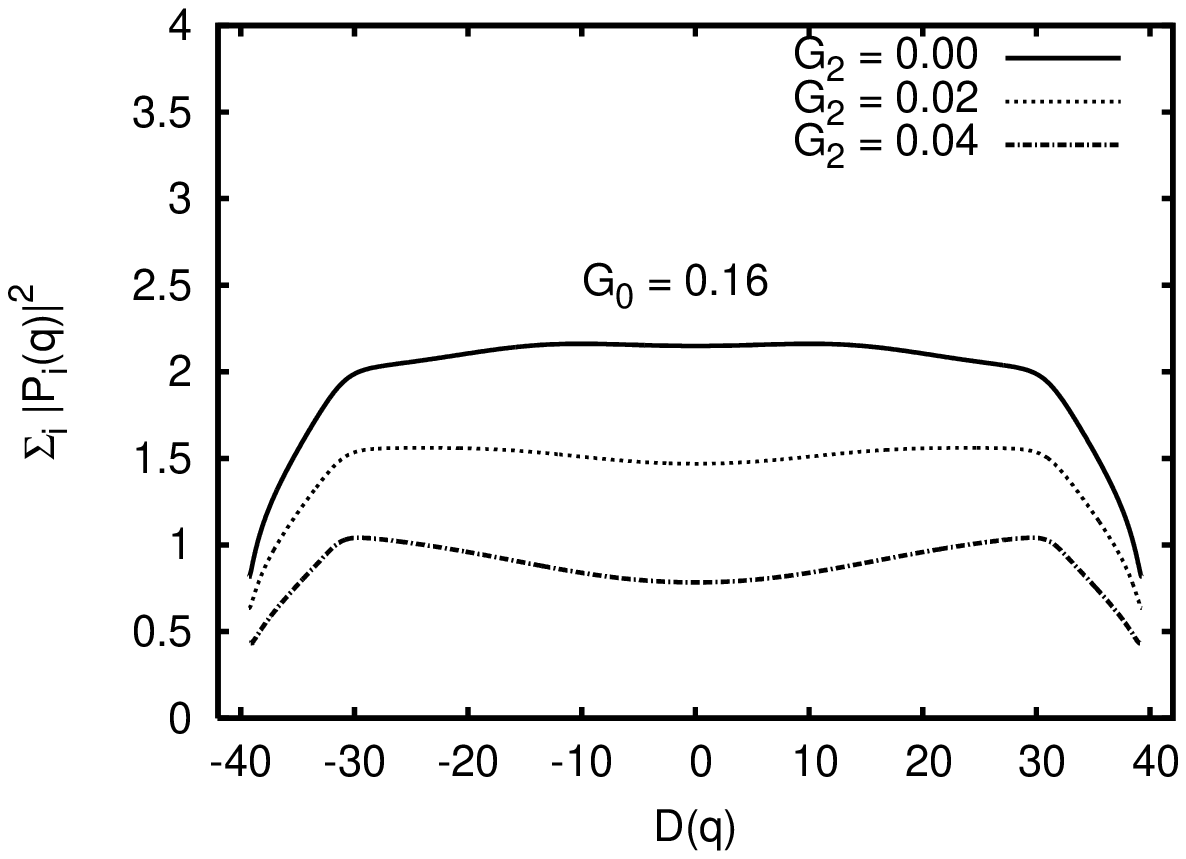} \\
\includegraphics[width=70mm]{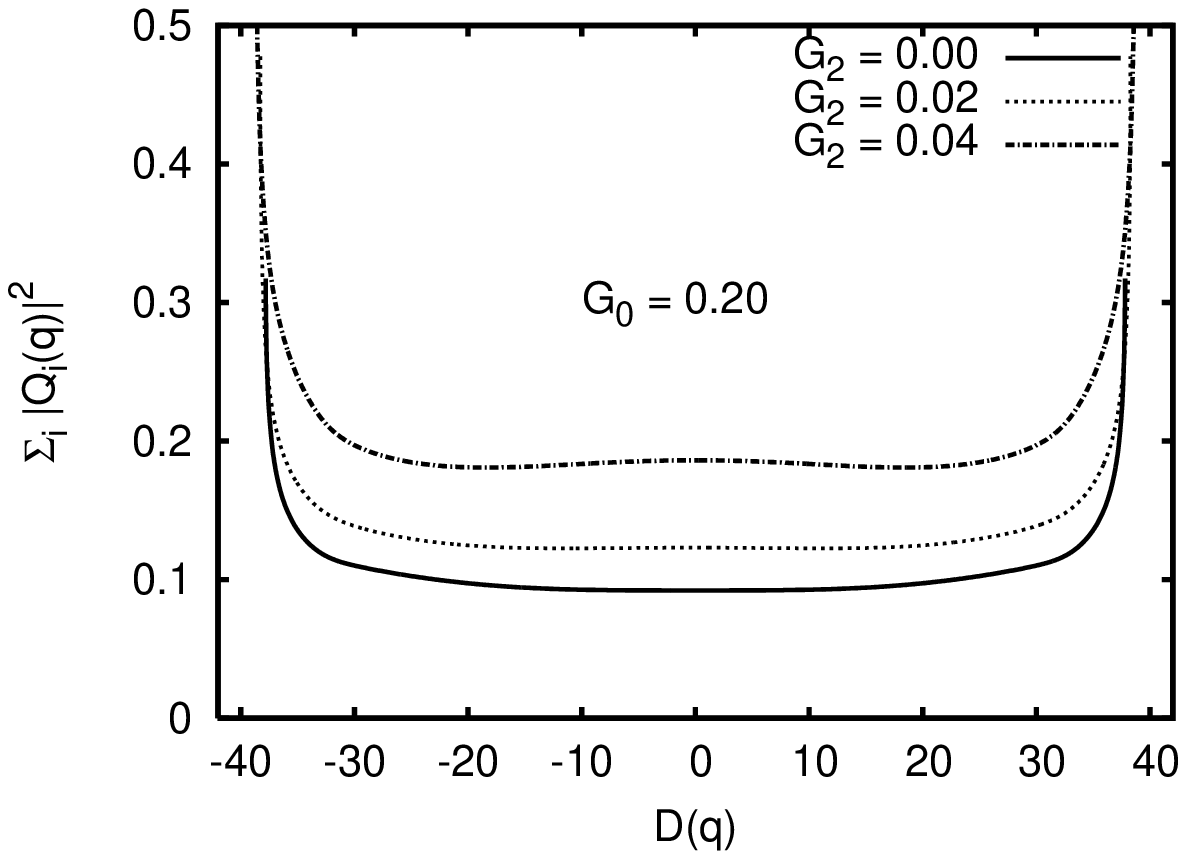} &
\includegraphics[width=70mm]{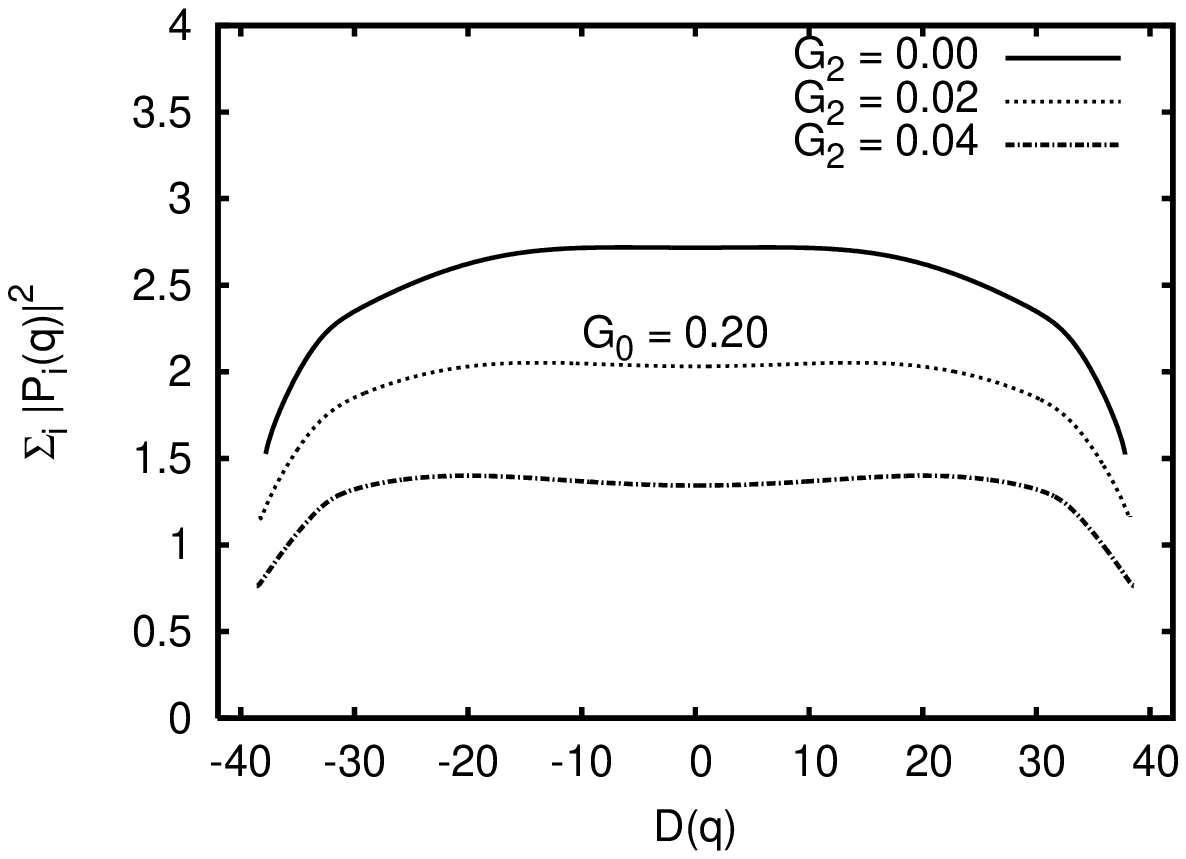} 
\end{tabular}
\end{center}
\caption{The sums, $\sum_{i>0} |Q_i(q)|^2$ and $\sum_{i>0} |P_i(q)|^2$, of
the two quasi-particle components, $Q_i(q)$ and $P_i(q)$, of the 
infinitesimal generators, $\hat Q(q)$ and $\hat P(q)$, 
plotted as functions of the quadrupole deformation $D$.
The upper, middle and lower rows display the results for
$G_0$ = 0.14, 0.16, and 0.20, respectively.
In each panel, the results for $G_2= 0.00, 0.02, 0.04$ are compared.
}
\label{fig:Q20}
\end{figure}

Significant effects of the quadrupole-pairing interaction are 
obviously seen in the solutions of the local harmonic equations.
The squared frequencies $\omega^2(q)$, representing the curvature of 
the collective potential, are shown in the right column of Fig.~\ref{fig:q-D},
while the sums, $\sum_{i>0} |Q_i(q)|^2$ and $\sum_{i>0} |P_i(q)|^2$, of
the two quasiparticle components, $Q_i(q)$ and $P_i(q)$, 
are displayed in Fig.~\ref{fig:Q20}.
They are again plotted as functions of $D$.
It is clearly seen that absolute magnitudes of $Q_i(q)$ significantly 
increase, while those of $P_i(q)$ decrease with increasing $G_2$.
(To avoid complication, their sums rather 
than individual values are presented in Fig.~\ref{fig:Q20}.)
These changes in the microscopic structure of ${\hat Q(q)}$ and ${\hat P(q)}$ 
lead to the increase of the derivative $dq/dD$ with increasing $G_2$
(Fig.~\ref{fig:q-D}).
Then, this will result in a significant enhancement of the collective mass
$M(D(q))$ with respect to the coordinate of
the quadrupole deformation parameter $D$.
These interesting properties of the collective mass are a main subject of
this paper and will be discussed in the next subsection.

We have also carried out a constrained HB (CHB) calculations
choosing the quadrupole
operator $\Dhat$ as the constraint. Calculated values of 
$V(D)$, $\Delta_0(D)$ and $\Delta_2(D)$ 
were almost indistinguishable from those obtained by the ASCC method 
displayed in Figs.~\ref{fig:VD0D2a} and \ref{fig:VD0D2b}.
Thus, concerning these static mean-field quantities,
we see no difference between the ASCC and the CHB calculations.
It should be noted, however, that such a good agreement is due to a simplicity
of the multi-$O(4)$ model, 
i.e., it contains only one degree of freedom,
$\Dhat$, relevant to the large-amplitude collective motion.
In reality, many degrees of freedom   
(a variety of particle-hole excitations associated with shell structures, 
triaxial deformations, various multipolarities, 
low and high frequency excitations, etc.) 
would be interwoven to generate large-amplitude collective motions.
In fact, it has been shown that the self-consistently determined collective
coordinate operators for low-frequency quadrupole-type collective vibrations
are significantly different from the quadrupole operators 
(see, for instance, Ref.~\citen{nak00}).
Let us now turn to a discussion on collective mass 
where we can clearly see the merit of the ASCC method.

\subsection{Collective mass}

As mentioned in subsection \ref{sec:LHE}, we set the scale of the collective 
coordinate $q$ so as to make the collective mass unity.
In order to compare the collective mass obtained by the ASCC method 
with the conventional cranking mass, let us evaluate the collective mass 
as a function of the quadrupole deformation $D$. This quantity, $M(D(q))$,
is readily obtained by transforming the collective kinetic energy as a function
of the velocity $\dot{D}$;
\begin{align}
 \frac{1}{2}B(q)p^2 &= \frac{1}{2}p^2 = \frac{1}{2}\dot{q}^2 = 
 \frac{1}{2}M(D(q))\dot{D}^2, \\
 M(D(q)) &= \left( \frac{dq}{dD}\right)^2 
 = ( 4\sum_{i>0} d_i \sigma_i u_i v_i P_i(q))^{-2}.
\end{align}

As is well known, the Inglis-Belyaev cranking mass is derived 
by means of the adiabatic perturbation theory\cite{rin80} and written as
\begin{equation}
 M_{\rm cr}(D) = 2 \sum_n
\frac{|\bra{\phi_n(D)}\displaystyle\frac{\del}{\del D}\ket{\phi_0(D)}|^2 }
     {E_n(D) - E_0(D)}. 
\end{equation}
Here, $\ket{\phi_0(D)}$ and $E_0(D)$, respectively,
represent the ground state and 
its energy at the deformation $D$, while $\ket{\phi_n(D)}$ 
and $E_n(D)$ the (two-quasiparticle) excited states and their energies.
We can use this formula either treating the deformation $D$ 
as a phenomenological parameter
or a self-consistently determined quantity.
If the ground states are calculated at every points of $D$ 
by means of the CHB method, 
\begin{align}
 \delta\bra{\phi_0(D)}\Hhat - \lambda(D)\Nhat -
 \mu(D)\Dhat\ket{\phi_0(D)} = 0,
\end{align}
with the self-consistency conditions for the particle number
and the quadrupole deformation, 
\begin{align}
 \bra{\phi_0(D)}\Nhat\ket{\phi_0(D)}= N_0,\quad &
 \bra{\phi_0(D)}\Dhat\ket{\phi_0(D)} = D,
\end{align}
then the following explicit expression for $M_{\rm cr}(D)$ is valid: 
\begin{equation}
 M_{\rm cr}(D) = 2 \sum_{i>0} \frac{\Big\arrowvert 2u_i(D)v_i(D)
 ((\chi + \displaystyle\frac{\del\mu}{\del D} )d_i\sigma_i 
 + \displaystyle\frac{\del\lambda}{\del D})+
 (u_i^2(D)-v_i^2(D))(\frac{\del\Delta_0}{\del
 D}+d_i\sigma_i\frac{\del\Delta_2}{\del D})\Big\arrowvert^2}{(2E_i(D))^3}.
\end{equation}
Hereafter, we call this ``CHB-cranking mass."
Note that we used a slightly different definition of the cranking mass
in Ref.~\citen{kob03}, in which the CHB self-consistency is ignored.

\begin{figure}[htbp]
\begin{center}
\begin{tabular}{cc}
\includegraphics[width=70mm]{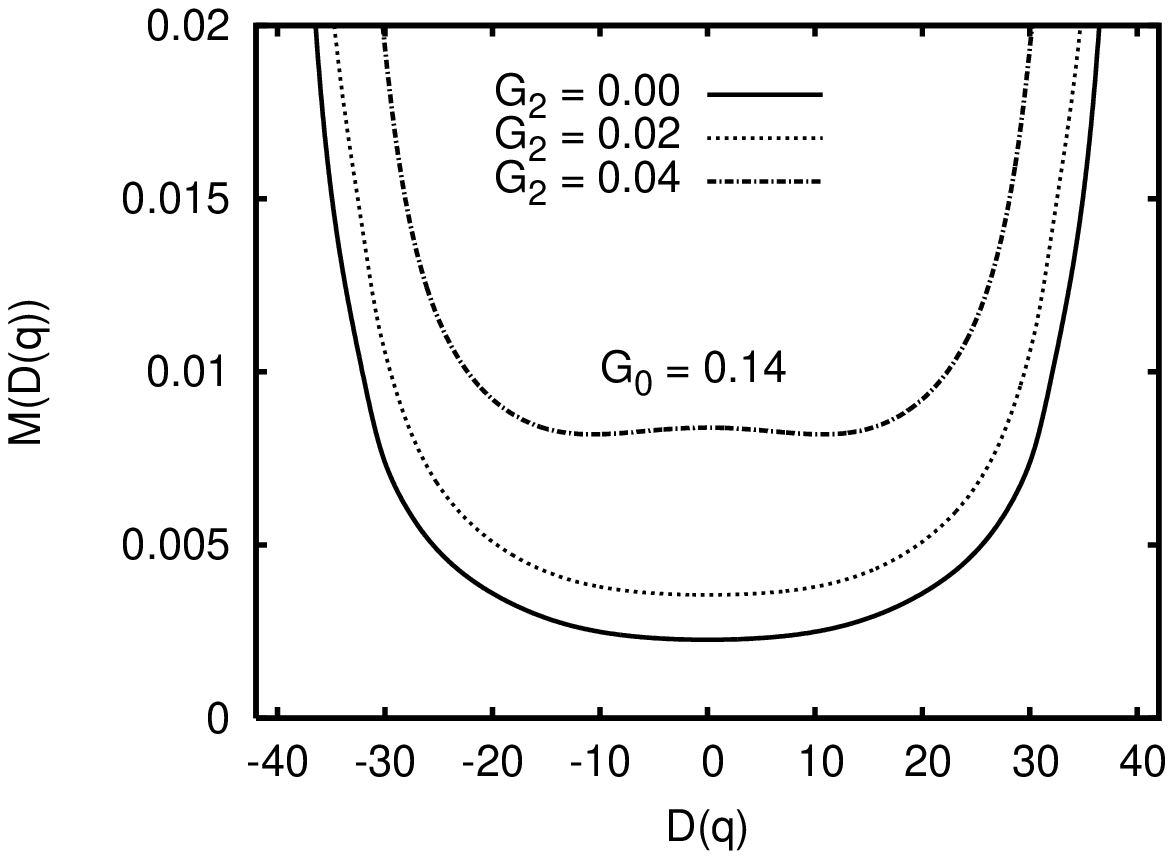} &
\includegraphics[width=70mm]{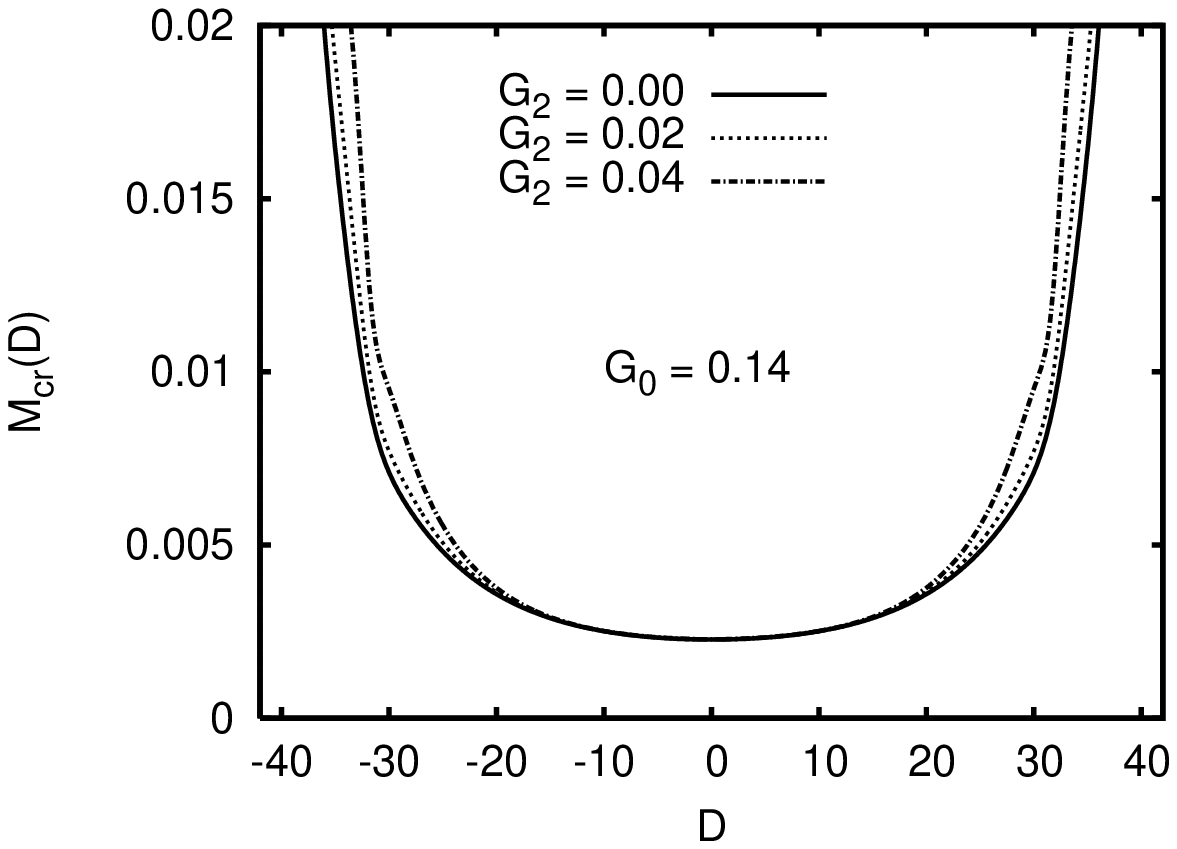} \\
\includegraphics[width=70mm]{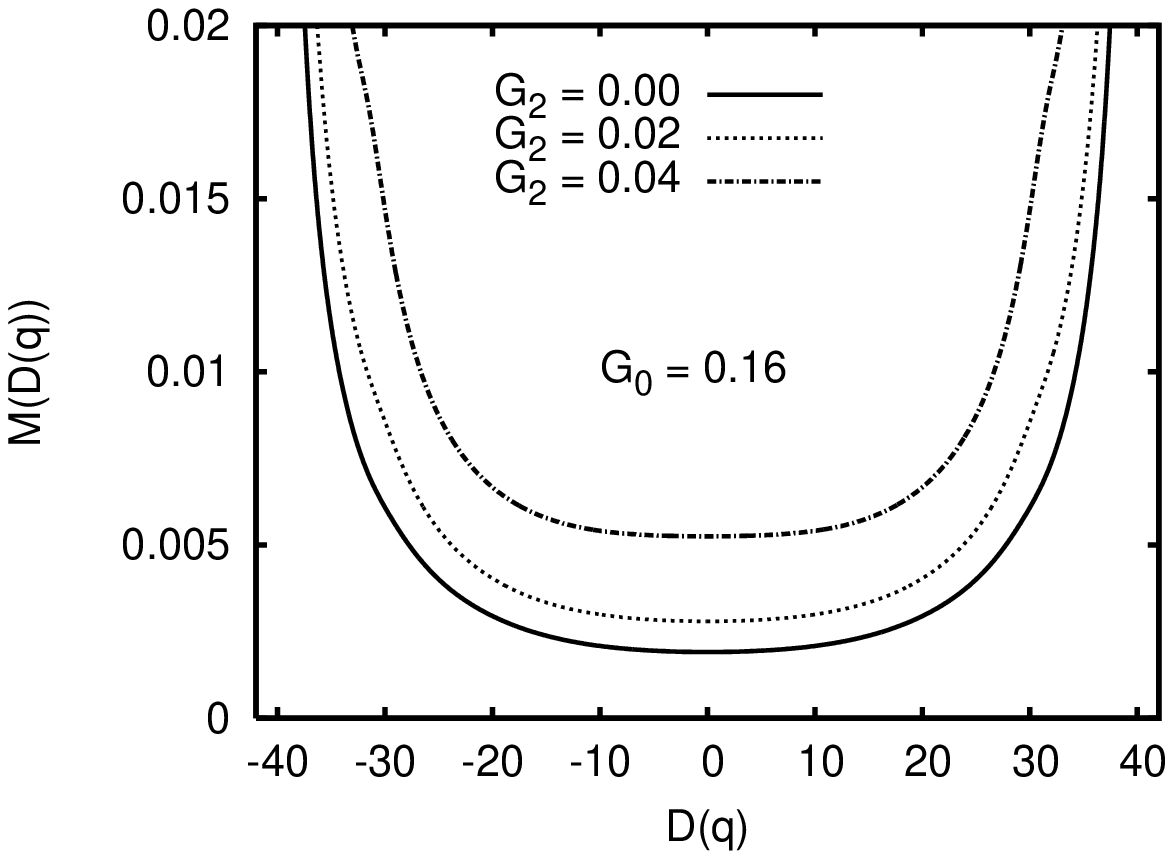} &
\includegraphics[width=70mm]{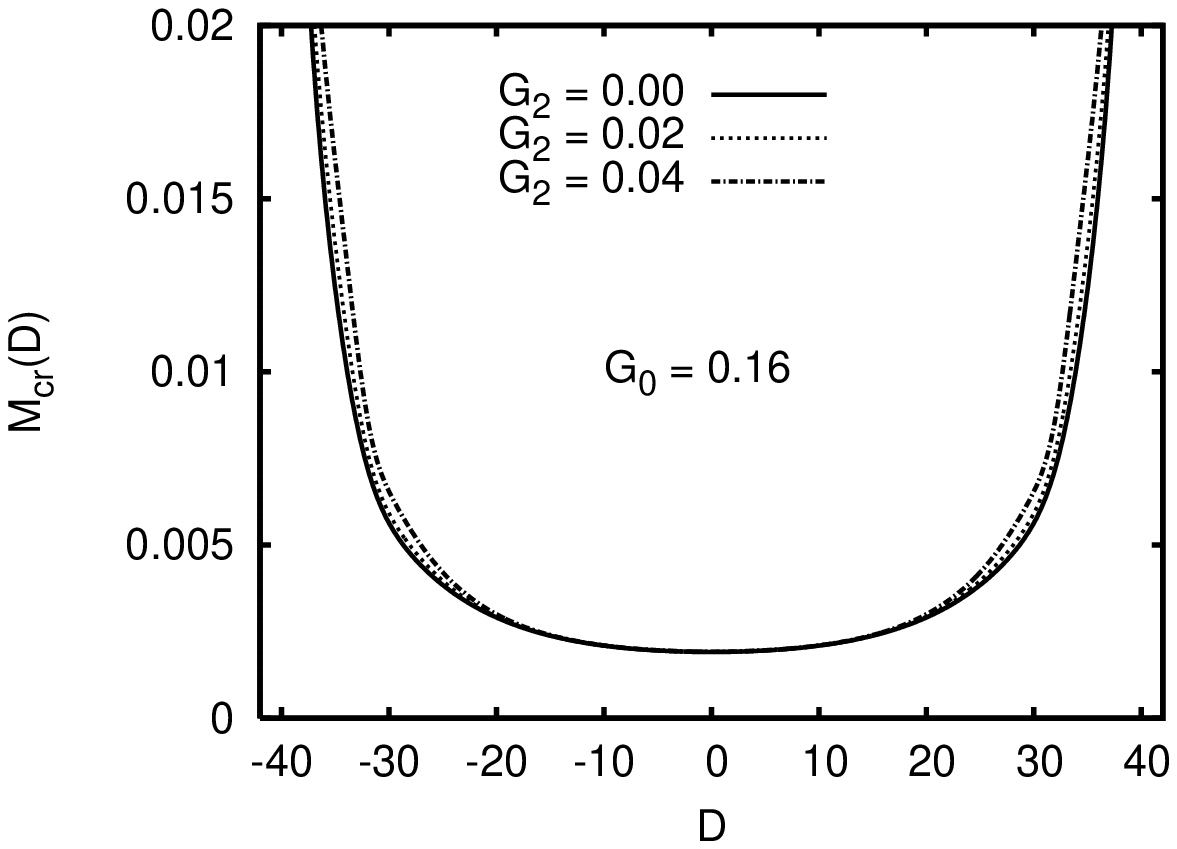} \\
\includegraphics[width=70mm]{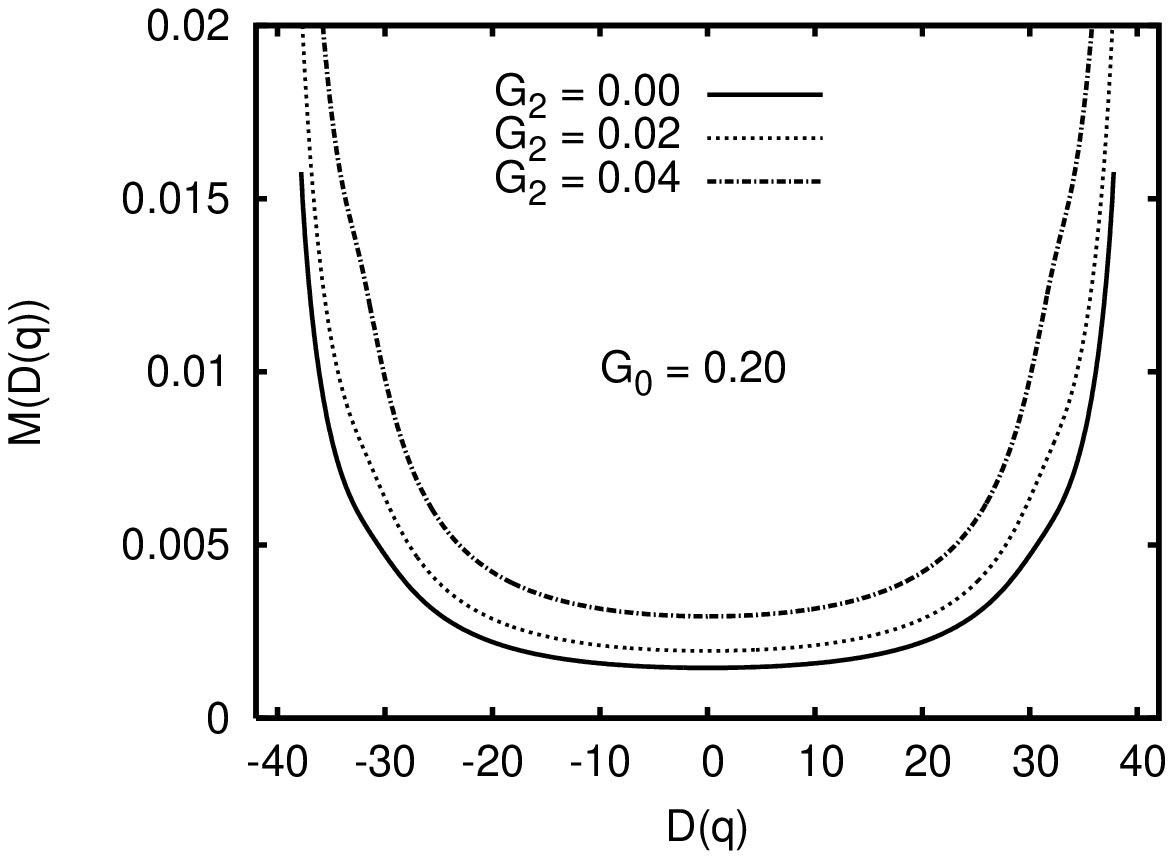} &
\includegraphics[width=70mm]{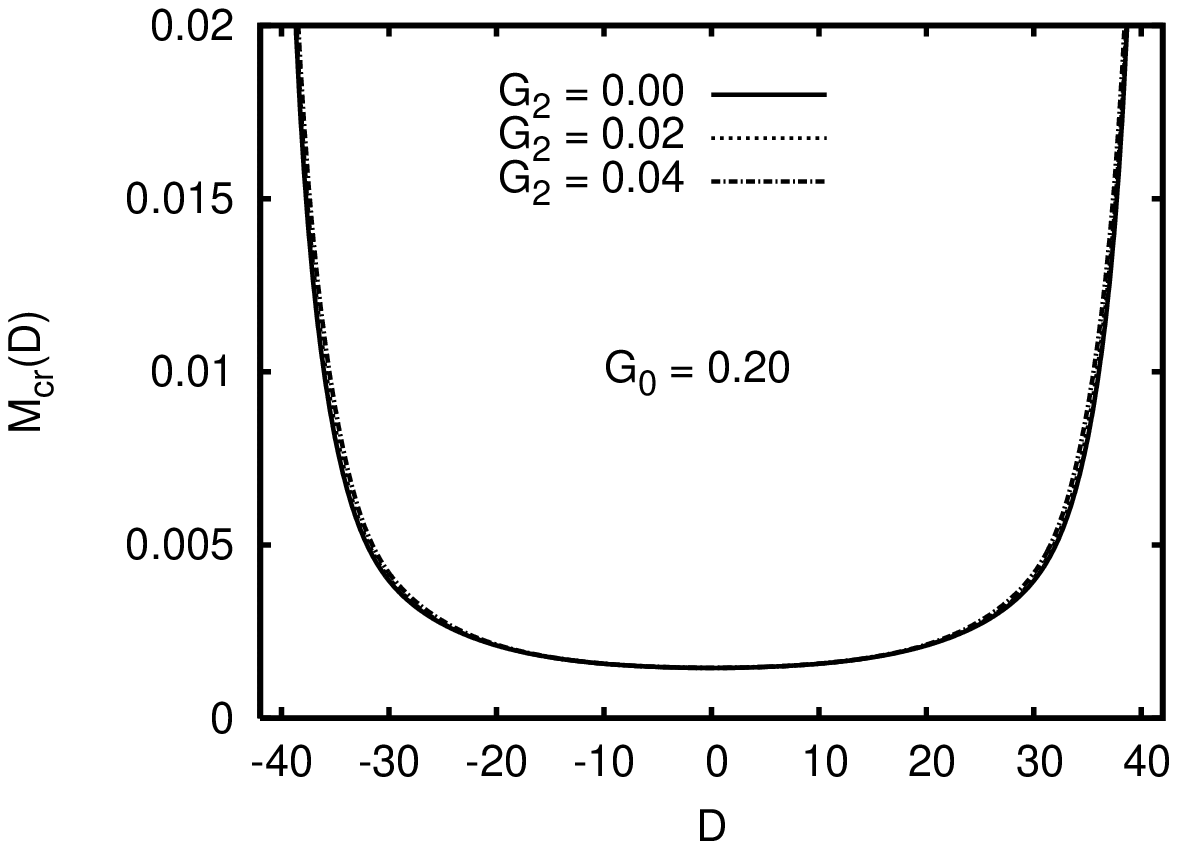}
\end{tabular}
\end{center}
\caption{The ASCC collective mass $M(D(q))$ ({\it left column})
and the CHB-cranking mass $M_{\rm cr}(D)$ ({\it right column}) 
as functions of deformation $D$.
The upper, middle and lower rows display the results for
$G_0=0.14, 0.16$, and $0.20$, respectively.
In each panel, the results for $G_2= 0.00, 0.02, 0.04$ are compared.
}
\label{fig:M}
\end{figure}

In Fig.~\ref{fig:M}, the ASCC mass $M(D(q))$ are shown as a function of $D$
and compared with the CHB-cranking mass  $M_{\rm cr}(D)$ 
for various combinations of the pairing-interaction strengths, 
$G_0$ and $G_2$.
One may notice that they diverge near $D_{\rm min}=-42$ and 
$D_{\rm max}=42$.
This behavior indicates that, in the multi-$O(4)$ model 
under consideration, it becomes harder and harder to increase $D$ 
as approaching its limit. 
Let us then focus our attention on the middle region of $D$.
This is the region important for the quantum-mechanical tunneling motion 
through the barrier (small $G_0/\chi$ case)
and for the vibrational motion about the spherical equilibrium
(large $G_0/\chi$ case).
It is seen that, when the quadrupole-pairing interaction is absent ($G_2=0$), 
the magnitudes of the ASCC mass are almost the same as those of
the CHB-cranking mass. 
They also exhibit similar deformation dependence. 
When the quadrupole-pairing interaction is switched on, however, 
a significant difference between the ASCC and the CHB-cranking mass appears: 
The ASCC mass significantly increases as $G_2$ increases, 
while the CHB-cranking mass hardly changes.

The origin of this different behavior between $M(D(q))$ and $M_{\rm cr}(D)$
may be understood in the following way:
The CHB-cranking mass is derived by neglecting the contribution
of the residual interaction to the collective mass,
so that effects of the quadrupole-pairing interaction 
are taken into account only in the static quantities like 
the quadrupole-pairing gap $\Delta_2(D)$. 
On the other hand, the effects of the residual interaction on dynamics 
are taken into account in the ASCC method through the time-odd components
of the mean-field that change sign under the time reversal $p \to -p$. 
The time-odd part of the mean-field Hamiltonian  $\hhat(t)$  
represents the change in the self-consistent mean-field associated with
the dynamical motion. 
It is well known that this part is indispensable for reproducing 
the correct center of mass for the translational motions\cite{bar78}. 
On the other hand, it is also known that both 
the monopole-pairing and the quadrupole particle-hole interactions 
do not contribute to the time-odd part\cite{bar78,dob81,bel65}.
Thus, only the quadrupole-pairing interaction contributes to it 
in the case of the multi-$O(4)$ model Hamiltonian under consideration.

The remarkable difference between the ASCC mass and the CHB-cranking mass 
displayed in Fig.~\ref{fig:M} clearly indicates the importance of the 
quadrupole-pairing interaction on collective dynamics.
This shows a striking contrast to its effect on static properties.
Namely, the major properties of the collective potential energy curve 
$V(D)$ are determined by the competition between the monopole-pairing 
and quadrupole particle-hole correlations.
There, $\Delta_2(D)$ is much smaller than $\Delta_0(D)$, and
the quadrupole pairing plays only a minor role.
However, it turns out to play a major role
for dynamical properties of collective motion.

\subsection{Excitation spectra and transition matrix elements}

We calculate the eigen-energies $E_k$ and wave functions $\Psi_k(q)$ 
for the quantized large-amplitude collective motion 
by solving the collective Schr\"odinger equation:
\begin{align}
 \left( - \frac{1}{2}\frac{\del^2}{\del q^2} + V(q)\right ) \Psi_k(q) 
 = E_k \Psi_k(q),
\end{align}
with the orthonormalization 
\begin{align}
 \int_{q_{\rm min}}^{q_{\rm max}} \Psi^\ast_k(q)\Psi_l(q) dq = \delta_{kl},
\end{align}
and the boundary conditions, $\Psi_k(q_{\rm min})=\Psi_k(q_{\rm max})=0$,
where $q_{\rm min}$ and $q_{\rm max}$ are the minimum and maximum values of 
$q$ along the collective path obtained (see Fig. 3).
The quadrupole transition matrix elements are evaluated by 
\begin{align}
 \bra{\Psi_k}\Dhat\ket{\Psi_l} = 
 \int_{q_{\rm min}}^{q_{\rm max}} \Psi^\ast_k(q)D(q)\Psi_l(q) dq,
\end{align}
where the deformation $D(q)$ on the collective path is defined 
by Eq. (\ref{eq:deformation}).

\begin{figure}[htbp]
\begin{center}
\begin{tabular}{c}
\includegraphics[width=100mm]{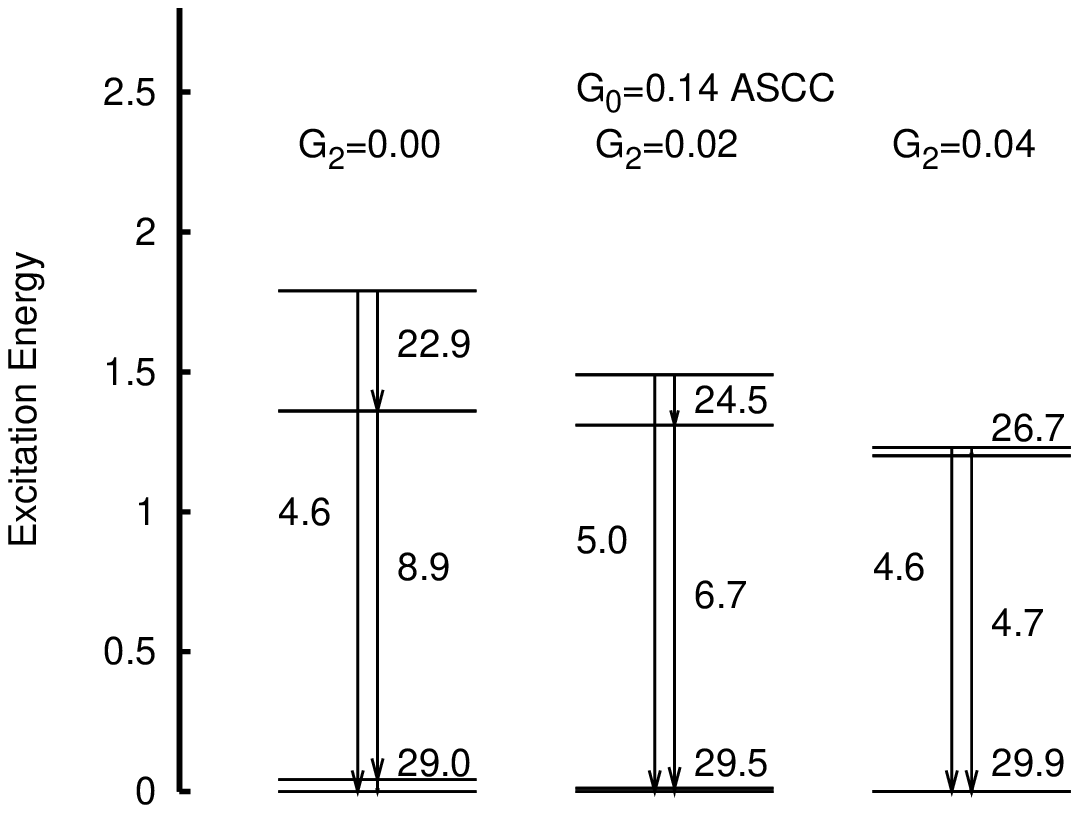} \\
\includegraphics[width=100mm]{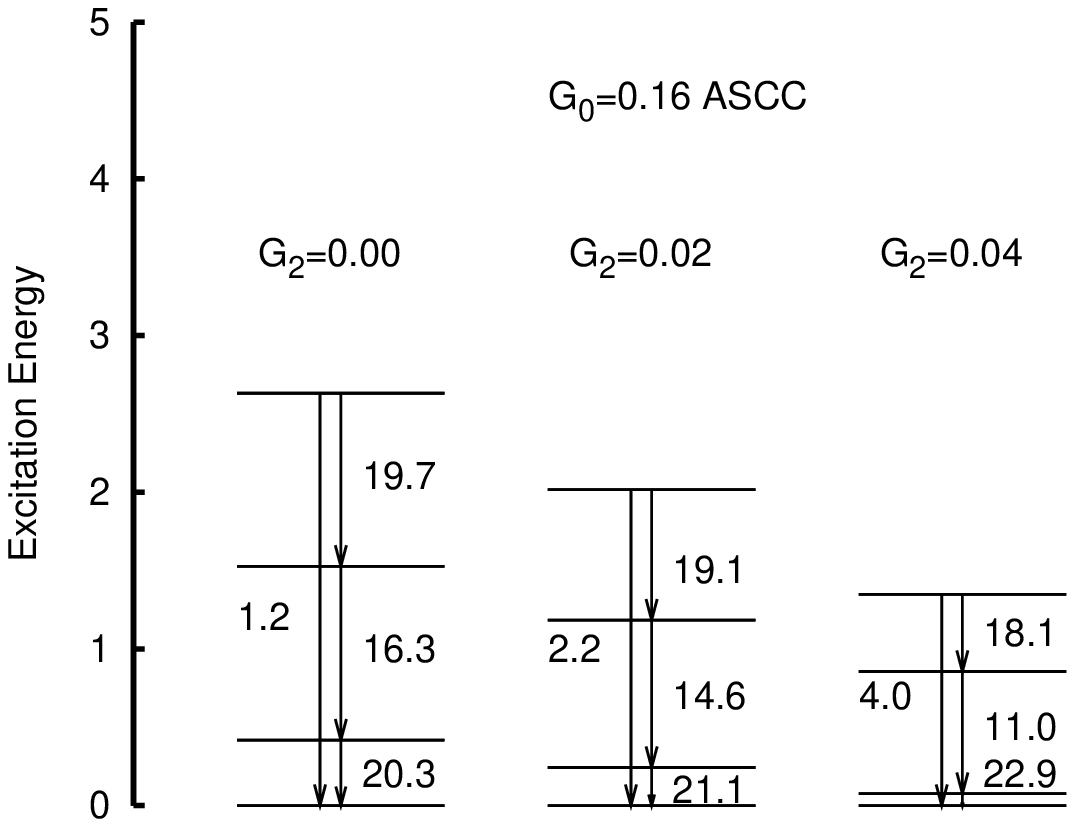} \\
\includegraphics[width=100mm]{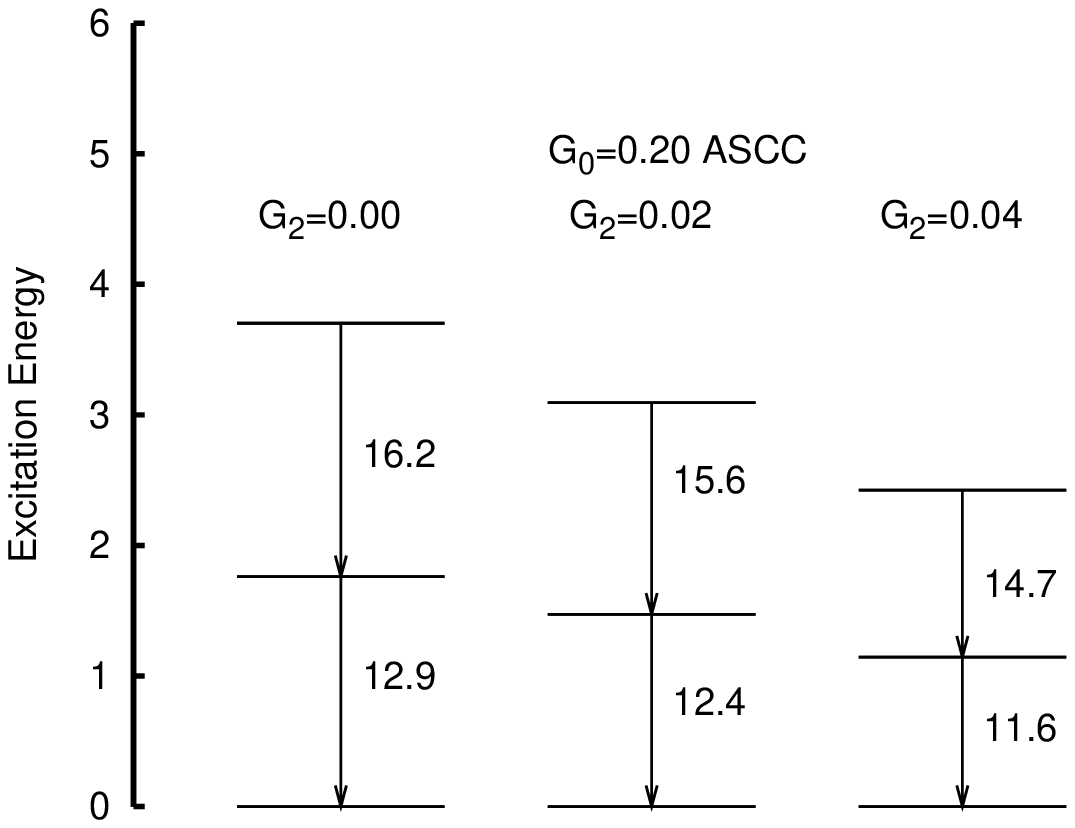} \\
\end{tabular}
\end{center}
\caption{Excitation spectra calculated with the ASCC method.
The upper, middle and lower rows display the results for
$G_0$ = 0.14, 0.16, and 0.20, respectively.
In each row, the results for $G_2= 0.00, 0.02, 0.04$ are compared.
The numbers adjacent to vertical lines indicate absolute values of 
the transition matrix elements.
The matrix elements between the doublets are indicated beside them.
}
\label{fig:spectra.ascc}
\end{figure}

Figure~\ref{fig:spectra.ascc} displays the results of the ASCC calculation 
for excitation spectra and quadrupole transition matrix elements 
between low-lying states.
For $G_0=0.20$ we obtain anharmonic vibrational spectra 
about the spherical equilibrium. 
In contrast, the collective potential $V(D)$ for $G_0=0.14$ and 0.16 
possesses two local minima corresponding to the oblate and prolate shapes, 
and the spherical point $(D=0)$ becomes the barrier top. 
In the $G_0=0.14$ case, this barrier is high. Consequently, 
a ground state doublet similar to the well-known parity doublet 
in the double well potential appears.
In the present multi-$O(4)$ model, the doublet corresponds to the 
symmetric and anti-symmetric superpositions of the oblate and prolate 
ground states, and its energy splitting provides a sensitive measure
of the quantum tunneling effect through the potential barrier. 
On the other hand, in the $G_0=0.16$ case, the barrier is rather low 
so that the spectrum exhibits a transient feature toward the doublet 
pattern mentioned above.  
In the quantum spectra of Fig.~\ref{fig:spectra.ascc},
we can clearly identify the effects of the increase of the collective mass 
due to the quadrupole pairing. 
First, the vibrational excitation energy decreases as $G_2$ increases. 
Second, the energy splitting of the doublet decreases 
with increasing $G_2$ indicating that 
the tunneling motion becomes harder as the collective mass increases. 
Thus, for the combination of $G_0=0.14$ and $G_2=0.04$,
we obtain a doublet of excited states 
in addition to the ground-state doublet.
The wave functions of the excited-state doublet as well as the ground-state 
doublet are displayed in Fig.~9.
This figure clearly indicates that the excited doublet corresponds 
to the symmetric and anti-symmetric linear combinations of 
vibrational excitations about the oblate and prolate local minima.
We can also confirm that the amplitude in the barrier region 
indeed decreases with increasing $G_2$.

\begin{figure}[htbp]
\begin{center}
\begin{tabular}{c}
\includegraphics[width=100mm]{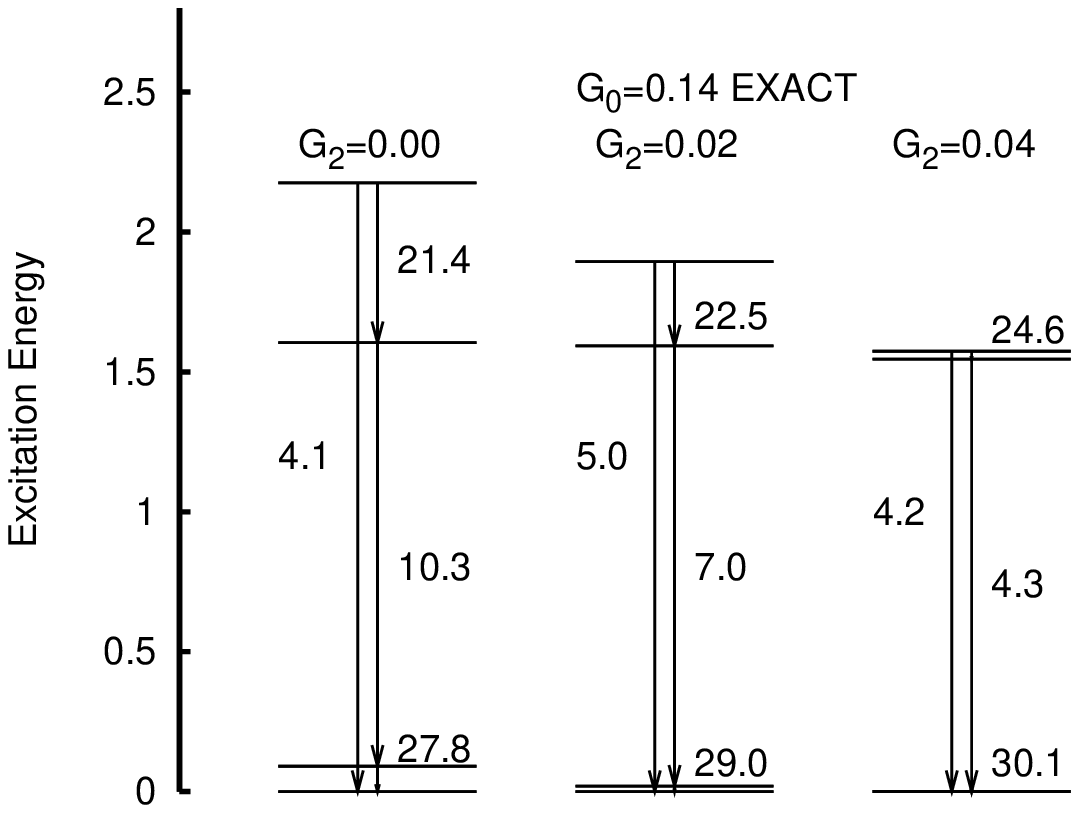} \\
\includegraphics[width=100mm]{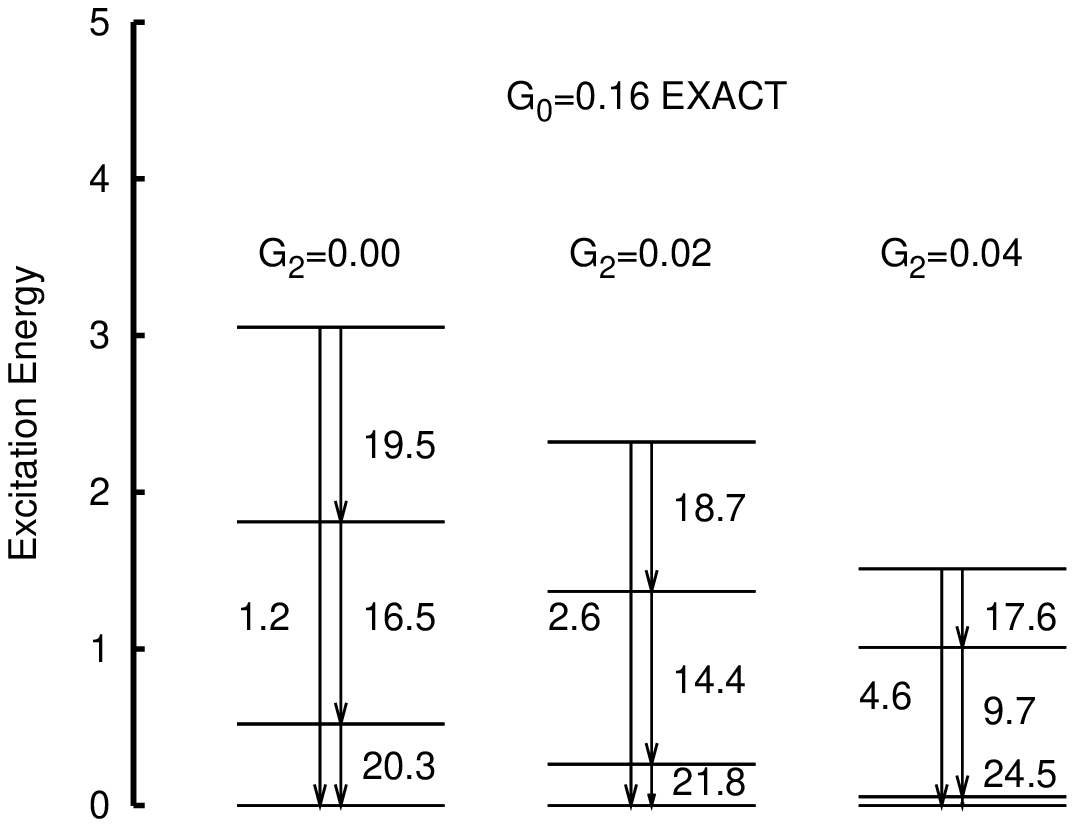} \\
\includegraphics[width=100mm]{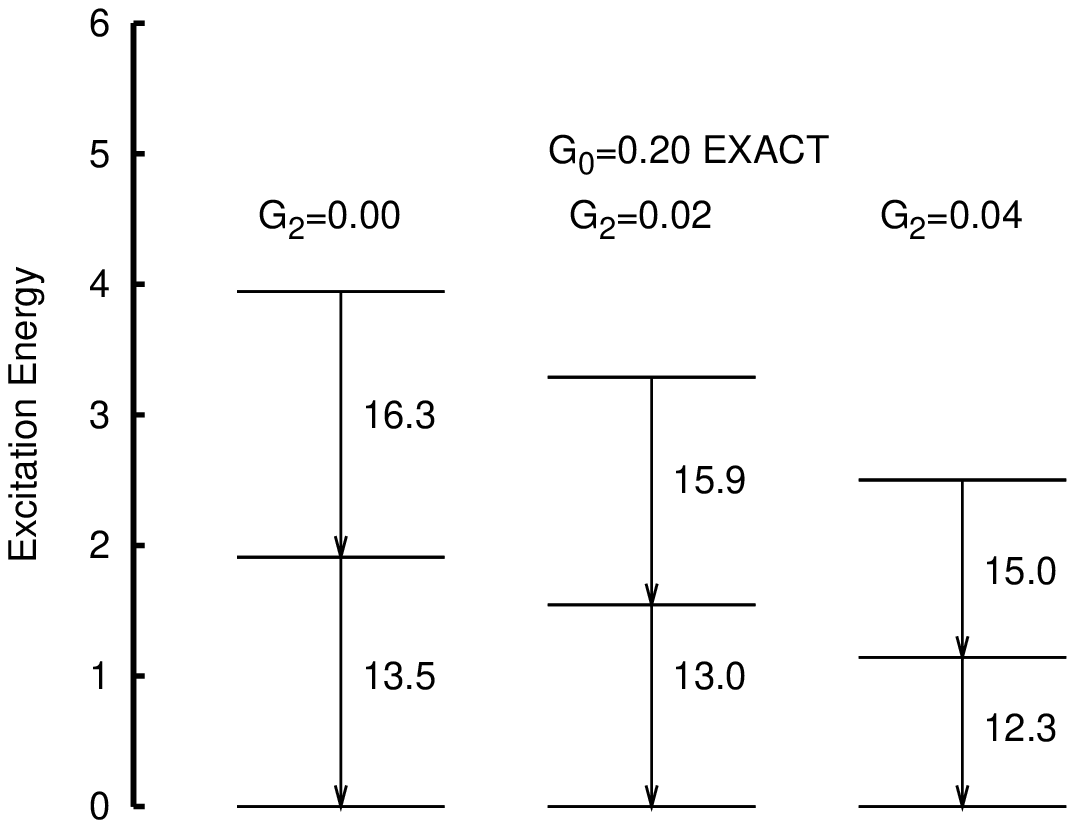} \\
\end{tabular}
\end{center}
\caption{Excitation spectra calculated with the exact diagonalization.
See the caption of Fig.~\ref{fig:spectra.ascc}.
}
\label{fig:spectra.exact}
\end{figure}

In Fig.~\ref{fig:spectra.exact}, 
the results of the exact matrix diagonalization 
of the microscopic multi-$O(4)$ Hamiltonian are presented. 
For every combinations of the interaction strengths, $G_0, G_2$ and $\chi$,
the excitation spectra and the transition matrix elements 
obtained by the ASCC method agree in a very good approximation 
with the results of the exact calculation.
It is remarkable that the ASCC calculation succeeds in describing the 
gradual change of the quantum spectra associated with the 
phase transition of the finite system from the spherical shape to 
the oblate-prolate shape coexistence.

Let us make a severe quantitative comparison concerning the energy 
splitting of the ground-state doublet in the $G_0=0.14$ case.
The splittings obtained in the ASCC method are 
0.043, 0.012, and 5$\times 10^{-4}$ for
$G_2=0.00, 0.02$, and $0.04$, respectively.
The corresponding values obtained by the exact diagonalization 
are 0.091, 0.020, and 3$\times 10^{-4}$.
It should be noted here that the energy splitting under discussion 
is a very small quantity associated with the barrier penetration for which 
even a slight error in the collective mass will result in an error of 
the order of magnitude. Therefore, the agreement within the factor of two 
indicates that the collective mass evaluated by the ASCC method is very
reliable. It should also be emphasized that this tunneling motion
is a large-amplitude collective motion associated with the major 
rearrangement of microscopic configurations of many particles and that
the collective mass represents the inertia of this motion of many-body system 
as a whole. 
Thus, the correct evaluation of the collective mass is a highly non-trivial
task.

\begin{figure}[htbp]
\begin{center}
\begin{tabular}{c}
\includegraphics[width=100mm]{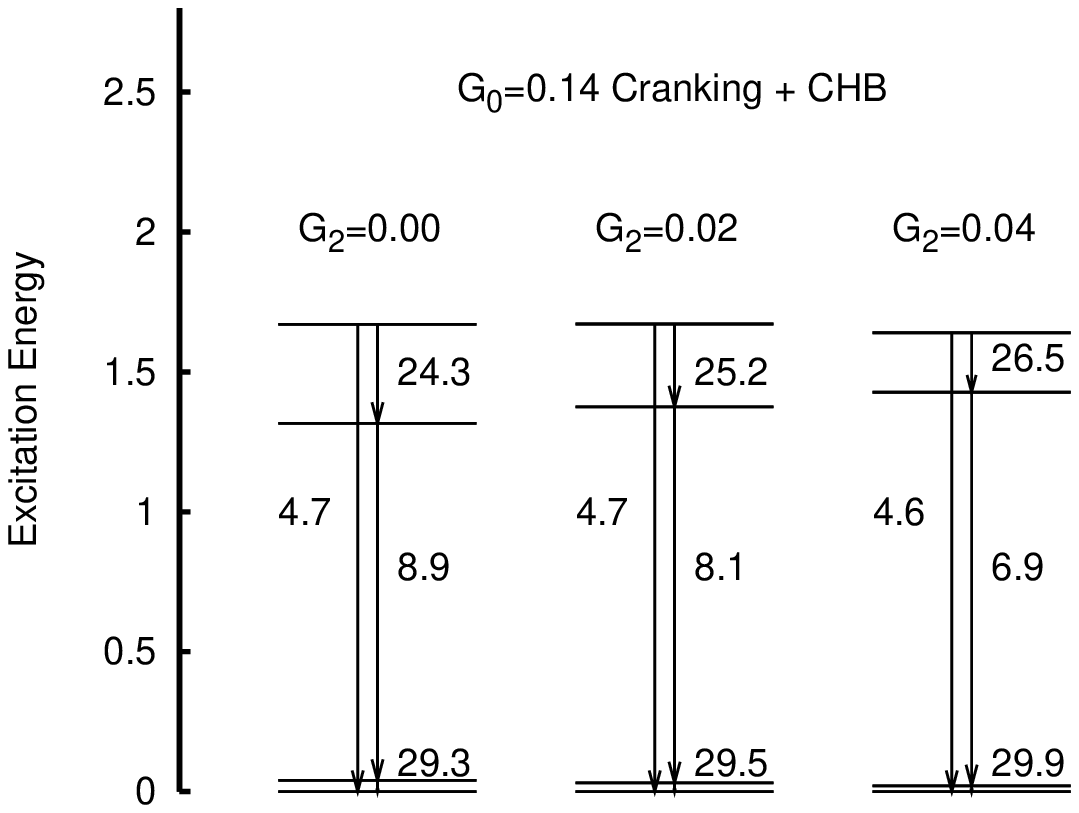} \\
\includegraphics[width=100mm]{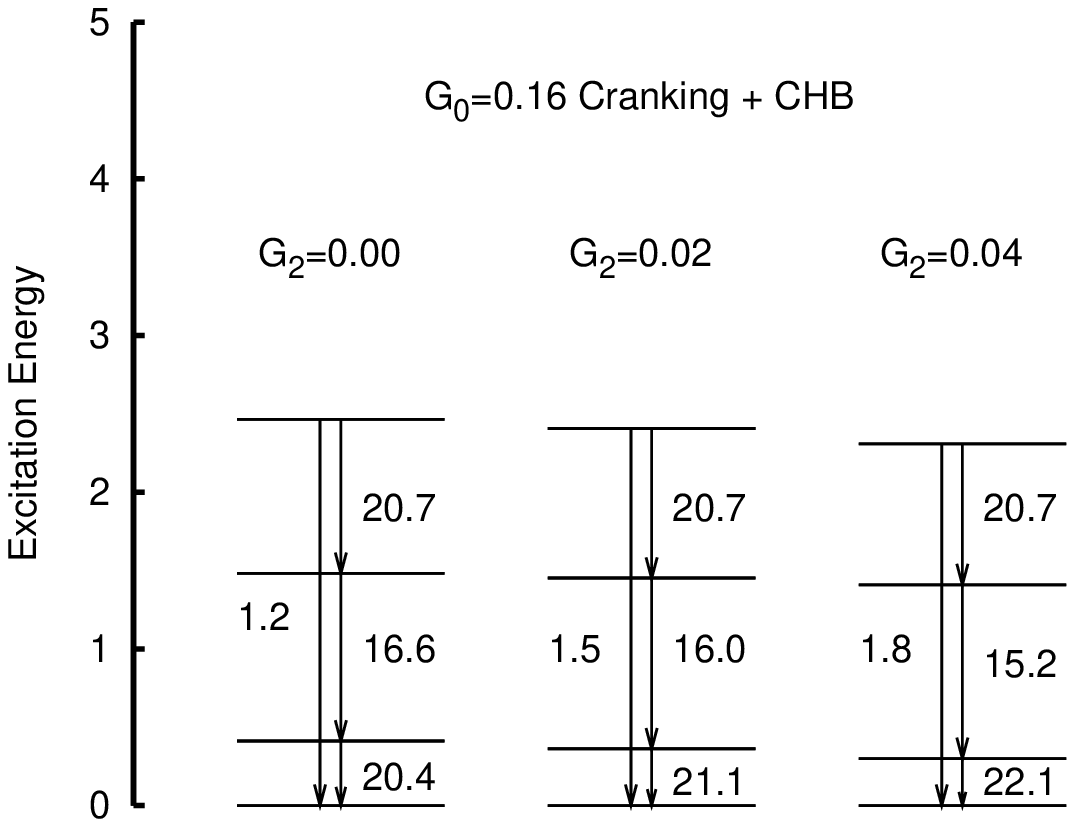} \\
\includegraphics[width=100mm]{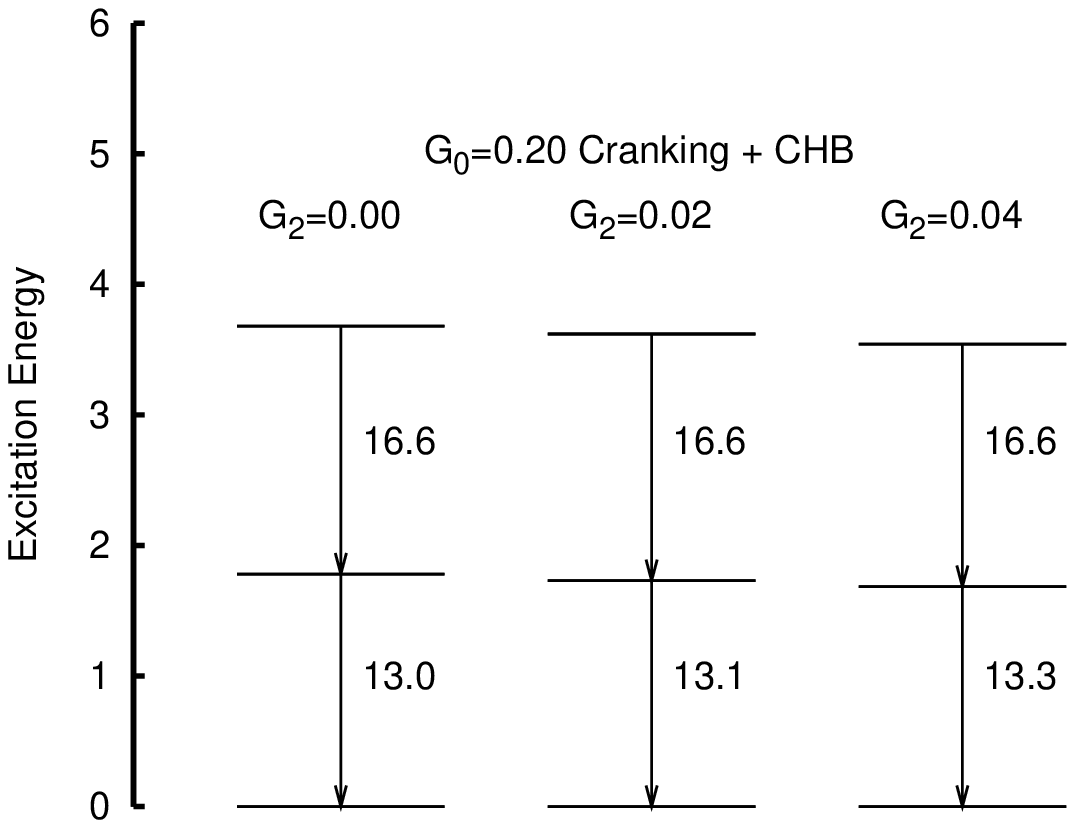} \\
\end{tabular}
\end{center}
\caption{Excitation spectra calculated with the CHB-cranking procedure.
See the caption of Fig.~\ref{fig:spectra.ascc}.
}
\label{fig:spectra.crank}
\end{figure}

\begin{figure}[htbp]
\begin{center}
\begin{tabular}{cc}
\includegraphics[width=70mm]{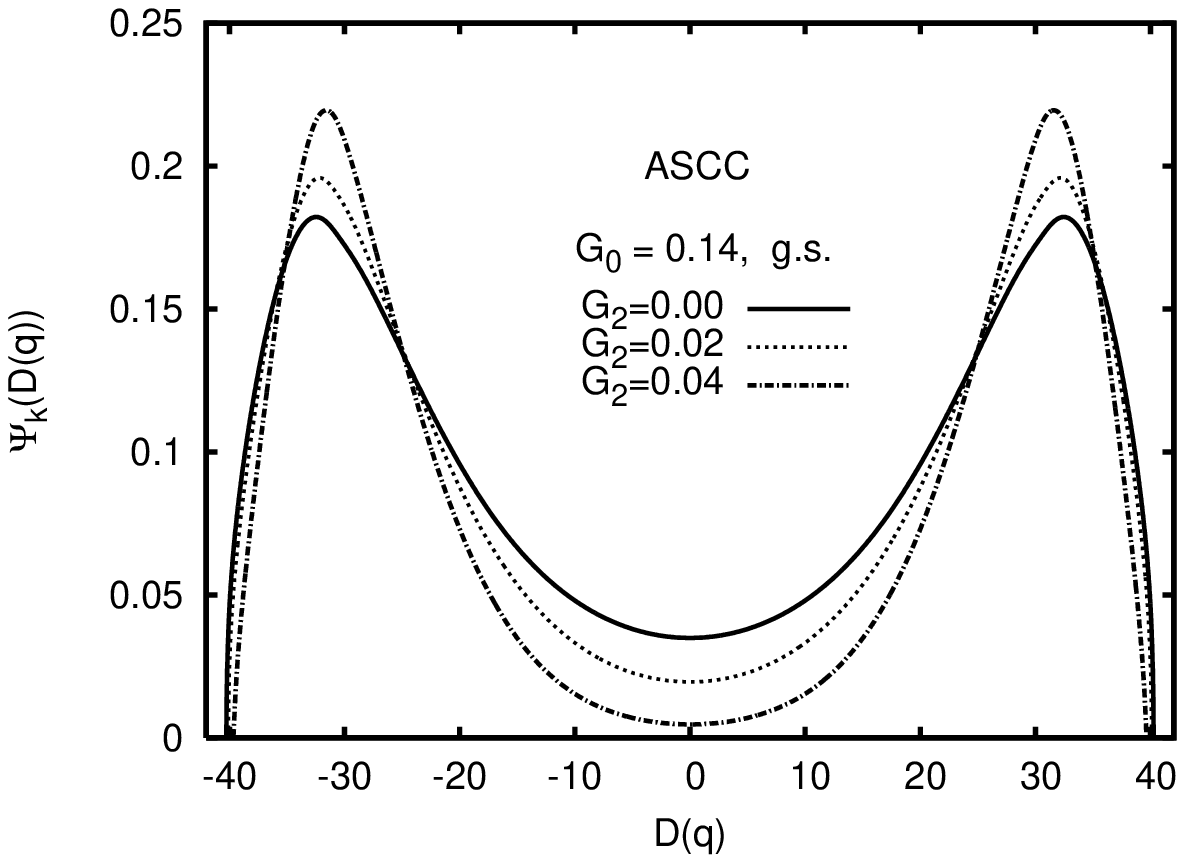} &
\includegraphics[width=70mm]{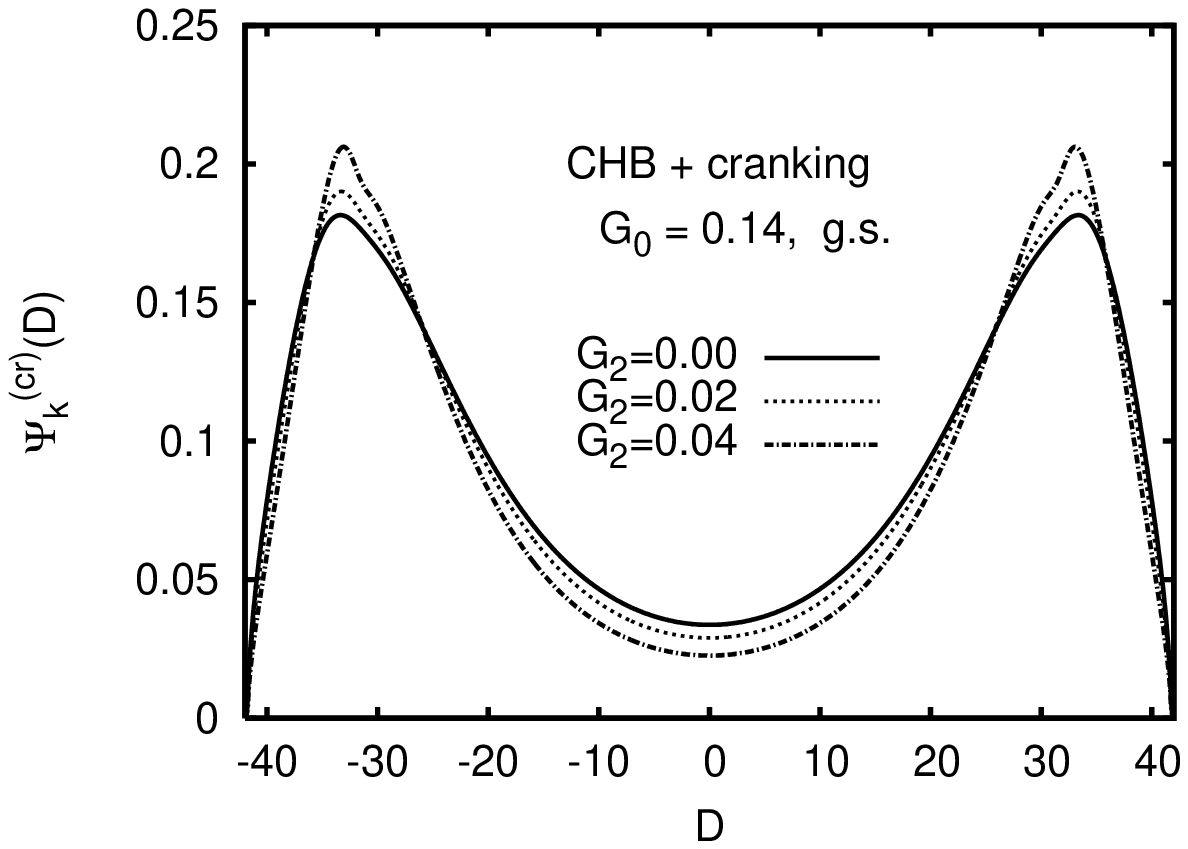} \\
\includegraphics[width=70mm]{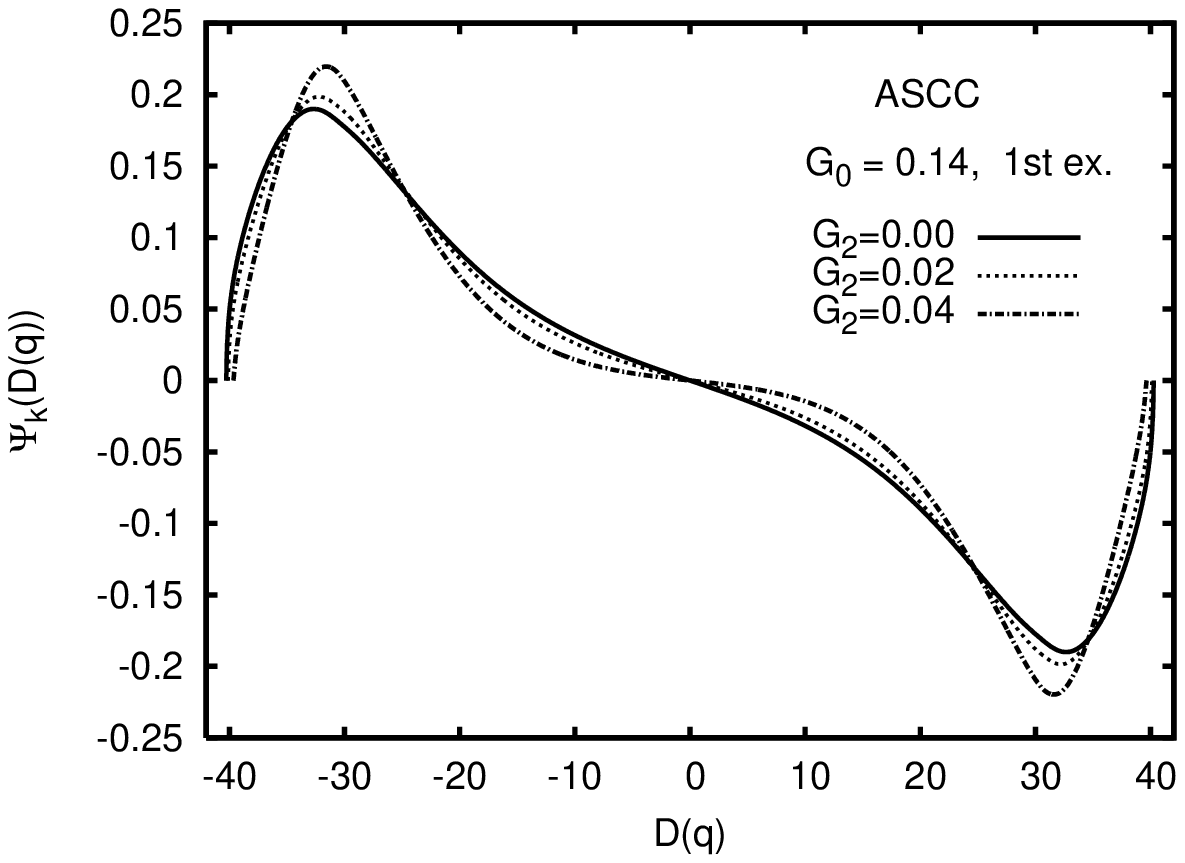} &
\includegraphics[width=70mm]{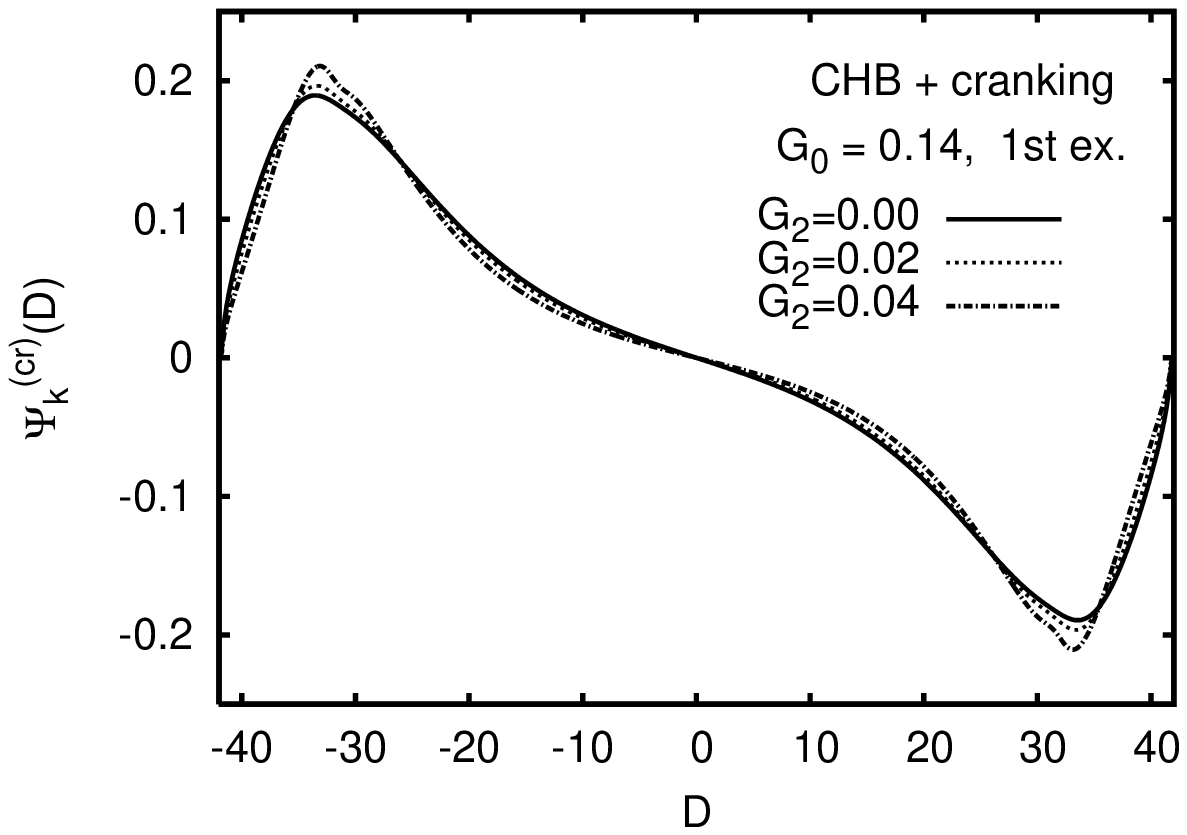} \\
\includegraphics[width=70mm]{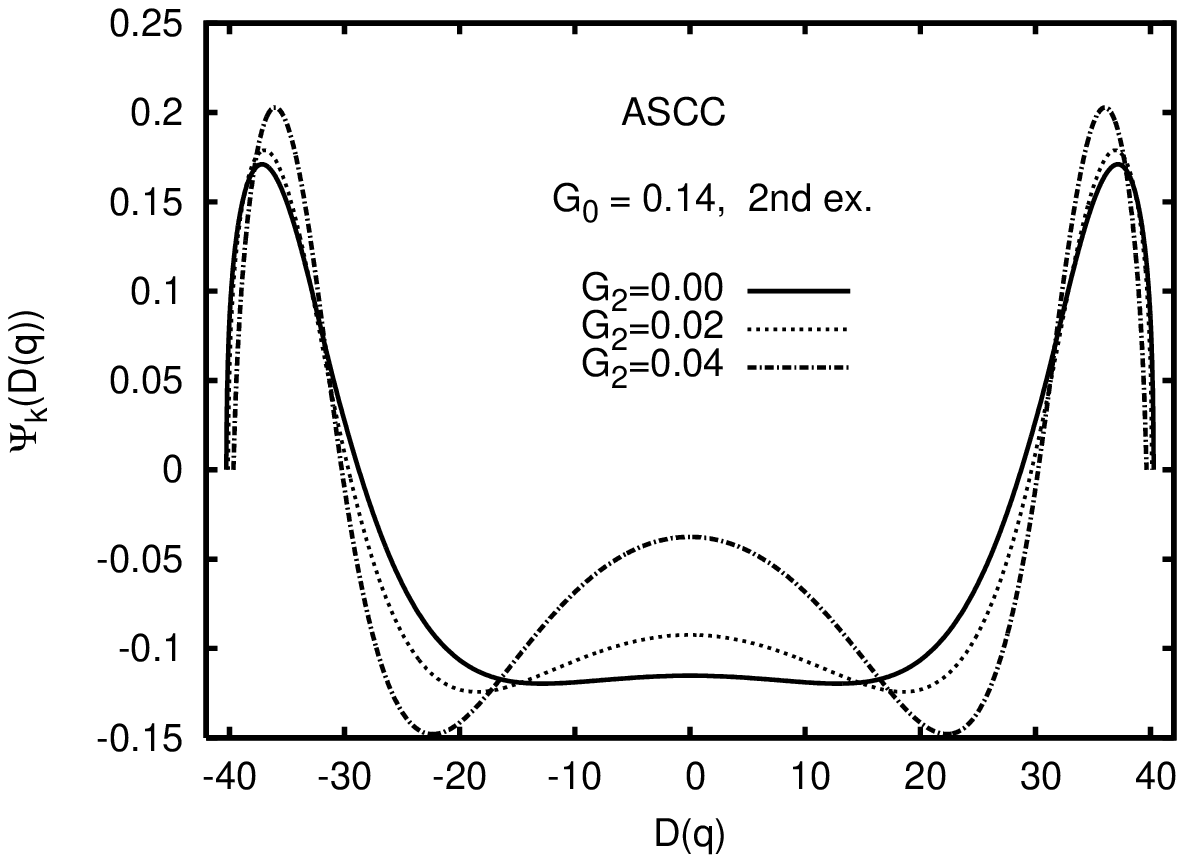} &
\includegraphics[width=70mm]{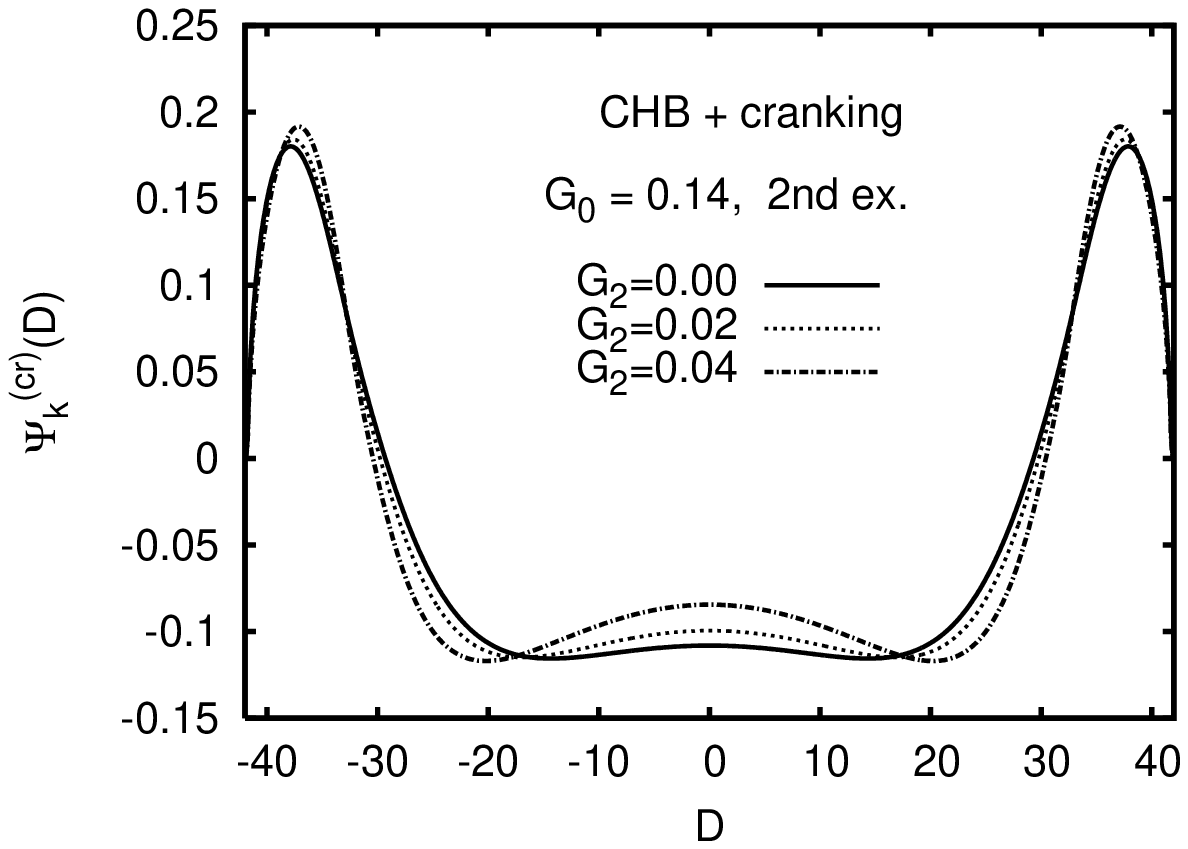} \\
\includegraphics[width=70mm]{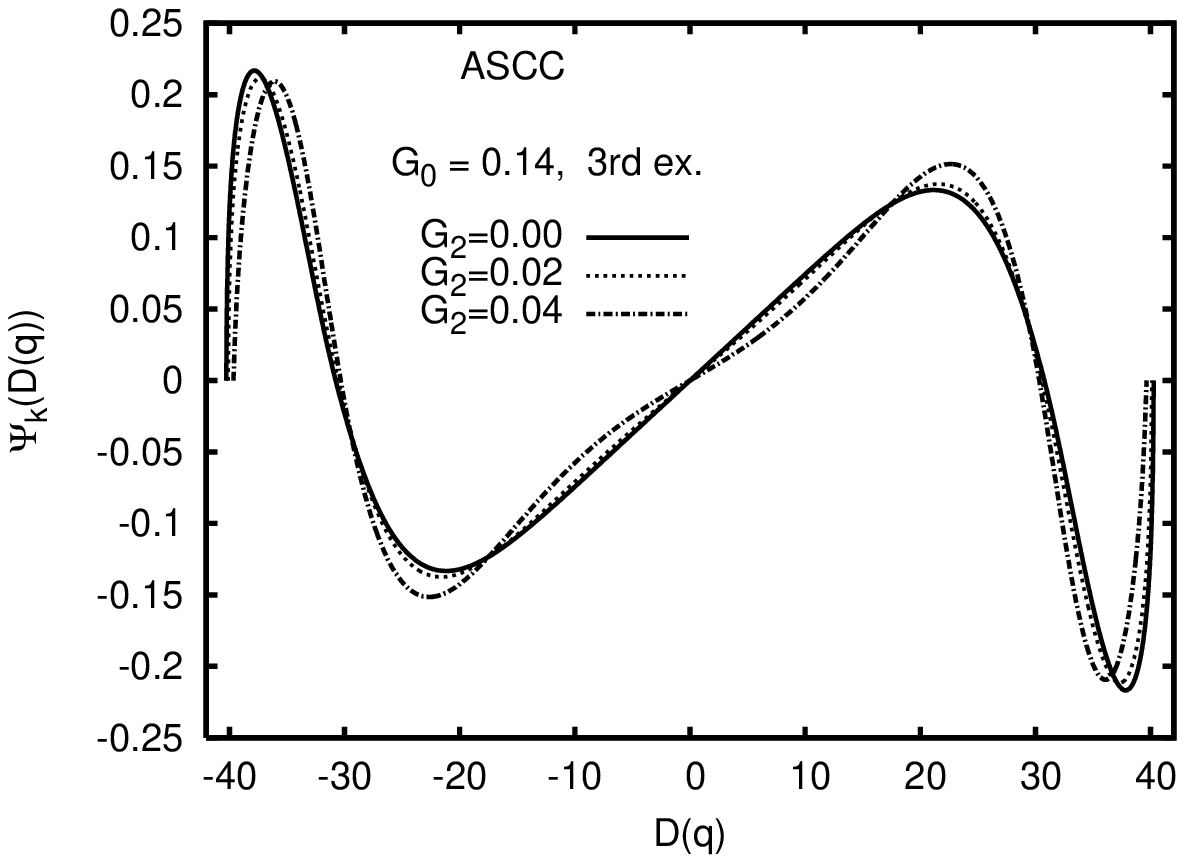} &
\includegraphics[width=70mm]{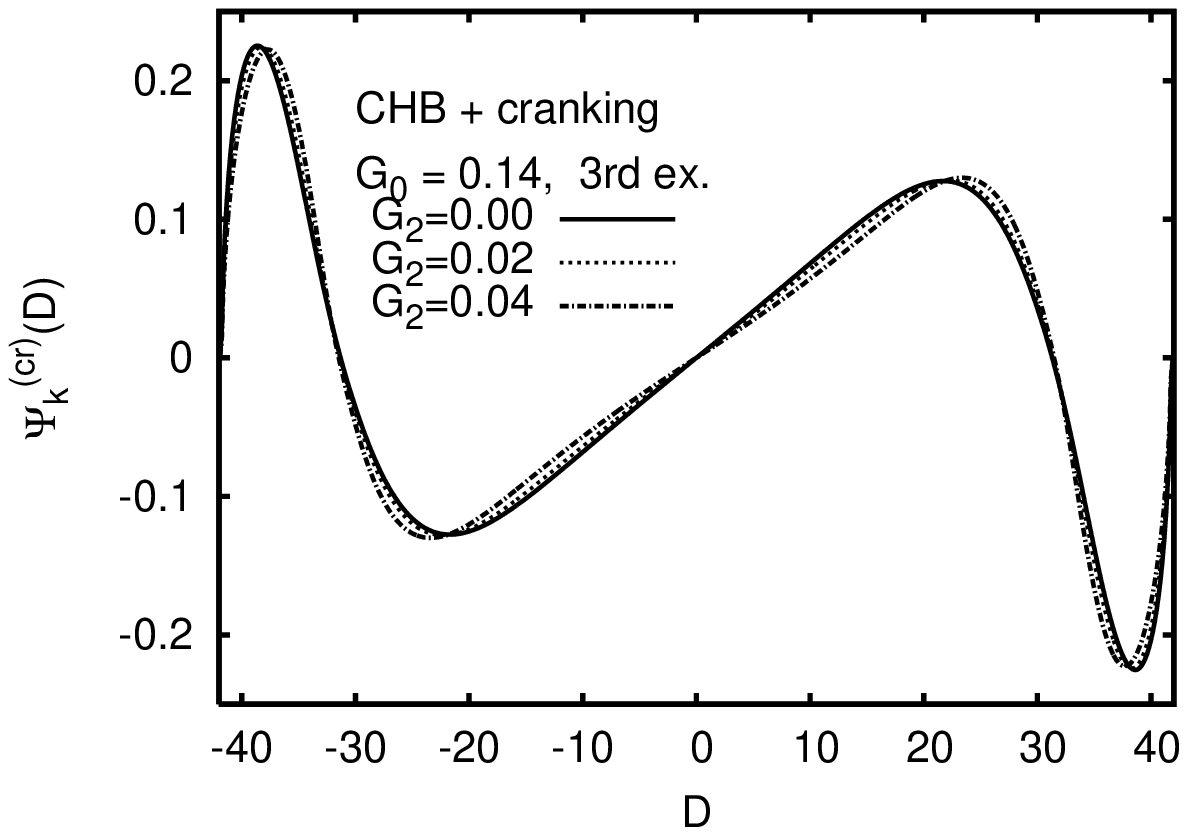} 
\end{tabular}
\end{center}
\caption{
Wave functions of eigenstates of the collective Schr\"odinger equation,
Eq. (\ref{eq:collective-Schroedinger-eq}),
for $G_0=0.14$.
Those obtained with the ASCC (CHB-cranking) method are shown
in the left (right).
In order to make a comparison,
they are plotted as functions of a common parameter $D$.
Namely, those of the ASCC method are defined by  
$\Psi_k(D(q)) \equiv \Psi_k(q) \sqrt{dq/dD} = \Psi_k(q) M(D(q))^{1/4}$
and normalized as $\int |\Psi_k(D)|^2 dD =1$.
The first, second, third and fourth rows display those 
of the ground state, the first, second, and third excited states,
respectively. 
In each panel, the results for $G_2= 0.00, 0.02, 0.04$ are compared.
}
\label{fig:wave}
\end{figure}

Let us now investigate how the difference between the ASCC mass
and the CHB-cranking mass discussed in the previous subsection  
affects the excitation spectra and transition matrix elements.
Adopting the Pauli quantization prescription, 
we obtain the collective Schr\"odinger equation 
\begin{align}
\label{eq:collective-Schroedinger-eq}
 \left(- \frac{1}{2M_{\rm cr}(D)^{1/4}}
         \frac{\del}{\del D}
         \frac{1}{\sqrt{M_{\rm cr}(D)}}
         \frac{\del}{\del D} 
         \frac{1}{M_{\rm cr}(D)^{1/4}}
       + V_{\rm CHB}(D)\right)
       \Psi_k^{\rm (cr)}(D) = E_k^{\rm (cr)} \Psi_k^{\rm (cr)}(D),
\end{align}
for the wave functions $\Psi_k^{\rm (cr)}(D)$ 
that incorporate the metric factor $M_{\rm cr}(D)^{1/4}$ 
such that the orthonormalizations are given by\cite{eis70}
\begin{align}
 \int_{D_{\rm min}}^{D_{\rm max}} 
 \Psi_k^{\rm (cr)\ast}(D)\Psi_l^{\rm (cr)}(D) dD=\delta_{kl}.
\end{align}
We solve this Schr\"odinger equation under the boundary conditions, 
$\Psi_k^{\rm (cr)}(D_{\rm min})=\Psi_k^{\rm (cr)}(D_{\rm max})=0$.
The quadrupole transition matrix elements are evaluated by 
\begin{align}
 \bra{\Psi_k^{\rm (cr)}}\Dhat\ket{\Psi_l^{\rm (cr)}} = 
 \int_{D_{\rm min}}^{D_{\rm max}} 
 \Psi_k^{{\rm (cr)}\ast}(D) D\Psi_l^{\rm (cr)}(D) dD.
\end{align} 
As mentioned in subsection \ref{sec:path},
the ASCC collective potential $V(q)$ almost coincides 
with the potential $V_{\rm CHB}(D)$ calculated by means of the 
CHB method, so that the major difference in the
quantum spectra can be attributed to the difference in the collective mass.
The results are displayed in Fig.~\ref{fig:spectra.crank}.
When the quadrupole pairing is absent ($G_2=0$), 
we see rather good agreements 
with the ASCC and with the exact solutions.
However, the excitation spectra and transition matrix elements are 
almost unchanged when 
the quadrupole-pairing interaction is switched on and $G_2$ increases. 
This is very different from the ASCC results and the exact solutions.
We can confirm this point also in the wave functions displayed 
in Fig.~\ref{fig:wave}.
The amplitudes in the barrier region change only little even when $G_2$ 
increases, in contrast to the ASCC wave functions.
It is obvious that this failure in taking into account the effect
of the quadrupole pairing originates from the fact that  
the CHB-cranking procedure ignores the time-odd mean-field contribution
to the collective mass.

\section{Conclusions}
\label{sec:Conclusions}

The multi-$O(4)$ model is a simple model to simulate the phase transitions 
in finite quantum systems from the spherical shape to the oblate-prolate 
shape coexistence.
We have applied the ASCC method to this model and studied the collective mass 
of the many-body tunneling motion through the potential barrier between the
oblate and prolate local minima. 
Comparing with the results of exact diagonalization,
we have shown that the ASCC method succeeds in describing the gradual
change of the excitation spectra from the anharmonic vibration about the
spherical equilibrium to the doublet pattern associated 
with the deformed double-well potential possessing the oblate-prolate symmetry.
It was found that the collective mass significantly increases
due to the quadrupole-pairing contribution to the time-odd component of
the moving mean field.
We have also shown that the CHB-cranking procedure 
underestimates the collective mass, because the contribution from 
the time-odd component is disregarded there.

Along the approach developed in this paper,
we shall evaluate in a forthcoming paper\cite{hin06} 
the contribution of the time-odd components to the collective mass of 
large amplitude collective motion associated with the oblate-prolate shape 
coexistence in the $^{68}$Se and $^{72}$Kr region.

\section*{Acknowledgements}

This work is supported by 
the Grant-in-Aid for the 21st Century COE ``Center for Diversity and
Universality in Physics" from the Ministry of Education, Culture, Sports,
Science and Technology (MEXT) of Japan and also 
by the Grant-in-Aid  for Scientific Research (Nos. 16540249, 17540231
and 17540244) from the Japan Society for the Promotion of Science.
We thank the Yukawa Institute for Theoretical Physics at
Kyoto University: Discussions during the YITP workshop YITP-W-05-01 on
"New Developments in Nuclear Self-Consistent Mean-Field Theories" were
useful to complete this work.
We also thank the Institute for Nuclear Theory at the University of Washington
for its hospitality and the Department of Energy for partial support during 
the completion of this work.


\appendix


\section{Exact diagonalization of the multi-$O(4)$ model Hamiltonian}

It is possible to construct the following two sets of $SU(2)$ generators
for each $j$-shell from the basic operators of the multi-$O(4)$ model:
\begin{align}
 \Kp_{j}&=\frac{1}{2}(\Ahatdag_j + \Bdag_j), &\quad
 \Lp_{j}&=\frac{1}{2}(\Ahatdag_j - \Bdag_j),\\
 \Km_{j}&=\frac{1}{2}(\Ahat_j + \Bhat_j), &\quad
 \Lm_{j}&=\frac{1}{2}(\Ahat_j - \Bhat_j),\\
 \Kz_{j}&=\frac{1}{2}(\Nhat_j + \Dhat_j -\Omega_j), &\quad
 \Lz_{j}&=\frac{1}{2}(\Nhat_j - \Dhat_j -\Omega_j).
\end{align}
The operators $\{K_{j+},K_{j-},K_{j0}\}$ and $\{L_{j+},L_{j-},L_{j0}\}$
satisfy the commutation relations of the $SU(2)$ algebra:
\begin{align}
 [\Kp_j, \Km_{j'}] &= 2\Kz_j \delta_{jj'}, &\quad
 [\Lp_j, \Lm_{j'}] &= 2\Lz_j \delta_{jj'}, \\
 [\Kz_j, K^{\pm}_{j'}] &= \pm K^{\pm}_j \delta_{j\jp}, &\quad
 [\Lz_j, L^{\pm}_{j'}] &= \pm L^{\pm}_j \delta_{j\jp}.
\end{align}
All commutation relations between $K$'s and $L$'s are zero.
Thus, we can construct the orthogonal basis vectors of the model space 
as
\begin{align}
| \nKb,\nLb \rangle \equiv
\prod_j | n_{K_j}, n_{L_j} \rangle \equiv
\prod_j (\Kp_{j})^{n_{K_j}} (\Lp_{j})^{n_{L_j}} | 0 \rangle ,
\label{eq:o4base}
\end{align}
where ($n_{K_j}, n_{L_j}$) are integer numbers satisfying
$0 \le n_{K_j}, n_{L_j}\le \Omega_j / 2$ and
$ \sum_j (n_{K_j}+ n_{L_j}) = N_0 / 2$.

In terms of these $SU(2)$ generators,
the multi-$O(4)$ Hamiltonian is expressed as
 \begin{multline}
  \Hhat = \sum_j e^0_j \{2(\Kz_j + \Lz_j) +  \Omega_j \} - 2\chi\sum_{ij}
  d_i d_j (\Kz_i - \Lz_i)(\Kz_j - \Lz_j) \\
  - \frac{1}{2}\sum_{ij}(G_0+G_2 d_i d_j)(
  \Kp_i\Km_j + \Kp_i\Km_j + \Lp_i\Lm_j + \Lm_i\Lp_j ) \\
  -\frac{1}{2}\sum_{ij}(G_0-G_2 d_i
  d_j)(\Km_i\Lp_j+\Kp_i\Lm_j+\Lm_i\Kp_j+\Lp_i\Km_j).
\label{eq:HinKL}
 \end{multline}
The operators, $\Kp_{j}, \Km_{j}$ and  $\Kz_{j}$, 
act on the basis $\ket{n_{K_j},n_{L_j}}$ as
\begin{subequations}
\begin{align}
 \Kz_j\ket{\nK{j},\nL{j}} &= \left(\nK{j} -
 \frac{\Omega_j}{4}\right)\ket{\nK{j},\nL{j}}, \\
 \Kp_j\ket{\nK{j},\nL{j}} &=
 \sqrt{(\nK{j}+1)\left(\frac{\Omega_j}{2}-\nK{j}\right)}\ket{\nK{j}+1, \nL{j}}, \\
 \Km_j\ket{\nK{j},\nL{j}} &=
 \sqrt{\nK{j}\left(\frac{\Omega_j}{2}-\nK{j}+1\right)}\ket{\nK{j}-1, \nL{j}}.
\end{align} 
\label{eq:koperator}
\end{subequations}
Similar equations hold for $\Lp_{j}, \Lm_{j}$, and $\Lz_{j}$.
The matrix elements of the multi-$O(4)$ Hamiltonian, 
$\bra{\nKb', \nLb'} \Hhat\ket{\nKb, \nLb}$,
can be calculated by using Eqs.~(\ref{eq:HinKL}) and (\ref{eq:koperator}).
Diagonalizing this matrix, we obtain exact eigen-energies and eigen-states. 
For the parameters given in the text,
the dimension of this Hamiltonian matrix is 1894. 
The quadrupole transition matrix elements between eigen-states,
$\ket{\phi_\alpha}$ and $\ket{\phi_\beta}$, are given by 
\begin{align}
 \bra{\phi_\alpha} \Dhat \ket{\phi_\beta} &= \sum_{\nKb, \nLb} 2 d_j (\nK{j} -
 \nL{j}) C^{\alpha\ast}_{\nKb, \nLb} C^{\beta}_{\nKb, \nLb},
\end{align}
where $C^{\alpha}_{\nKb,\nLb}$ are expansion coefficients
in the $SU(2)$ basis,
\begin{align}
 \ket{\phi_\alpha} &= \sum_{\nKb, \nLb} C^{\alpha}_{\nKb,\nLb}
 \ket{\nKb, \nLb}.
\end{align}

\section{Gradient method for the moving-frame HB equation}
\label{sec:grad}

We solve the variational equation of the following form 
using the gradient method:\cite{rin80} 
\begin{align}
 \delta \bra{\phi(q)} \Hhat - \lambda \Nhat - \mu \Qhat\ket{\phi(q)} = 0,
\end{align}
with the constraint conditions for the number operator $\Nhat$ 
and a one-body operators $\Rhat$,
\begin{align}
 \bra{\phi(q)}\Nhat\ket{\phi(q)} = N_0 \quad  
 \bra{\phi(q)}\Rhat\ket{\phi(q)} &= R_0.
 \label{eq:grad-const}
\end{align}
Here, $\Rhat$ is an arbitrary one-body operator which, in general, 
may be different from $\Qhat$.

Let $\ket{\phi^{(k)}(q)}$ be the state vector at the iterative step $k$.
Using the quasiparticle bilinear operators,
$\Abdag_i(q)$ and $\Ab_i(q)$ which satisfy $\Ab_i(q)\ket{\phi^{(k)}(q)}=0$,
we then generate the state vector at the $(k+1)$-th step 
in a form of unitary transform of $\ket{\phi^{(k)}(q)}$
as follows: 
\begin{align}
 \ket{\phi^{(k+1)}(q)} = e^{\Zhat(q)} \ket{\phi^{(k)}(q)}
\end{align}
with the anti-Hermitian operator 
\begin{align}
 \Zhat(q) = \sum_{i>0} Z^{20}_i(q)(\Abdag_i(q) - \Ab_i(q)).
\end{align}
It should be noted that
the normalization is preserved during the iteration.
Assuming that $\Zhat(q)$ is small, we expand the energy difference between 
$\ket{\phi^{(k+1)}(q)}$ and $\ket{\phi^{(k)}(q)}$ as follows:
\begin{align}
 \Delta E &= \bra{\phi^{(k+1)}(q)}\Hhat - \lambda\Nhat -
 \mu\Qhat\ket{\phi^{(k+1)}(q)}
-  \bra{\phi^{(k)}(q)}\Hhat - \lambda\Nhat -
 \mu\Qhat\ket{\phi^{(k)}(q)} \\
 &= \bra{\phi^{(k)}(q)}[\Hhat-\lambda\Nhat-\mu\Qhat,\Zhat(q)]\ket{\phi^{(k)}(q)}
 + O(\Zhat^2) \\
 &= \sum_{i>0} (H^{20}_{i}(q)-\lambda N^{20}_{i}(q)-\mu Q^{20}_{i}(q))
 Z^{20}_{i}(q) + O(\Zhat^2).
\end{align}
If $Z^{20}_i(q)$ is chosen as
\begin{align}
 Z^{20}_i(q) = - \Delta T (H^{20}_{i}(q)-\lambda N^{20}_i(q) -\mu Q^{20}_i(q)),
\label{eq:grad_Z}
\end{align}
with a positive step size $\Delta T$, 
$\Delta E$ is negative in each iteration.
The constraint conditions (\ref{eq:grad-const}) can also be expanded 
up to the first order in $\Zhat(q)$,
\begin{subequations}
\label{eq:grad_const}
\begin{align}
 \bra{\phi^{(k+1)}(q)}\Rhat\ket{\phi^{(k+1)}(q)} =&
\bra{\phi^{(k)}(q)}\Rhat\ket{\phi^{(k)}(q)} +
 \bra{\phi^{(k)}(q)}[\Rhat,\Zhat(q)]\ket{\phi^{(k)}(q)}\nonumber\\
& + O(\Zhat^2), \\
 \bra{\phi^{(k+1)}(q)}\Nhat\ket{\phi^{(k+1)}(q)} =&
\bra{\phi^{(k)}(q)}\Nhat\ket{\phi^{(k)}(q)} +
 \bra{\phi^{(k)}(q)}[\Nhat,\Zhat(q)]\ket{\phi^{(k)}(q)} \nonumber\\
& + O(\Zhat^2).
\end{align}
\end{subequations}
Substituting Eq.~(\ref{eq:grad_Z}) into Eq.~(\ref{eq:grad_const}),
we obtain the following equation which determine
the Lagrange multipliers $\lambda$ and $\mu$:
\begin{align}
\begin{bmatrix}
 \displaystyle\sum_{i} R^{20}_{i} N^{20}_{i} &  
 \displaystyle\sum_{i} R^{20}_{i} Q^{20}_{i} \\ \\
 \displaystyle\sum_{i} N^{20}_{i} N^{20}_{i} & 
 \displaystyle\sum_{i} N^{20}_{i} Q^{20}_{i}
\end{bmatrix}
 \begin{bmatrix} \\
\ \ \lambda \ \ \\ \\
\ \ \mu     \ \ \\
\
\end{bmatrix}
 =
\begin{bmatrix}
\\
\displaystyle\frac{R_0 - \bar{R}}{2\Delta T} + \sum_{i} R^{20}_{i} H^{20}_{i} \\
\displaystyle\frac{N_0 - \bar{N}}{2\Delta T} + \sum_{i} N^{20}_{i} H^{20}_{i} \\
\end{bmatrix},
\label{eq:constraints}
\end{align}
where $N^{20}_{i}, R^{20}_{i}, Q^{20}_{i}$ and $H^{20}_{i}$ denote the
coefficients of the two-quasiparticle creation and annihilation parts of 
the operators $\Nhat, \Rhat, \Qhat$ and $\Hhat$, respectively, while
the quantities $\bar{R}$ and $\bar{N}$ represent
$\bra{\phi^{(k)}(q)}\Rhat\ket{\phi^{(k)}(q)}$ and 
$\bra{\phi^{(k)}(q)}\Nhat\ket{\phi^{(k)}(q)}$, respectively.

When solving the moving-frame HFB equation (\ref{eq:ascc1_sep}) 
with the constraint conditions, (\ref{eq:Ncond}) and (\ref{eq:Qcond}), 
we use $\Qhat(q-\delta q)$ for $\Rhat(q)$. 
This operator is defined at $q-\delta q$
in terms of the quasiparticle bilinear operators, 
$\Abdag_i(q-\delta q)$ and $\Ab_i(q-\delta q)$,
which satisfy $\Ab_i(q-\delta q)\ket{\phi(q-\delta q)}=0$.
Thus, at each iteration step, it is necessary to rewrite
$\Qhat(q-\delta q)$ in terms of
$\Abdag_i(q)$, $\Ab_i(q)$ and $\Nb_i(q)$, 
defined with respect to $\ket{\phi^{(k)}(q)}$ at $q$.
\begin{align}
 \Rhat(q) = \Qhat(q-\delta q) &= \sum_{i>0} 
 Q^{20}_i (q-\delta q)(\Abdag_i(q-\delta q)+\Ab_i(q-\delta q)) \\
 &= R^{00}(q) + \sum_{i>0}  R^{11}_i(q) \Nb_i(q) 
 + R^{20}_i(q)(\Abdag_i(q) + \Ab_i(q)).
\end{align}
The explicit expression for the coefficients $R^{20}_i(q)$, 
which we need in solving Eq.~(\ref{eq:constraints}), is 
\begin{align}
 R^{20}_i(q) = Q^{20}_i(q-\delta q) &\left[
(u_i(q-\delta q)^2 - v_i(q-\delta q)^2)(u_i^2(q) - v_i^2(q)) \right. \nonumber\\
&\left.\ +  4 u_i(q-\delta q) v_i(q-\delta q) u_i(q) v_i(q) \right].
\end{align}
The above procedure is repeated until convergence is attained.
In fact, this implements {\it Step 2} in the iterative algorithm 
described in subsection \ref{sec:algorithm}.

\newpage



\newpage

\newpage

\newpage

\newpage

\newpage

\newpage


\newpage

\newpage

\newpage

\end{document}